\documentclass{pinchcr}

\usepackage[dvips]{graphicx}
\usepackage{bbold}

\input amssym.def
\input amssym.tex

\title{Computational Strategies in Lattice QCD}

\author{Martin L\"uscher}
\affiliation{Physics Department,
             CERN, 1211 Geneva 23, Switzerland}
\authors{1}

\begin{document}

\maketitle

\preface

Numerical lattice QCD has seen many important innovations 
over the years. In this course an introduction to some of
the basic techniques is provided, emphasizing their
theoretical foundation rather than their implementation
and latest refinements.

The development of computational strategies in
lattice QCD requires physical insight to be combined with 
an understanding of modern numerical mathematics 
and of the capabilities of massively parallel computers. 
When a new method is proposed, it should ideally be 
accompanied by a theoretical analysis that
explains why it is expected to work out. However,
in view of the complexity of the matter,
some experimenting is often required.
The field thus retains a certain empirical character.

At present
numerical lattice QCD is still in a developing phase to some extent.
The baryon spectrum, for example, remains to be difficult to compute
reliably, because the signal-to-noise ratio of the associated
two-point functions decreases exponentially at large distances.
There is certainly ample room for improvements and it may
also be necessary to radically depart from the known techniques in
some cases. Hopefully these lectures will encourage some of the
problems to be studied and to be solved eventually.

\acknowledgements

I am indebted to the organizers for inviting me to lecture at this
Summer School and for providing a very pleasant and stimulating 
atmosphere. During my lectures many questions were asked that 
helped to clarify some subtle points.
I have rarely had such an
attentive audience and would like to thank the students for
patiently going with me through the rather technical material
covered in the course. Finally, I wish to thank the University of Grenoble,
the director of the school and the support staff for running this
wonderful place in a perfect manner.

\tableofcontents
\maintext



\def\rmd{{\rm d}}
\def\rmD{{\rm D}}
\def\rme{{\rm e}}
\def\rmO{{\rm O}}
\def\rmU{{\rm U}}
\def\rmo{{\rm o}}


\def\rz{{\Bbb R}}
\def\gz{{\Bbb Z}}
\def\nz{{\Bbb N}}
\def\Im{{\rm Im}\,}
\def\Re{{\rm Re}\,}


\def\defeq{\mathrel{\mathop=^{\rm def}}}
\def\proof{\noindent{\sl Proof:}\kern0.6em}
\def\endproof{\hskip0.6em plus0.1em minus0.1em
\setbox0=\null\ht0=5.4pt\dp0=1pt\wd0=5.3pt
\vbox{\hrule height0.8pt
\hbox{\vrule width0.8pt\box0\vrule width0.8pt}
\hrule height0.8pt}}
\def\frac#1#2{\hbox{$#1\over#2$}}
\def\dual{\mathstrut^*\kern-0.1em}
\def\mod{\;\hbox{\rm mod}\;}
\def\ring{\mathaccent"7017}
\def\lvec#1{\setbox0=\hbox{$#1$}
    \setbox1=\hbox{$\scriptstyle\leftarrow$}
    #1\kern-\wd0\smash{
    \raise\ht0\hbox{$\raise1pt\hbox{$\scriptstyle\leftarrow$}$}}
    \kern-\wd1\kern\wd0}
\def\rvec#1{\setbox0=\hbox{$#1$}
    \setbox1=\hbox{$\scriptstyle\rightarrow$}
    #1\kern-\wd0\smash{
    \raise\ht0\hbox{$\raise1pt\hbox{$\scriptstyle\rightarrow$}$}}
    \kern-\wd1\kern\wd0}
\def\slash#1{\setbox0=\hbox{$#1$}\setbox1=\hbox{$\kern1pt/$}
    #1\kern-\wd0\kern1pt/\kern-\wd1\kern\wd0}


\def\nab#1{{\nabla_{#1}}}
\def\nabstar#1{{\nabla\kern0.5pt\smash{\raise 4.5pt\hbox{$\ast$}}
               \kern-5.5pt_{#1}}}
\def\drv#1{{\partial_{#1}}}
\def\drvstar#1{{\partial\kern0.5pt\smash{\raise 4.5pt\hbox{$\ast$}}
               \kern-6.0pt_{#1}}}
\def\ldrv#1{{\lvec{\,\partial}_{#1}}}
\def\ldrvstar#1{{\lvec{\,\partial}\kern-0.5pt\smash{\raise 4.5pt\hbox{$\ast$}}
               \kern-5.0pt_{#1}}}


\def\MeV{{\rm MeV}}
\def\GeV{{\rm GeV}}
\def\TeV{{\rm TeV}}
\def\fm{{\rm fm}}
\def\MSbar{\overline{\rm MS\kern-0.5pt}\kern0.5pt}


\def\euler{\gamma_{\rm E}}


\def\psibar{\overline{\psi}}
\def\psitilde{\widetilde{\psi}}
\def\ubar{\bar{u}}
\def\dbar{\bar{d}}


\def\dirac#1{\gamma_{#1}}
\def\diracstar#1#2{
    \setbox0=\hbox{$\gamma$}\setbox1=\hbox{$\gamma_{#1}$}
    \gamma_{#1}\kern-\wd1\kern\wd0
    \smash{\raise4.5pt\hbox{$\scriptstyle#2$}}}
\def\dirachat{\hat{\gamma}_5}


\def\SUtwo{{\rm SU(2)}}
\def\SUthree{{\rm SU(3)}}
\def\SUn{{\rm SU}(N)}
\def\tr{{\rm tr}}
\def\Tr{{\rm Tr}}
\def\Ad{{\rm Ad}\,}
\def\Group{{\rm SU}(3)}
\def\Lie{\frak{su}(3)}


\def\Sw{S_{\rm w}}
\def\Sg{S_{\rm g}}
\def\Spf{S_{\rm pf}}
\def\Spfs{S_{\rm pf,s}}
\def\Obs{{\mathcal O}}
\def\Fnc{{\mathcal F}}


\def\Dee{D_{\rm ee}}
\def\Deo{D_{\rm eo}}
\def\Doe{D_{\rm oe}}
\def\Doo{D_{\rm oo}}
\def\Dhat{\hat{D}}
\def\Ddag{D^{\dagger}}
\def\DdagD{D\Ddag}
\def\Ds{D_{\rm s}}
\def\DsdagDs{\Ds{\Ds}^{\kern-1pt\dagger}}
\def\Qs{Q_{\rm s}}

\def\om{\Omega}
\def\oms{\om^{\ast}}
\def\dom{\partial\om}
\def\doms{\partial\oms}
\def\Dom{D_{\om}}
\def\Doms{D_{\oms}}
\def\Ddom{D_{\dom}}
\def\Ddoms{D_{\doms}}

\def\amin{\alpha_{\rm min}}
\def\amax{\alpha_{\rm max}}
\def\krylov{{\mathcal K}}
\def\Proj{{\Bbb P}}


\def\dsp{{\mathcal S}}


\def\mom{\pi}
\def\eps{\epsilon}
\def\epst{\tilde{\epsilon}}


\def\cn{{\rm cn}}
\def\sn{{\rm sn}}


\def\msea{m_{\rm sea}}
\def\mval{m_{\rm val}}
\def\ms{m_{\rm s}}
\def\mpi{m_{\pi}}
\def\meff{m_{\rm eff}}
\def\meffhat{\hat{m}_{\rm eff}}
\def\gpi{g_{\pi}}
\def\mN{m_{N}}


\def\Pacc{P_{\rm acc}}
\def\Hspace{\mathcal H}
\def\ssum#1{\hbox{$\sum_{#1}$}}
\def\That{\hat{T}}
\def\avgavg#1{\langle\!\langle#1\rangle\!\rangle}
\def\avg#1{{\kern1.0pt\overline{\kern-1.0pt#1\kern-1.0pt}\kern1.0pt}}
\def\Imom{{\mathcal I}_0}
\def\Ifld{{\mathcal I}_U}
\def\Iint{{\mathcal I}}
\def\Jint{{\mathcal J}}


\def\src{\rm src}
\def\Nsrc{N_{\src}}
\def\Ssrc{S_{\src}}
\def\Obshat{\hat{\Obs}}
\def\Obsbar{\avg{\Obs}}
\def\Vol{V_3}
\def\Loop{{\mathcal C}}
\def\Wloop{{\mathcal W}}
\def\twolink{{\Bbb T}}
\def\lineop{{\Bbb L}}
\def\inslice{{\rm int}}


\def\con{{\rm c}}
\def\abar{\bar{a}}
\def\Zfun{{\mathcal Z}}
\def\Wfun{{\mathcal W}}
\def\phihat{\hat{\phi}}

\chapter{Computation of quark propagators}

Quark propagators in presence of a specified $\Group$ gauge field
are fundamental building blocks in lattice QCD.
Their computation amounts to solving
the Dirac equation
\begin{equation}
  D\psi(x)=\eta(x)
  \label{DiracEq}
\end{equation}
a number of times,
where $D$ denotes the massive lattice Dirac operator,
$\eta(x)$ a given quark field (the source field) and $\psi(x)$ the 
desired solution. 

Although the subject is nearly as old as lattice QCD itself,
there have been important advances in the last few years
that allow the quark propagators to be calculated much
more rapidly than was possible before. As a consequence,
the ``measurement'' of hadronic quantities
and the simulation of the theory with light sea quarks
are both accelerated significantly.

\section{Preliminaries}

\subsection{Accuracy \& condition number}

The Dirac equation (\ref{DiracEq})
is a large linear system that can only be solved
iteratively, i.e.~through some recursive procedure that 
generates a sequence 
$\psi_1,\psi_2,\psi_3,\ldots$ of increasingly accurate 
approximate solutions. 
A practical measure for the accuracy of an
approximate solution $\phi$ is the norm of the associated residue
\begin{equation}
  \rho=\eta-D\phi.
\end{equation}
If, say, $\|\rho\|<\eps\|\eta\|$ for some small value $\eps$,
an important question is then by how much $\phi$ deviates from
the exact solution $\psi$ of the equation.
Using standard norm estimates, the deviation is found to be bounded by
\begin{equation}
  \|\psi-\phi\|<\eps\kappa(D)\|\psi\|,
\end{equation}
where
\begin{equation}
  \kappa(D)=\|D\|\|D^{-1}\|
  \label{CondNum}
\end{equation}
is referred to as the condition number of $D$.
The relative error of $\phi$ thus tends to be
larger than $\eps$ by the factor $\kappa(D)$
(in this chapter, the standard scalar product 
of quark fields is used as well as 
the field and operator norms that derive from it).

The extremal eigenvalues of 
$\Ddag\kern-1pt D$, $\amin$ and $\amax$,
are proportional to the square of the quark mass $m$
and the square of the inverse of the lattice spacing $a$, respectively.
In particular, the condition number
\begin{equation}
  \kappa(D)=\left(\amax/\amin\right)^{1/2}\propto
  (am)^{-1}
\end{equation}
can be very 
large at small quark masses and lattice spacings.
One says that the Dirac operator is ``ill-conditioned'' in this case.
The important point to keep in mind is that 
the accuracy of the solution of the Dirac equation 
which can be attained on a given computer
is limited by the condition number.

\subsection{Iterative improvement}

If the residual error $\eps$ of the approximate
solution $\phi$ is still well above the limit set
by the machine precision and the condition number of the Dirac
operator, an improved solution
\begin{equation}
  \tilde\phi=\phi+\chi
  \label{IterImpSum}
\end{equation}
may be obtained by approximately solving the residual equation
\begin{equation}
  D\chi=\rho.
\end{equation}
It is straightforward to show that the residue of the solution is
reduced by the factor $\delta$ in this way
if $\chi$ satisfies $\|\rho-D\chi\|<\delta\|\rho\|$.
Note that $\chi$ is approximately
equal to the deviation of $\phi$ from the exact solution
of the Dirac equation and is therefore usually a small
correction to $\phi$.

Iterative improvement is used by all solvers
that need to be restarted after a while.
The GCR algorithm discussed below is an example of 
such a solver.
Another application of iterative improvement, known as  
``single-precision acceleration'', 
exploits the fact that modern processors 
perform 32 bit arithmetic operations significantly faster than
64 bit operations.
The idea is to solve the Dirac equation to 64 bit precision
by going through a few cycles of iterative improvement, where, in each 
cycle, 
the residual equation is solved to a limited precision using
32 bit arithmetic, while
the residue $\rho$ and the sum
(\ref{IterImpSum}) are evaluated using 64 bit arithmetic
\shortcite{InexactDeflation}.

\section{Krylov-space solvers}

The Krylov space $\krylov_n$ of dimension $n$ is 
the complex linear space spanned by the fields
\begin{equation}
  \eta,\,D\eta,\,D^2\eta,\ldots,\,D^{n-1}\eta.
\end{equation}
Many popular solvers, including the CG, BiCGstab and GCR algorithms, 
explicitly or implicitly build up a Krylov space
and search for the solution of the Dirac equation
within this space. The very readable book of
\shortciteN{Saad} describes these solvers in full detail. A
somewhat simpler discussion of the CG (conjugate gradient) algorithm
is given in the book of \shortciteN{GolubLoan} and useful
additional references for the BiCGstab algorithm
are \shortciteN{BiCGstabI} and \shortciteN{BiCGstabII}.
Here the GCR (generalized conjugate residual) 
algorithm is discussed as a representative case.

\subsection{The GCR algorithm}

\begin{figure}[t]
\begin{center}
  \includegraphics[width=5.0cm]{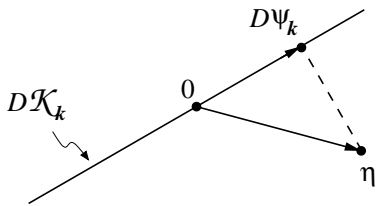}
\end{center}
\caption{The approximate solution $\psi_k\in\krylov_k$
of the Dirac equation
constructed by the GCR algorithm is such that 
the distance $\|\eta-D\psi_k\|$ is minimized. 
The field $D\psi_k$ therefore coincides with
the orthogonal projection of the source $\eta$ to the space 
$D\krylov_k$.}
\label{KrylovProj}
\end{figure}

The approximate solutions 
of the Dirac equation (\ref{DiracEq}) generated
by the GCR algorithm are the fields 
$\psi_k\in\krylov_k$, $k=1,2,3,\ldots$, that minimize
the norm of the residues
\begin{equation}
  \rho_k=\eta-D\psi_k.
\end{equation}
An equivalent requirement is that 
$D\psi_k$ coincides with the orthogonal
projection of the source field $\eta$ to the 
$k$-dimensional linear space $D\krylov_k$ (see Fig.~\ref{KrylovProj}).
The algorithm proceeds somewhat indirectly 
by first constructing an orthonormal basis 
$\chi_0,\chi_1,\chi_2,\ldots$ of these spaces
through a recursive process. Independently of the 
details of the construction, the orthogonality 
property mentioned above then implies that the fields
\begin{equation}
  \rho_k=\eta-\sum_{l=0}^{k-1}c_l\chi_l,
  \qquad c_l=(\chi_l,\eta),
  \label{GCRresidues}
\end{equation}
are the residues of the approximate solutions $\psi_k$.
The residues are thus obtained before the latter are known.

In each iteration of the recursion, the next basis field $\chi_k$ is 
constructed from 
the previous fields $\chi_0,\ldots,\chi_{k-1}$ and the source
$\eta$. First the residue $\rho_k$ is computed through
eqn~(\ref{GCRresidues}) and $\chi_k$ is then taken to be 
a (properly orthonormalized)
linear combination of $D\rho_k$ and the previous fields.
Since $\chi_{k-1}$ is a linear combination of $D\rho_{k-1}$
and the fields $\chi_0,\ldots,\chi_{k-2}$, and so on, it is
clear that the recursion also yields the coefficients $a_{lj}$ in
the equations
\begin{equation}
  \chi_l=\sum_{j=0}^la_{lj}D\rho_j,
  \qquad l=0,1,2,\ldots,
  \label{GCRmatrix}
\end{equation}
in which $\rho_0=\eta$.

Once the fields $\chi_0,\ldots,\chi_{n-1}$ are known
for some $n$, the last solution $\psi_n$ is obtained
starting from the orthogonality condition
\begin{equation}
  D\psi_n=\sum_{l=0}^{n-1}c_l\chi_l.
\end{equation}
After substituting eqn~(\ref{GCRmatrix}),
the equation may then be divided by $D$
and one finds that the solution is given by
\begin{equation}
  \psi_n=\sum_{l=0}^{n-1}\sum_{j=0}^lc_la_{lj}\rho_j.
  \label{GCRSolution}
\end{equation}
Note that the right-hand side of this equation can be
evaluated straightforwardly since
all entries are known at this point.

In total the computation of $\psi_n$
requires $n$ applications of the 
Dirac operator and the evaluation of some $\frac{1}{2}n^2$
linear combinations and scalar products of quark fields. Moreover,
memory space for about $2n$ fields is needed.
Choosing values of $n$ from, say, $16$ to $32$ proves
to be a reasonable compromise in practice, where
the computational effort must be balanced against
the reduction in the residue that is achieved.
If the last solution is not sufficiently accurate,
the algorithm can then always be restarted following the 
rules of iterative improvement.

\subsection{Convergence properties}

\begin{figure}[t]
\begin{center}
  \includegraphics[width=12cm]{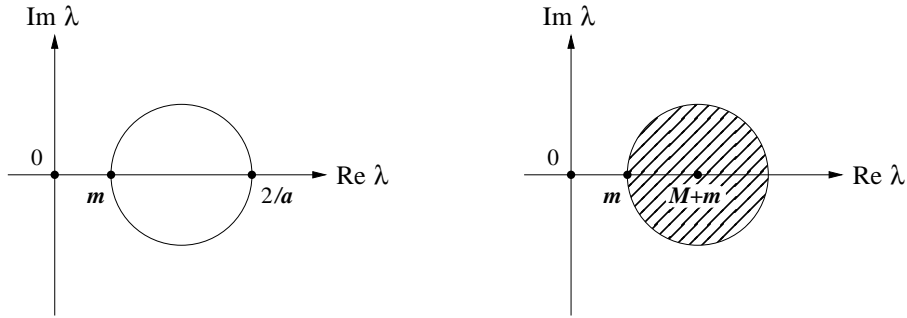}
\end{center}
\caption{The eigenvalues of the Neuberger--Dirac operator
with bare quark mass $m$ lie on a circle in the complex plane
(drawing on the left). For the convergence analysis of the 
GCR algorithm, the spectrum of the Dirac operator is assumed
to be contained in a disk $\Bbb D$ in the right-half plane with radius
$M$ and distance $m>0$ from the origin (drawing on the right).  
}
\label{SpectralDisk}
\end{figure}

The convergence of the GCR algorithm and related Krylov-space solvers can 
be proved rigorously if the (complex) spectrum
of the Dirac operator
is contained in the half-plane on the right of the imaginary axis.
In the case of the Neuberger--Dirac operator, for example, 
all eigenvalues of $D$
lie on the circle shown in Fig.~\ref{SpectralDisk}.
The spectrum of the Wilson--Dirac operator is 
rather more complicated, but is usually 
contained in an ellipsoidal region 
in the right half-plane.

For simplicity,
the Dirac operator is, in the following paragraphs, 
assumed to be diagonalizable
and to have all its eigenvalues in the shaded disk $\Bbb D$ shown 
in Fig.~\ref{SpectralDisk}. The convergence analysis of the
GCR algorithm then starts from the observation that
the residue $\rho_k$ is given by
\begin{equation}
  \rho_k=\eta-D\psi_k=p_k(D)\eta,
\end{equation}
where $p_k(\lambda)$ is a polynomial of degree $k$ that
satisfies $p_k(0)=1$.
Since the algorithm minimizes the residue, it follows that
\begin{equation}
  \|\rho_k\|=\min_{p_k}\|p_k(D)\eta\|\leq\min_{p_k}\|p_k(D)\|\|\eta\|,
\end{equation}
the minimum being taken over all such polynomials.

Lattice quark fields are large arrays of complex
numbers. In this language, the Dirac operator $D$ is just a complex 
square matrix. The assumption that $D$ is
diagonalizable then implies the existence of a diagonal
matrix $\Lambda$ and of an invertible matrix $V$ such
that $D=V\Lambda V^{-1}$. As a consequence
\begin{equation}
  \|p_k(D)\|=\|Vp_k(\Lambda)V^{-1}\|
  \leq\kappa(V)\|p_k(\Lambda)\|
\end{equation}
and therefore 
\begin{equation}
  \|\rho_k\|\leq\kappa(V)\max_{\lambda\in{\Bbb D}}|p_k(\lambda)|
  \|\eta\|
  \label{GCRbound}
\end{equation}
for any polynomial $p_k(\lambda)$ of degree $k$ satisfying $p_k(0)=1$.
One may, for example, insert
\begin{equation}
  p_k(\lambda)=\left(1-{\lambda\over M+m}\right)^k,
  \label{GCRpoly}
\end{equation}
in which case the inequality (\ref{GCRbound}) leads to
the bound 
\footnote{A rigorous mathematical result, known as Zarantello's lemma,
asserts that it is
not possible to obtain a more stringent bound by choosing a 
different polynomial, i.e.~the polynomial (\ref{GCRpoly}) is
the optimal one.}
\begin{equation}
  \|\rho_k\|\leq\kappa(V)\left(1+{m\over M}\right)^{-k}\|\eta\|.
  \label{GCRrate}
\end{equation}
The GCR algorithm thus converges roughly
like $\exp(-km/M)$ if $m/M\ll1$. Note that the convergence rate
$m/M\sim2/\kappa(D)$ can be quite small in practice. For 
$m=10$ MeV and $M=2$ GeV, for example,
the estimate (\ref{GCRrate})
suggests that values of $k$ as large as $4000$ are required
for a reduction of the residue by the factor $10^{-10}$.

\begin{figure}[t]
\begin{center}
  \includegraphics[width=12cm]{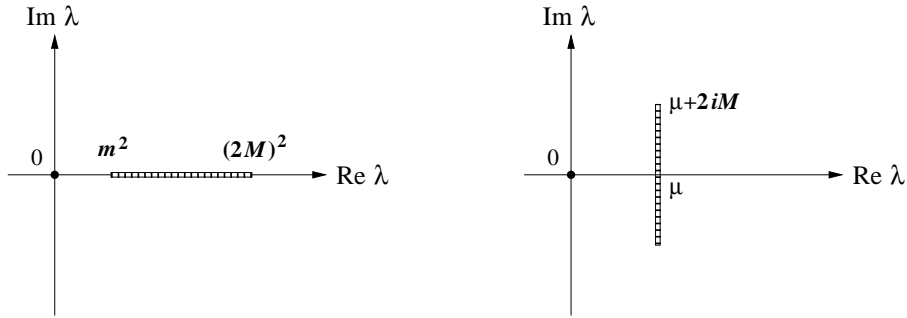}
\end{center}
\caption{The eigenvalues of the hermitian operator 
$D^{\dagger}\kern-1pt D$
and the twisted-mass Dirac operator
$i\dirac{5}D+\mu$ occupy the shaded line segments shown
in the left and right drawings, respectively.
}
\label{SpectralLine}
\end{figure}

The GCR algorithm can also be applied to the so-called normal equation
\begin{equation}
  D^{\dagger}\kern-1pt D\psi=\eta
\end{equation}
and to the Dirac equation 
\begin{equation}
  (i\dirac{5}D+\mu)\psi=\eta
\end{equation}
in ``twisted-mass'' QCD. A notable difference with respect to
the ordinary Dirac equation is that 
the operators on the left of these equations can be
diagonalized through unitary transformations. Moreover,
their spectra are contained in straight-line segments
in the complex plane (see Fig.~\ref{SpectralLine}).
Using Chebyshev polynomials in place of the power (\ref{GCRpoly}),
the estimate
\begin{equation}
  \|\rho_k\|\lesssim 2\rme^{-rk}\|\eta\|
\end{equation}
may be derived in these cases,
where $r=m/M$ for the normal equation and $r=\mu/2M$ for 
the twisted-mass Dirac equation.

The convergence of Krylov-space solvers is thus mainly 
determined by the properties of the spectrum
of the operator considered.
In QCD the fact that the masses of the light quarks are much
smaller than the inverse lattice spacing consequently
tends to slow down the 
computations enormously. One can do better, however, by
exploiting specific properties of the Dirac operator.

\subsection{Preconditioning}

Preconditioning is a general strategy that allows such properties
to be taken into account.
Let $L$ and $R$ be some invertible operators acting on
quark fields. Instead of the Dirac equation
(\ref{DiracEq}), one may then consider the so-called
preconditioned equation
\begin{equation}
  LDR\phi=L\eta.
  \label{PrecondSystem}
\end{equation}
Once this equation solved, using a Krylov-space solver for example,
the solution of the Dirac equation is obtained by setting $\psi=R\phi$. 
If $D\approx L^{-1}R^{-1}$, and if the application of $L$ and
$R$ to a given quark field is not too time-consuming, 
the total computer time required for the solution 
of the equation may be significantly reduced in this way. 

\begin{figure}[t]
\begin{center}
  \includegraphics[width=4.5cm]{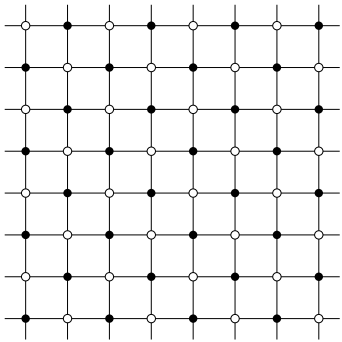}
\end{center}
\caption{Hyper-cubic lattices may be divided into
the sublattices of the even and the odd sites (black and white
points, respectively). Even-odd preconditioning effectively
amounts to ``integrating out'' the quark field on the odd 
sublattice.
}
\label{EoPrecnd}
\end{figure}

In lattice QCD, a widely used preconditioning method 
for the Wilson--Dirac operator is ``even-odd preconditioning''. 
A lattice point $x\in\gz^4$ is referred to as even or odd
depending on whether the sum of its coordinates $x_{\mu}$ is even or odd
(see Fig.~\ref{EoPrecnd}). If the points are ordered such that
the even ones come first, the Dirac operator assumes the 
block form
\begin{equation}
  D=\pmatrix{\Dee & \Deo \cr
             \Doe & \Doo \cr},
\end{equation}
where $\Deo$, for example, stands for the hopping terms 
that go from the odd to the even sites. The blocks on the diagonal, $\Dee$ and
$\Doo$, include the mass term and a Pauli term if the theory is 
O($a$)-improved. Since they do not couple different lattice points,
they can be easily inverted and it makes sense to consider
the preconditioners
\begin{equation}
  L=\pmatrix{1 & -\Deo\Doo^{-1} \cr
             0 & 1 \cr},
  \qquad
  R=\pmatrix{1 & 0 \cr
             -\Doo^{-1}\Doe & 1 \cr}.
\end{equation}
With this choice, the Dirac operator is block-diagonalized,
\begin{equation}
  LDR=\pmatrix{\Dhat & 0 \cr
                0 & \Doo \cr},
  \qquad
  \Dhat=\Dee-\Deo\Doo^{-1}\Doe,
\end{equation}
and the solution of eqn~(\ref{PrecondSystem}) thus amounts to  
solving a system in the space of quark fields on the even sublattice.
The condition number of $\Dhat$ is usually less than half the one of
$D$ and even-odd preconditioning consequently leads to
an acceleration of the solver by a factor $2$ to $3$ or so.

Other preconditioners used in lattice QCD are the successive
symmetric overrelaxation (SSOR) preconditioner 
\shortcite{BiCGstabII} and a domain-decomposition preconditioner
based on the Schwarz alternating procedure 
\shortcite{SchwarzSolver}. 
The latter is an example of an ``expensive'' preconditioner, whose 
implementation involves an iterative procedure and is therefore
inexact to some extent. 
Inaccuracies at this level of the algorithm are however not
propagated to the final results if the GCR solver is 
used for the preconditioned equation (\ref{PrecondSystem}). 
This algorithm actually always finds the best approximation to the solution
in the space generated by applying the preconditioner to the
residues of the previous solutions.
For the same reason, the GCR algorithm is also 
safe of rounding errors.

\section{Low-mode deflation}

The low modes of the Dirac operator are intimately related
to the spontaneous breaking of chiral symmetry and therefore
play a special r\^ole in QCD. Treating them separately from
the other modes seems appropriate from the physical
point of view and is recommended 
for technical reasons at small quark masses.

\subsection{Textbook deflation}

In the case of the hermitian system,
\begin{equation}
  A\psi=\eta,\qquad A=\Ddag\kern-1pt D,
  \label{HermitianSystem}
\end{equation}
there exists an orthonormal basis of eigenvectors $v_k$,
$k=1,2,3,\ldots$, such that
\begin{equation}
  A v_k=\alpha_kv_k,\qquad 0\leq\alpha_1\leq\alpha_2\leq\ldots
\end{equation}
The action on any quark field $\psi$ 
of the orthonormal projector $P$ to the $N$ lowest modes 
is then given by
\begin{equation}
  P\psi=\sum_{k=1}^Nv_k(v_k,\psi).
\end{equation}
Since $P$ commutes with $A$,
the linear system (\ref{HermitianSystem}) splits 
into the decoupled equations
\begin{eqnarray}
  A_{\parallel}\psi_{\parallel}&=&\eta_{\parallel},
  \qquad
  \psi_{\parallel}=P\psi,
  \label{LittleSystem}
  \\[2ex]
  A_{\perp}\psi_{\perp}&=&\eta_{\perp},
  \qquad
  \psi_{\perp}=(1-P)\psi,
  \label{DeflatedSystem}
\end{eqnarray}
where $A_{\parallel}=P AP$ and $A_{\perp}=(1-P)A(1-P)$
are, respectively, referred to as the ``little operator'' and 
the ``deflated operator''.

If the eigenvectors $v_1,\ldots,v_N$
are known, and if there are no zero-modes,
the solution of the little system (\ref{LittleSystem}) can
be obtained exactly through
\begin{equation}
  \psi_{\parallel}=\sum_{k=1}^N{1\over\alpha_k}v_k(v_k,\eta).
\end{equation}
The deflated system (\ref{DeflatedSystem}), 
on the other hand, can only be solved iteratively
using the CG algorithm, for example. 
With respect to the full system,
the associated condition number
\begin{equation}
  \kappa(A_{\perp})={\alpha_1\over\alpha_{N+1}}\kappa(A)
\end{equation}
is however reduced and one therefore expects the solver 
to be accelerated by the factor
$(\alpha_{N+1}/\alpha_1)^{1/2}$ or so.

The deflation of the hermitian system
(\ref{HermitianSystem}) along these lines
is straightforward to implement, but the method tends to be limited
to small lattices, because the computer time required for the 
calculation of the low eigenvectors 
grows rapidly with the lattice volume. In the past few years,
it was nevertheless further developed and improved in
various directions (for a review, see \shortciteN{Wilcox}, for example).

\subsection{The Banks--Casher relation}

In the continuum theory (which is considered here for simplicity),
the eigenvalues of the Dirac operator $D$ in presence of 
a given gauge field are of the form $m+i\lambda_k$,
where $m$ denotes the quark mass and $\lambda_k\in\rz$,
$k=1,2,3,\ldots$, the eigenvalues of the massless operator.
The associated average spectral density,
\begin{equation}
  \rho(\lambda,m)={1\over V}\sum_{k=1}^{\infty}
  \left\langle\delta(\lambda-\lambda_k)\right\rangle,
\end{equation}
is conventionally normalized by the space-time volume $V$ 
so that it has a meaningful infinite-volume limit.

In a now famous paper, \shortciteN{BanksCasher} showed many years ago 
that the density at the origin,
\begin{equation}
  \lim_{\lambda\to0}\lim_{m\to0}\lim_{V\to\infty}
  \rho(\lambda,m)={1\over\pi}\Sigma,
  \label{BanksCasherRelation}
\end{equation}
is proportional to the quark condensate
\begin{equation}
  \Sigma=-\lim_{m\to0}\lim_{V\to\infty}\langle\ubar u\rangle,
  \qquad\hbox{$u$: up-quark field},
\end{equation}
in the chiral limit.
The spontaneous breaking
of chiral symmetry in QCD is thus linked to the presence of 
a non-zero density of eigenvalues 
at the low end of the spectrum of the Dirac operator.

\begin{figure}[t]
\begin{center}
  \includegraphics[clip,scale=0.52]{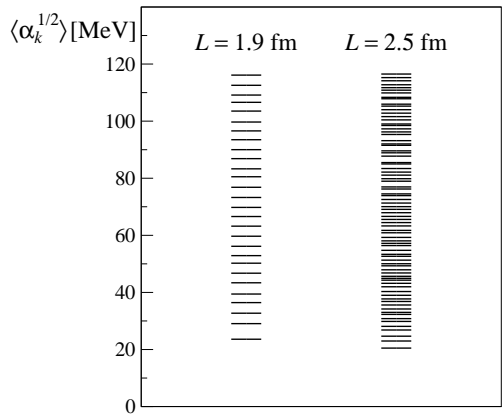}
\end{center}
\caption{
Expectation value of the eigenvalues 
of $(\Ddag\kern-1pt D)^{1/2}$ below $116$ MeV
in O($a$)-improved two-flavour QCD on a 
$2L\times L^3$ lattice at two values of $L$.
Both spectra were obtained at lattice spacing 
$a=0.08$ fm and renormalized 
sea-quark mass $m=26$ MeV. 
From the smaller to the larger lattice,
the number of modes per MeV increases approximately as
predicted by the Banks--Casher relation.
}
\label{EvMed}
\end{figure}

Since the eigenvalues $\alpha_k=m^2+\lambda_k^2$ of 
$\Ddag\kern-1pt D$ are simply 
related to the eigenvalues of $D$, the Banks--Casher relation
(\ref{BanksCasherRelation}) 
immediately leads to the estimate
\begin{equation}
  \nu(M,m)\simeq{2\over\pi}\Lambda\Sigma V,
  \qquad\Lambda^2=M^2-m^2,
\end{equation}
for the number of low modes of $\Ddag\kern-1pt D$ 
with eigenvalues $\alpha_k\leq M^2$. If one sets
$m=0$, $M=100$ MeV and $\Sigma=(250\,\MeV)^3$, for example,
and considers a space-time volume of size $V=2L^4$, the mode numbers 
are estimated to be $21$, $106$ and $336$ for 
$L=2$, $3$ and $4$ fm, respectively.
As illustrated by Fig.~\ref{EvMed}, the low-mode condensation 
is readily observed in numerical simulations of lattice QCD.

Since $\nu(M,m)$ increases proportionally
to space-time volume $V$, an effective deflation of
the Dirac equation requires O($V$) modes 
to be deflated. The associated computational effort
increases like $V^2$ and straightforward deflation 
consequently tends to become inefficient or impractical
at large volumes. However, as explained in the following, 
the $V^2$-problem can be overcome using inexact deflation 
\shortcite{InexactDeflation}
and domain-decomposed deflation subspaces
\shortcite{LocalDeflation}.

\subsection{Deflation w/o eigenvectors}

Returning to the lattice Dirac equation (\ref{DiracEq}),
a more general form of deflation will now be described,
which starts from an unspecified set
$\phi_1,\ldots,\phi_N$ of $N$ orthonormal
quark fields.
The linear space $\dsp$ spanned by these fields
will play the r\^ole of the deflation subspace, 
but is not assumed to be an eigenspace of 
the Dirac operator.

The action on any quark field $\psi$ of the orthogonal 
projector $P$ to $\dsp$ is given by
\begin{equation}
  P\psi=\sum_{k=1}^N\phi_k\left(\phi_k,\psi\right).
\end{equation}
As before, the restriction $PDP$ of the Dirac operator to $\dsp$ 
will be referred to as the ``little Dirac operator''. Its action
is encoded in the complex $N\times N$ matrix
\begin{equation}
  A_{kl}=\left(\phi_k,D\phi_l\right),
  \quad k,l=1,\ldots,N,
\end{equation}
through
\begin{equation}
  PDP\psi=\sum_{k,l=1}^N\phi_k A_{kl}\left(\phi_l,\psi\right).
\end{equation}
A technical assumption made in the following is
that $A$ (and thus $PDP$ as an operator acting in $\dsp$) is invertible. 

The form of inexact deflation discussed below is based on 
the projectors
\begin{eqnarray}
  P_L&=&1-DP(PDP)^{-1}P,
  \\[2ex]
  P_R&=&1-P(PDP)^{-1}PD.
\end{eqnarray}
It is not difficult to prove that
\begin{eqnarray}
  P_L^2&=&P_L,\qquad P_R^2=P_R,
  \\[2ex]
  PP_L&=&P_RP=(1-P_L)(1-P)=(1-P)(1-P_R)=0,
  \\[2ex]
  P_LD&=&DP_R.
  \label{CommProp}
\end{eqnarray}
In particular, the operator 
$P_L$ projects any quark field to the orthogonal complement
$\dsp^{\perp}$ of the deflation subspace.
Note, however, that $P_L$ and $P_R$ are not hermitian and therefore
not ordinary orthogonal projectors.

The Dirac equation (\ref{DiracEq}) may now be split into
two decoupled equations,
\begin{equation}
  D\psi_{\parallel}=\eta_{\parallel},
  \qquad
  D\psi_{\perp}=\eta_{\perp},
\end{equation}
where
\begin{eqnarray}
  \psi_{\parallel}&=&(1-P_R)\psi,\qquad \psi_{\perp}=P_R\psi,
  \\[2ex]
  \eta_{\parallel}&=&(1-P_L)\eta,\qquad \eta_{\perp}=P_L\eta.
\end{eqnarray}
The use of a different projector for the splitting of 
the solution $\psi$ and the source $\eta$ is entirely
consistent in view of
the commutator relation (\ref{CommProp}) and merely reflects
the fact that the Dirac operator is not hermitian.

The solution of the little system,
\begin{equation}
  \psi_{\parallel}=P(PDP)^{-1}P\eta=
  \sum_{k,l=1}^N\phi_k(A^{-1})_{kl}\left(\phi_l,\eta\right),
\end{equation}
is easily found, but the other equation can only be solved
using an iterative procedure such as the GCR algorithm.
However, the operator on the left of the equation is
the deflated operator
\begin{equation}
  \Dhat=DP_R=P_LD(1-P),
\end{equation}
which acts on quark fields in $\dsp^{\perp}$ and
which may have a much smaller condition number than $D$, particularly so
at small quark masses.
An acceleration of the computation is then achieved
since the solution is obtained in fewer iterations
than in the case of the unmodified Dirac equation.

\begin{figure}[t]
\begin{center}
  \includegraphics[clip,scale=0.285]{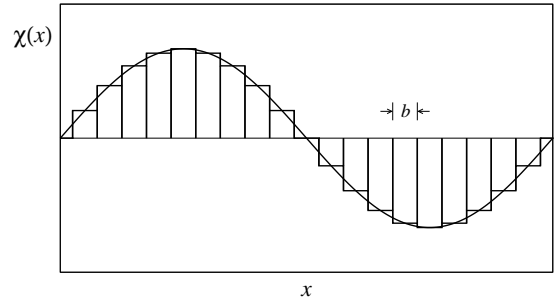}
\end{center}
\caption{
The low modes $\chi(x)$ of the lattice 
Dirac operator in the free-quark theory
are plane waves with small momenta $p$. These can be well
approximated in the norm by functions that are constant
on blocks of lattice points if the block size $b$ 
satisfies $bp\ll 1$. Efficient deflation subspaces can thus be
constructed using fields that are discontinuous and therefore 
far from being approximate
eigenmodes of the Dirac operator.
}
\label{BlkProj}
\end{figure}

The condition number $\kappa(\Dhat)=\|\Dhat\|\|\Dhat^{-1}\|$ 
of $\Dhat$ depends on the quark mass 
mainly through the factor
\begin{equation}
  \|\Dhat^{-1}\|
  =\|(1-P)D^{-1}(1-P)\|
  \leq\|(1-P)(\Ddag\kern-1pt D)^{-1}(1-P)\|^{1/2}.
\end{equation}
It is then quite obvious that $\kappa(\Dhat)$ will be 
much smaller than $\kappa(D)$
if $1-P$ effectively ``projects away'' 
the low modes of $\Ddag\kern-1pt D$,
i.e.~if they are well approximated by the deflation subspace.
However, contrary to what may be assumed, this requirement
is fairly weak and does not imply that the deflation
subspace must be spanned by approximate eigenvectors of 
$\Ddag\kern-1pt D$ (see Fig.~\ref{BlkProj}).

\begin{figure}[t]
\begin{center}
  \includegraphics[clip,scale=0.1]{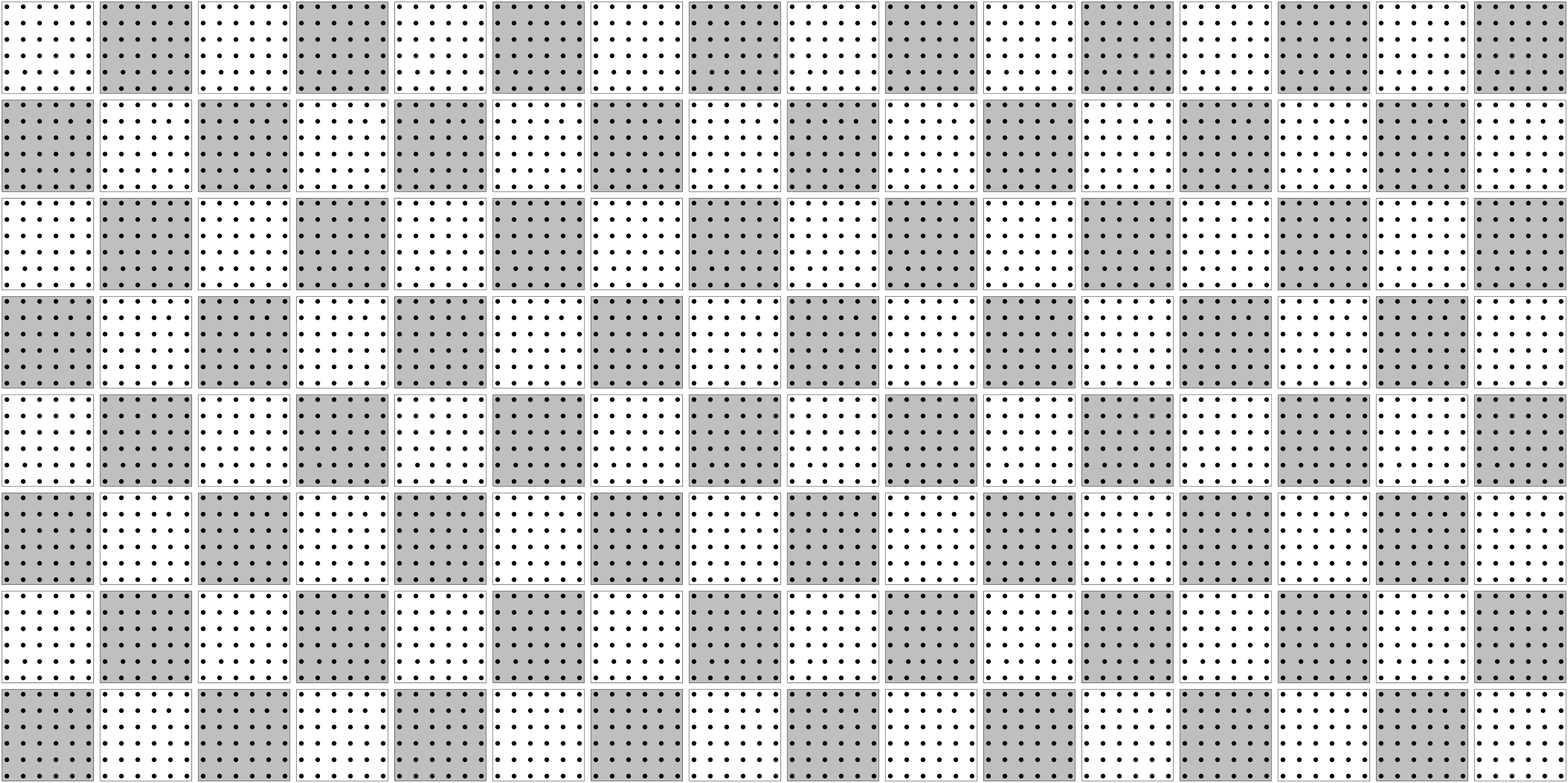}
\end{center}
\caption{
Two-dimensional view of a $96\times 48^3$ lattice, divided 
into 8192 non-overlapping blocks of size $6^4$.
}
\label{BlkGrid}
\end{figure}

\subsection{Domain-decomposed deflation subspaces}

In the following, a decomposition of the lattice
into rectangular blocks $\Lambda$ such as the one shown
in Fig.~\ref{BlkGrid} will be considered. 
A special kind of deflation subspace $\dsp$
is then obtained by choosing a set of orthonormal fields
$\phi^{\Lambda}_1,\ldots,\phi^{\Lambda}_{N_s}$ on each block
$\Lambda$ and by taking $\dsp$ to be the linear span of all
these fields. The associated projector is
\begin{equation}
  P=\sum_{\Lambda}P_{\Lambda},
  \qquad
  P_{\Lambda}\psi=\sum_{k=1}^{N_s}\phi^{\Lambda}_k
  \left(\phi^{\Lambda}_k,\psi\right).
\end{equation}
Evidently, this construction fits the general scheme discussed in 
Section~1.3.3 except perhaps 
for the labeling of the fields that span the deflation subspace.

The size of the blocks $\Lambda$ 
is a tunable parameter of domain-decomposed subspaces.
Usually the blocks are taken to be fairly small
($4^4$, $6^4$ or $8\times4^3$, for example), but the 
exact choice should eventually be based on the measured
performance of the deflated solver.
A key feature of domain-decomposed subspaces is the fact
that their dimension is proportional to the volume $V$ of the lattice,
while the application of the projector $P$ to a given quark field requires
only $\rmO(N_s V)$ floating-point operations (and not $\rmO(N_s V^2)$
operations as would normally be the case). Such subspaces
may thus allow the $V^2$-problem to be overcome, provided
high deflation efficiencies can be achieved for some 
volume-independent number $N_s$
of block modes.

In the case of the free-quark theory, Fig.~\ref{BlkProj} suggests
that this strategy will work out if the block modes are chosen to be
constant. Since quark fields have 12 complex components, one needs
$N_s=12$ such modes per block.
In presence of an arbitrary gauge field,
the choice of the block modes is less obvious, however,
because the notion of smoothness ceases to have a well-defined meaning.

\subsection{Local coherence \& subspace generation}

\begin{figure}[t]
\begin{center}
  \includegraphics[clip,scale=0.22]{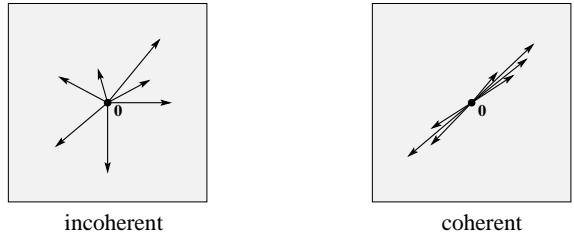}
\end{center}
\caption{
When restricted to a small block of lattice points, the O($V$)
low modes of the Dirac operator tend to align to 
a relatively low-dimensional linear space, a property referred to as
local coherence.
}
\label{LocalCoherence}
\end{figure}

It is helpful to note at this point that the deflation deficits
\begin{equation}
  \|(1-P)\chi\|^2=\sum_{\Lambda}\|(1-P_{\Lambda})\chi\|^2
\end{equation}  
of the low modes $\chi$ of $\Ddag\kern-1pt D$
can only be small if they are small on all blocks $\Lambda$.
Since the local deflation subspace has fixed dimension $N_s$, and since
the number of low modes is proportional to $V$ and thus
tends to be much larger than $N_s$, this condition cannot in 
general be met unless the low modes happen to collapse to 
a lower dimensional space on each block, i.e.~unless they are 
``locally coherent'' (see Fig.~\ref{LocalCoherence}).

The local coherence of the low modes is numerically well established.
On a $64\times32^3$ lattice with spacing $a=0.08$ fm, for example,
deflation deficits as small as a few percent are achieved when using
$4^4$ blocks and $N_s=12$ block fields. Moreover, $N_s$ does not
need to be adjusted when the lattice volume increases. 
So far, however, no theoretical explanation of why the modes are locally
coherent has been given. In particular, it is unclear whether
the property has anything to do with chiral symmetry.

Efficient domain-decomposed 
deflation subspaces can now be constructed fairly easily
\shortcite{LocalDeflation}.
One first notes that any quark field $\psi$ 
satisfying $\|D\psi\|\leq M\|\psi\|$ for some sufficiently small 
value of $M$ (say, $M=100$ MeV)
is well approximated by a linear combination of low modes and
is therefore locally coherent with these. 
By generating a set $\psi_1,\ldots,\psi_{N_s}$ of independent 
fields of this kind, using inverse iteration,
for example,
and by applying
the Gram--Schmidt orthonormalization process to the 
projected fields
\begin{equation}
  \psi_k^{\Lambda}(x)=
  \left\{\begin{array}{c@{\qquad}l}
                       \psi_k(x)  & \hbox{if $x\in\Lambda$},\\[0.5ex]
                          0       & \hbox{otherwise,}
                       \end{array}\right.
\end{equation}
one thus obtains a basis $\phi^{\Lambda}_1,\ldots,\phi^{\Lambda}_{N_s}$
of block fields with large projections to the low modes
\footnote{
The deflation subspace can alternatively be generated
``on the fly'' while solving the Dirac equation, exploiting
the fact that the residue of an approximate solution
tends to align to the low modes of the Dirac operator
\shortcite{MultigridI,MultigridII}.
}. The generation of the deflation subspace along these
lines requires a modest amount of computer time and certainly
far less than would be needed for an approximate 
computation of the low modes of the Dirac operator. Moreover,
since $N_s$ can be held fixed, the computational effort
scales like $V$ rather than $V^2$.

\subsection{Solving the deflated system}

Once the deflation subspace is constructed, the deflated equation
$D\psi_{\perp}=\eta_{\perp}$ can in principle be solved using 
the GCR algorithm with $D$ replaced by $\Dhat=P_LD$. 
However, the application of the projector $P_L$ 
requires the little system to be solved for a given
source field, which is not a small task in general. 
Note that the little Dirac operator acts on fields on the block lattice
with $N_s$ complex components. 
An exact solution of the little system is therefore not practical
on large lattices.

\begin{figure}[t]
\begin{center}
  \includegraphics[clip,scale=0.38]{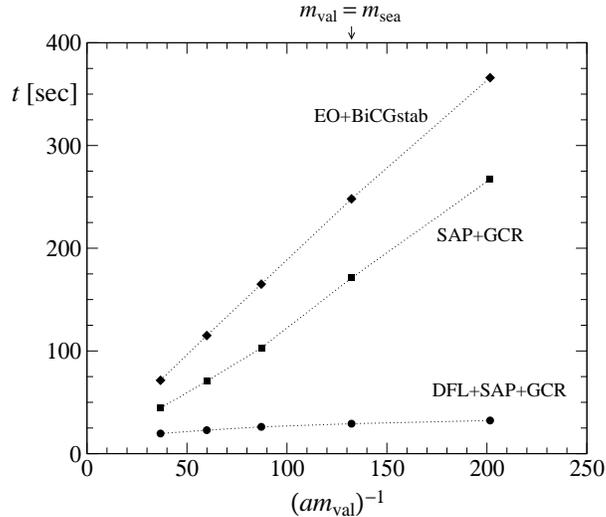}
\end{center}
\caption{
Computer time needed for the solution of the O($a$)-improved Wilson--Dirac 
equation in two-flavour QCD on a $64\times32^3$ lattice 
with spacing $a=0.08$ fm. In these tests,
the sea-quark mass $\msea$ was $26$ MeV, the 
valence-quark mass $\mval$ ranged from about $15$ to $90$ MeV
and the relative residue of the solution 
was required to be $10^{-10}$.
All timings were taken on a PC cluster with $64$ (single-core)
processors.
}
\label{DflPerf}
\end{figure}

In the case of the Wilson--Dirac operator and its relatives,
the little Dirac operator has only nearest-neighbour
couplings among the blocks.
The solution of the little system
may then be obtained iteratively using the even-odd preconditioned
GCR algorithm, for example.
The effort required for the solution of the little system
is nevertheless not completely negligible and it is advisable to 
consider solving the deflated right-preconditioned system
\begin{equation}
  P_LDR\phi=\eta_{\perp},
  \qquad
  \psi_{\perp}=P_RR\phi,
\end{equation}
instead of the deflated equation directly,
the operator $R$ being a suitable preconditioner for $D$.
A preconditioner that has
been used in this context is the Schwarz alternating procedure 
\shortcite{SchwarzSolver}.
The important point to note is that the preconditioner
tends to reduce the high-mode components of the residue
of the current approximate solution,
while the low-mode component of the residue is
projected away by the projector $P_L$.
Deflation and right-preconditioning thus tend to complement one another.

The performance figures plotted in Fig.~\ref{DflPerf}
show that local deflation works very well in lattice QCD.
In this study, the block size was taken to be $4^4$ and $N_s$ was
set to $20$. With respect to the even-odd preconditioned
BiCGstab algorithm (points labeled EO+BiCGstab in the figure), 
the deflated Schwarz-preconditioned GCR algorithm (DFL+SAP+GCR)
achieves an acceleration by more than
an order of magnitude at the smallest quark masses considered.

On other lattices, the
deflated Schwarz-preconditioned 
solver for the O($a$)-improved Wilson--Dirac equation
performs as well as in the case
reported in Fig.~\ref{DflPerf}.
In particular,
the accumulated experience unambiguously shows that
the algorithm overcomes the $V^2$-problem and that
the critical slowing down 
towards the chiral limit, which previously hampered
quark-propagator computations, is nearly eliminated.

\chapter{Simulation algorithms}

Lattice QCD simulations are based on 
Markov chains and the concept of importance sampling.
More specifically, most large-scale simulations 
performed today rely on some variant of 
the so-called Hybrid Monte Carlo algorithm
\shortcite{HMC}.
An exception to this rule are simulations of the 
pure $\Group$ gauge theory, where
link-update algorithms are usually preferred for reasons
of efficiency.

QCD simulation algorithms
have a long history and
incorporate many ideas
and improvements.
Some important deficits remain, however, 
and further progress in algorithms will no doubt be required
to be able to perform accurate simulations of a significantly 
wider range of lattices than is possible at present.

\section{Importance sampling}

\subsection{Statistical interpretation of the functional integral}

The quark fields in the QCD functional integral take values
in a Grassmann algebra. So far no practical method has been devised
that would allow such fields to be simulated directly. 
The theory simulated is therefore always the one
obtained after integrating out the quark fields.

In the case of two-flavour lattice QCD with 
mass-degenerate Wilson quarks, the partition function
then assumes the form
\begin{equation}
  {\mathcal Z}=\int\rmD[U]\left\{\det D(U)\right\}^2\rme^{-\Sg(U)},
  \qquad
  \rmD[U]=\prod_{x,\mu}\rmd U(x,\mu),
\end{equation}
where $D(U)$ denotes the massive Wilson--Dirac operator
in presence of the gauge field $U$,
$\Sg(U)$ the gauge action
and $\rmd U(x,\mu)$ the $\Group$-invariant integration measure
for the link variable $U(x,\mu)$. 
Since $D^{\dagger}=\dirac{5}D\dirac{5}$, the quark determinant $\det D$ is real
and the product
\begin{equation}
  p(U)={1\over\mathcal Z}
  \left\{\det D(U)\right\}^2\rme^{-\Sg(U)}
  \label{ProbDensity}
\end{equation}
is therefore a normalized 
probability density on the space of all gauge fields.
The physics described by the theory is eventually extracted from
expectation values
\begin{equation}
  \langle\Obs\rangle=
  \int\rmD[U]\,p(U)\Obs(U)
\end{equation}
of observables $\Obs(U)$ such as a Wilson loop
or a quark-line diagram. From this point of view,
lattice QCD thus looks like a classical statistical system, where the states
(the gauge-field configurations) occur with a 
certain probability and where one is 
interested in the expectation values of some properties of the states.

The simulation algorithms discussed in the following depend on the 
existence of a probabilistic representation of the theory.
In particular, 
they do not apply to two-flavour QCD with non-degenerate quark masses
and three-flavour QCD unless the product
of the quark determinants is guaranteed to be non-negative
(as is the case if the lattice Dirac operator
preserves chiral symmetry).

\subsection{Representative ensembles}

Representative ensembles $\{U_1,\ldots,U_N\}$ of gauge fields
are obtained by choosing the fields randomly with probability
$\rmD[U]\,p(U)$, i.e.~such that
\begin{equation}
  \hbox{no.~of fields in $\frak R$}=\int_{\frak R}\rmD[U]\,p(U)
  +\rmO(N^{-1/2})
\end{equation}
for any open region $\frak R$ in field space,
the term 
of order $N^{-1/2}$ being a statistical error that
depends on $\frak R$ and the generated ensemble
of fields.
The high-probability regions thus contain many fields $U_i$ and
are therefore sampled well, while in other areas of field space
there may be only a few fields or none at all (if $N$ is not
astronomically large).

Given a representative ensemble of fields,
the expectation values of the observables of interest
can be estimated through
\begin{equation}
  \langle\Obs\rangle=
  {1\over N}\sum_{i=1}^N\Obs(U_i)+\rmO(N^{-1/2}).
\end{equation}
A bit surprising may be the fact that results with small 
statistical errors
can often be obtained in this way even if the ensemble 
contains only $100$ or perhaps $1000$ field configurations.
Naive estimates, taking the dimension of field space into account,
actually suggest that an accurate numerical evaluation of 
the functional integral requires some $k^{32n}$ configurations,
where $n$ is the number of 
lattice points and $k$ at least $10$ or so.

The apparent paradox is resolved by
noting that small ensembles of field configurations 
can only capture some aspects of the theory.
That is, one should not expect to obtain
the expectation values of all possible observables
with small statistical errors. Typically
any quantity sensitive to the correlations of the field variables
at large distances tends to have large statistical errors, sometimes
to the extent that the results of the computation are completely
useless.

\subsection{Translation symmetry and the infinite-volume limit}

In order to minimize finite-volume effects, 
periodic boundary conditions are usually imposed on all fields,
an exception being the quark fields, which are often taken to be anti-periodic
in time. Since the translation symmetry of the theory
is preserved by this choice of boundary conditions,
representative ensembles of gauge fields are
expected ``to look the same'' in distant regions of 
a large lattice.

The meaning of this statement is best explained by 
dividing the lattice into blocks $\Lambda$, as
in Section 1.3.4, and by
considering an extensive quantity
\begin{equation}
  {\mathcal E}=\sum_x\Obs(x),
  \label{ExtensiveQuantity}
\end{equation}
where $\Obs(x)$ is some local gauge-invariant field.
Translation symmetry then implies that the contributions
$\langle\Obs_{\Lambda}\rangle$ to the sum
\begin{equation}
  \langle{\mathcal E}\rangle=
  \sum_{\Lambda}\langle\Obs_{\Lambda}\rangle,
  \qquad\Obs_{\Lambda}=\sum_{x\in\Lambda}\Obs(x),
  \label{BlockSum}
\end{equation}
are all equal. Moreover,
the statistical fluctuations of $\Obs_{\Lambda}$
and $\Obs_{\Lambda'}$ are practically
uncorrelated and tend to cancel one another
in the sum (\ref{BlockSum}) if the 
blocks $\Lambda$ and $\Lambda'$ are separated by a distance
larger than the 
range of the connected correlation function of $\Obs(x)$.
For a fixed ensemble size, the statistical error of 
the density $\langle{\mathcal E}\rangle/V$ therefore
decreases like $V^{-1/2}$ for $V\to\infty$, 
i.e.~the statistics is effectively 
multiplied by a factor proportional to $V$.

The discussion also illustrates the fact that the 
efficiency of importance
sampling depends on the observable considered and on
how its expectation value is calculated. 
Rather than from eqn~(\ref{ExtensiveQuantity}),
one might actually start from the identity
$\langle{\mathcal E}\rangle=V\langle\Obs(0)\rangle$, in which
case the information contained in the field ensemble
away from the origin $x=0$ remains unused.
Both calculations yield the correct expectation value, 
but only the first profits from the available data on the 
full lattice
and consequently obtains the result with much better
statistical precision.

\subsection{Simulating gaussian distributions}

Representative ensembles of fields can be easily constructed
in the case of a complex field $\phi(x)$ distributed according to the
gaussian probability density
\begin{equation}
  p_{A}(\phi)\propto\exp\Bigl\{-\sum_x\phi(x)^{\dagger}
  (A\phi)(x)\Bigr\},
\end{equation}
where $A$ is a hermitian, strictly positive linear operator.
If one sets
\begin{equation}
  A=-\Delta+m_0^2,
  \qquad\hbox{$\Delta$: lattice laplacian},
\end{equation}
for example,
the theory describes a free scalar field with bare mass $m_0$.
Another possible choice of $A$ is
\begin{equation}
  A=(DD^{\dagger})^{-1},
  \qquad\hbox{$D$: lattice Dirac operator}.
\end{equation}
The field $\phi(x)$ must be a
pseudo-fermion field in this case,
i.e.~a complex-valued and therefore bosonic quark field
(see Section 2.5.1).

A representative ensemble of fields $\phi_1,\ldots,\phi_N$
may be generated for any gaussian distribution by randomly choosing a set
$\chi_1,\ldots,\chi_N$ of fields with normal distribution $p_1(\chi)$
and by setting
\begin{equation}
  \phi_i(x)=(B\chi_i)(x),
  \qquad i=1,\ldots,N,
\end{equation}
where $B$ is an operator satisfying
\begin{equation}
  A=(BB^{\dagger})^{-1}.
\end{equation}
In the case of the pseudo-fermion
fields, for example, one can simply take $B=D$, while a possible
choice in the case of the free scalar field is
\begin{equation}
  B=(-\Delta+m_0^2)^{-1/2}.
\end{equation}
It is then straightforward to check that the fields $\phi_i$ 
generated in this way are correctly distributed. 

Since the fields $\chi_i(x)$ are normally distributed, their
components at different lattice points are decoupled and can be 
drawn randomly one after another.
The generation of random numbers on a computer
is a complicated subject, however, 
with many open ends and a vast literature
(for an introduction, see \shortciteN{KnuthII}, for example). 
For the time being, one of
the simulation-quality random number generators
included in the GNU Scientific Library
({\tt http://www.gnu.org/software/gsl}) may be used,
among them the {\tt ranlux} generator
\shortcite{ranluxI,ranluxII}, which is based on a strongly
chaotic dynamical system and thus comes with some
theoretical understanding of why the generated
numbers are random. An efficient ISO C code for this
generator can be downloaded from {\tt http://cern.ch/luscher/ranlux}.

\section{Markov chains}

Gaussian distributions are an exceptionally
simple case where representative ensembles of fields can be 
generated instantaneously.
In general, however, representative ensembles 
are generated through some recursive procedure (a Markov process)
which obtains the field configurations one after another
according to some stochastic algorithm.

This section is devoted to a theoretical discussion of such Markov
processes. Rather than QCD or the $\Group$ gauge theory,
an abstract discrete system will be considered 
in order to avoid some technical complications, which 
might obscure the mechanism on which Markov-chain
simulations are based.
The discrete system is left unspecified, but is assumed to 
have the following properties:
\begin{itemize}
\itemsep=0.6ex\it
\item[\rm(a)]{There is a finite number $n$ of states $s$.}
\item[\rm(b)]{The equilibrium distribution $P(s)$ 
              satisfies $P(s)>0$ for all $s$\\ and $\sum_s P(s)=1$.}
\item[\rm(c)]{The observables are real-valued functions
              $\Obs(s)$ of the states $s$.}
\end{itemize}
One is then interested in calculating the expectation values
\begin{equation}
   \langle\Obs\rangle=\ssum{s}\Obs(s)P(s)
\end{equation}
of the observables $\Obs(s)$.

\subsection{Transition probabilities}

A Markov chain is a random sequence 
$s_1,s_2,s_3,\ldots,s_N$ of states,
where $s_k$ is obtained from $s_{k-1}$ 
through some stochastic algorithm.
The chain thus depends on the initial state $s_1$ and the 
transition probability $T(s\to s')$ to go from the current state
$s$ to the next state $s'$. 
In the following, the basic idea is to choose
the latter such that the Markov chain provides 
a representative ensemble of states for large $N$.
The expectation value of any observable $\Obs(s)$
is then given by
\begin{equation}
   \langle\Obs\rangle={1\over N}\sum_{k=1}^N\Obs(s_k)+\rmO(N^{-1/2}),
\end{equation}
where the error term is dominated by the random fluctuations
of the chain.

When trying to construct such transition probabilities,
one may be guided by the following plausible requirements:

\begin{itemize}
\itemsep=1.5ex\it
\item[\rm 1.]{$T(s\to s')\geq0$ for all $s,s'$ and
           $\sum_{s'}T(s\to s')=1$ for all $s$.}
\item[\rm 2.]{$\sum_s P(s)T(s\to s')=P(s')$ for all $s'$.}
\item[\rm 3.]{$T(s\to s)>0$ for all $s$.}
\item[\rm 4.]{If $S$ is a non-empty proper subset of states, there exist
          two states\\ $s\in S$ and $s'\notin S$ such that
          $T(s\to s')>0$.}
\end{itemize}

\noindent
Property~1 merely guarantees that $T(s\to s')$ is a probability 
distribution in $s'$ for any fixed $s$, while property~2 says that
the equilibrium distribution should be preserved by the update process. 
The other properties ensure
that the Markov process
does not get trapped in cycles (property~3, 
referred to as ``aperiodicity'') or in subsets of
states (property~4, ``ergodicity'').

Later it will be shown that any transition probability $T(s\to s')$ 
satisfying $1-4$ generates Markov chains that simulate the system
in the way explained above. These properties are thus sufficient
to guarantee the correctness of the procedure.

\subsection{The acceptance-rejection method}

For illustration, an explicit example of a valid transition
probability will now be constructed. In order to simplify the
notation a little bit, the states are labeled from $0$ to $n-1$ and 
are thought to be arranged on a circle so that the neighbours of 
the state $i$ are the states $i\pm1\mod n$.
The construction
then starts from the transition probability
\begin{equation}
  T_0(i\to j)=
  \left\{\begin{array}{c@{\qquad}l}
           \frac{1}{3}  & \hbox{if $j=i$ or $j=i\pm1\mod n$},\\[0.5ex]
                0       & \hbox{otherwise,}
         \end{array}\right.
\end{equation}
which generates a random walk on the circle.
This transition probability
satisfies $1-4$, the equilibrium distribution being
the flat distribution $P_0(i)=1/n$.

\begin{figure}[t]
\begin{center}
  \includegraphics[clip,scale=0.50]{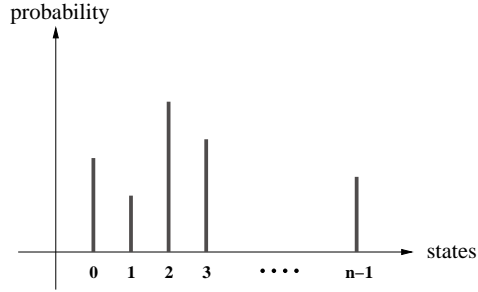}
\end{center}
\caption{
Example of an equilibrium probability distribution
of the abstract discrete system considered in this section. 
The update algorithm implementing the transition probability
(\ref{AcceptReject}) generates a random walk in the space of states,
where, in each step, one moves from the current state $i$
to one of its neighbours $j$
with probability $\frac{1}{3}\Pacc(i,j)$
or else stays at $i$.
}
\label{ProbDist}
\end{figure}

Now if the equilibrium 
distribution is not flat, as the one shown in Fig.~\ref{ProbDist},
a valid transition probability is given by
\shortcite{Metropolis}
\begin{equation}
  T(i\to j)=T_0(i\to j)\Pacc(i,j)+\delta_{ij}
  \ssum{k}T_0(i\to k)\left(1-\Pacc(i,k)\right),
  \label{AcceptReject}
\end{equation}
where
\begin{equation}
  \Pacc(i,j)=\min\left\{1,P(j)/P(i)\right\}
\end{equation}
is the so-called acceptance probability. In other words, 
starting from the current state $i$, the next state
$j$ is proposed with probability $T_0(i\to j)$.
A random number $r\in[0,1]$ is then chosen, with uniform distribution, 
and $j$ is accepted if $P(j)\geq rP(i)$. If the proposed state
is not accepted, the next state is taken to be the state $i$.

The acceptance-rejection method is widely used in various incarnations.
In general, the challenge is to find an a priori transition
probability $T_0(i\to j)$ where the proposed states
are accepted with high probability, as otherwise the simulation
will be very slow and therefore inefficient.
The transition probability (\ref{AcceptReject}) incidentally satisfies
\begin{equation}
  P(i)T(i\to j)=P(j)T(j\to i)\quad\hbox{for all}\quad i,j,
\end{equation}
a property referred to as ``detailed balance''. 
In the literature,
detailed balance is sometimes required for acceptable algorithms, but
the condition is quite strong and not needed to ensure the correctness
of the simulation.

\subsection{Implications of properties 1--4}

In the following, it is assumed that $T(s\to s')$ 
is a given transition probability satisfying the conditions 
1--4 listed in Section~2.2.1. 
The associated Markov process
is then shown to have certain mathematical properties,
which will allow, in Section 2.2.4,
to determine its asymptotic behaviour at large times.

Let $\Hspace$ be the linear space of real-valued functions $f(s)$
defined on the set of all states $s$. A useful norm of such functions is
given by
\begin{equation}
  \|f\|_1=\ssum{s}|f(s)|.
\end{equation}
The transition probability $T(s\to s')$ defines a linear operator $T$
in $\Hspace$ through
\begin{equation}
  (Tf)(s')=\ssum{s}f(s)T(s\to s').
\end{equation}
Note that, by properties 1 and 2,
the equilibrium distribution $P(s)$ has unit norm and 
is an eigenfunction of $T$ with eigenvalue $1$.

\begin{lemma}
\label{lemmaA}
For all $f\in\Hspace$, the bound $\|Tf\|_1\leq\|f\|_1$ holds. Moreover,
if $Tf=f$, there exists $c\in\rz$ such that $f(s)=cP(s)$ for all states $s$.
\end{lemma}

\vspace{0.0ex minus 1.0ex}
\begin{proof}
Any given function $f\in\Hspace$ may be decomposed 
into positive and negative parts according to
\begin{equation}
   f(s)=f_{+}(s)-f_{-}(s), 
   \qquad f_{\pm}(s)=\frac{1}{2}\{|f(s)|\pm f(s)\}\geq0.
\end{equation}
Property 1 then implies
\begin{equation}
   \|Tf_{\pm}\|_1=
   \ssum{s'}\bigl|\ssum{s}f_{\pm}(s)T(s\to s')\bigr|=
   \ssum{s}f_{\pm}(s)\ssum{s'}T(s\to s')=
   \|f_{\pm}\|_1.
\end{equation}
Using the triangle inequality, this leads to
\begin{eqnarray}
   \|Tf\|_1&=&\|Tf_{+}-Tf_{-}\|_1
   \nonumber\\[1.5ex]
   &\leq&\|Tf_{+}\|_1+\|Tf_{-}\|_1=
   \|f_{+}\|_1+\|f_{-}\|_1=
   \|f\|_1,
   \label{Tfnorm}
\end{eqnarray}
which proves the first statement made in the lemma.
Moreover, eqn~(\ref{Tfnorm})
shows that the equality $\|Tf\|_1=\|f\|_1$ holds
if and only if $Tf_{+}$ and $Tf_{-}$ have disjoint support.

Now let $f\in\Hspace$ be such that $Tf=f$. 
The functions 
$Tf_{+}$ and $Tf_{-}$ must have disjoint support in this case 
and are both non-negative.
Since the decomposition $Tf=f_{+}-f_{-}$
in positive and negative parts is unique, it follows that
$Tf_{+}=f_{+}$ and $Tf_{-}=f_{-}$. 
Properties~3 and 4 however imply that the support of a non-negative
function grows when $T$ is applied, unless the support is empty
or the whole set of states.
Either $f_{+}$ or $f_{-}$ must therefore be equal to zero.
The function $f$ thus has a definite sign.
 
Another function $g\in\Hspace$ may now be defined through
\begin{equation}
   g(s)=f(s)-cP(s),
   \qquad 
   c=\ssum{s}f(s).
\end{equation}
This function must also have a definite sign since $Tg=g$.
Moreover, $\sum_s g(s)=0$ by construction, which can only
be true if $g=0$, i.e.~if $f(s)=cP(s)$.\end{proof}

\vspace{1.0ex plus 0.5ex}
Lemma \ref{lemmaA} shows that the equilibrium distribution 
is the only distribution that is stationary under the action of
the operator $T$.
Some control over the complementary space
\begin{equation}
   \Hspace_0=\left\{f\in\Hspace\bigm|\ssum{s}f(s)=0\right\}
\end{equation}
of non-stationary functions will later be required as well.
To this end, it is helpful to introduce a funny scalar product
in $\Hspace$,
\begin{equation}
   (f,g)=\ssum{s}f(s)P(s)^{-1}g(s),
   \label{ScalarProduct}
\end{equation}
and the associated norm $\|f\|=(f,f)^{1/2}$.

\begin{lemma}
\label{lemmaB}
There exists $0\leq\rho<1$ such
that $\|Tf\|\leq\rho\|f\|$ for all $f\in\Hspace_0$.
\end{lemma}

\vspace{0.0ex minus 1.0ex}
\begin{proof}
With respect to the scalar product (\ref{ScalarProduct}), the adjoint
$T^{\dagger}$ of the operator $T$ may be defined and thus also
the symmetric operator $\That=T^{\dagger}T$. The action of 
$\That$ on any function $f\in\Hspace$ is given by
\begin{eqnarray}
   (\That f)(s')&=&\ssum{s}f(s)\That(s\to s'),\\[1.5ex]
   \That(s\to s')&=&\ssum{r}T(s\to r)P(r)^{-1}T(s'\to r)P(s').
   \label{That}
\end{eqnarray}
A moment of thought then shows that $\That(s\to s')$
is a transition probability satisfying conditions 1--4. 
In particular, Lemma \ref{lemmaA} holds for $\That$ as well.

Since $\That$ is symmetric, there exists a complete set of eigenfunctions
$v_i\in\Hspace$ of $\That$ with real eigenvalues $\lambda_i$
($i=0,1,\ldots,n-1$).
Without loss one may assume that
\begin{equation}
   \lambda_0\geq\lambda_1\geq\ldots\geq\lambda_{n-1},
   \qquad
   (v_i,v_j)=\delta_{ij}.
\end{equation}
Moreover, noting $\lambda_i=(v_i,\That v_i)=\|Tv_i\|^2\geq0$,
lemma \ref{lemmaA} (for $\That$) implies
\begin{equation}
   \lambda_0=1,\quad
   v_0=P\quad\hbox{and}\quad\lambda_1<1.
\end{equation}
Now since the states $v_i$, $i\geq1$, satisfy
\begin{equation}
   0=(v_0,v_i)=\ssum{s}v_i(s),
\end{equation}
they form a basis of the subspace $\Hspace_0$, i.e.~$\That$ maps
$\Hspace_0$ into itself and its largest eigenvalue in this subspace
is $\lambda_1$. As a consequence,
\begin{equation}
   \|Tf\|^2=(f,\That f)\leq\lambda_1\|f\|^2
   \quad\hbox{for all}\quad f\in\Hspace_0.
\end{equation}
This proves the lemma and also shows that 
the smallest possible value of the constant 
$\rho$ is $\lambda_1^{1/2}$.\end{proof}

\subsection{Statistical properties of Markov chains}

When analyzing Markov chains, an important conceptual point
to note is that one cannot reasonably speak of the statistical
properties of a single chain. However, one can ask what the average
properties of the generated sequences of states are if many
independent chains are considered. In practice, this
corresponds to running several simulations in parallel,
with different streams of random numbers. 

In the following theoretical discussion, it will be assumed
that an infinite number of independent simulations of length $N$
have been performed, with the same initial state $s_1$ and
a transition probability $T(s\to s')$ that has properties 1--4.
As before, the expectation value of an observable $\Obs$
will be denoted by $\langle\Obs\rangle$, while 
\begin{equation}
  \avg{\Obs}={1\over N}\sum_{k=1}^N\Obs(s_k)
\end{equation}
stands for its average over the states $s_1,s_2,\ldots,s_N$
generated in the course of a given simulation. 
One may then also consider functions $\phi(s_1,\ldots,s_N)$
of these sequences of states and their average 
$\avgavg{\phi}$ over the infinitely many parallel simulations.

\subsubsection{(a) Probability distribution of the states} 
The state $s_k$ generated after $k-1$ steps 
changes randomly from one simulation to another.
It is not difficult to show that the probability
$P_k(s)=\avgavg{\delta_{ss_k}}$ for $s_k$ to coincide with $s$
is given by
\begin{eqnarray}
  P_k(s)&=&\sum_{s_2,s_3,\ldots,s_{k-1}}T(s_1\to s_2)T(s_2\to s_3)\ldots
  T(s_{k-1}\to s)
  \nonumber\\[1.5ex]
  &=& (T^{k-1}P_0)(s),\qquad P_0(s)=\delta_{ss_1}.
\end{eqnarray}
Noting $P_0=P+f$, $f\in\Hspace_0$, property 2 and lemma \ref{lemmaB} 
then imply
\begin{equation}
  P_k(s)\mathrel{\mathop=_{k\to\infty}}
  P(s)+\rmO\bigl(\rme^{-k/\tau}\bigr),
  \label{PkLimit}
\end{equation}
where $\tau=-1/\ln\rho>0$
is the so-called exponential autocorrelation time of the Markov process.
The state $s_k$ is thus distributed according to the equilibrium
distribution if $k$ is much larger than $\tau$. In particular, after
so many steps, there is no memory of the initial state $s_1$ anymore
and one says that the simulation has ``thermalized''.

\subsubsection{(b) Calculation of expectation values}
Together with the generated states, the 
average $\avg{\Obs}$ of the ``measured'' values of an
observable $\Obs$ fluctuates randomly about the mean value
\begin{equation}
  \avgavg{\avg{\Obs}}
  ={1\over N}\sum_{k=1}^N\avgavg{\Obs(s_k)}
  =\sum_s\Obs(s){1\over N}\sum_{k=1}^NP_k(s).
\end{equation}
Recalling eqn~(\ref{PkLimit}), this formula shows that
\begin{equation}
  \avgavg{\avg{\Obs}}
  =\langle\Obs\rangle
\end{equation}
if the first $k\gg\tau$ measurements of $\Obs$ are dropped.
Averages of the measured values calculated after
thermalization thus coincide with the expectation value 
$\langle\Obs\rangle$ up to statistical fluctuations.

\subsubsection{(c) Autocorrelation functions}
The states $s_k$ in a Markov chain are
statistically dependent to some extent, because they are generated
one after another according to the
transition probability $T(s\to s')$. As a consequence, the 
measured values of an observable $\Obs$ are statistically correlated, 
i.e.~the autocorrelation
function
\begin{equation}
  \Gamma(t)=\avgavg{\Obs(s_k)\Obs(s_{k+t})}-
  \avgavg{\Obs(s_k)}\avgavg{\Obs(s_{k+t})}
\end{equation}
does not vanish.
For $k\gg\tau$ and $t\geq0$, the autocorrelation function is independent
of $k$ and given by
\begin{equation}
  \Gamma(t)=\sum_{s_k,s_{k+1},\ldots,s_{k+t}}
  P(s_k)\Obs(s_k)T(s_k\to s_{k+1})\ldots
  T(s_{k+t-1}\to s_{k+t})\Obs(s_{k+t})
  -\langle\Obs\rangle^2.
\end{equation}
Moreover, noting $P(s)\Obs(s)=P(s)\langle\Obs\rangle+f(s)$, $f\in\Hspace_0$,
it follows from this expression and lemma \ref{lemmaB}
that $\Gamma(t)$ falls off exponentially, 
like $\rme^{-t/\tau}$, at large separations $t$.
The measured values $\Obs(s_i)$ and $\Obs(s_j)$ are thus independently
distributed if $|i-j|\gg\tau$.  

\subsubsection{(d) Statistical fluctuations}
The statistical variance of the
averages $\avg{\Obs}$ of an observable $\Obs$ is,
after thermalization, given by 
\begin{equation}
  \avgavg{(\avg{\Obs}-\langle\Obs\rangle)^2}
  ={1\over N^2}\sum_{l,j=1}^N\Gamma(|l-j|)
  =\Gamma(0){2\tau_{\Obs}\over N}+\rmO(N^{-2}),
\end{equation}
where
\begin{equation}
  \tau_{\Obs}
  ={1\over 2}+\sum_{t=1}^{\infty}{\Gamma(t)\over\Gamma(0)}
\end{equation}
denotes the so-called integrated autocorrelation time of $\Obs$.
Up to terms of order $N^{-3/2}$, the standard deviation of 
$\avg{\Obs}$ from its expectation value 
$\langle\Obs\rangle$ is thus
\begin{equation}
  \sigma=\sigma_0\left({2\tau_{\Obs}\over N}\right)^{1/2},
  \qquad
  \sigma_0=\langle(\Obs-\langle\Obs\rangle)^2\rangle^{1/2},
  \label{StatError}
\end{equation}
i.e.~the statistical error of $\avg{\Obs}$ decreases proportionally
to $N^{-1/2}$. Note that
$\sigma_0$ is just the standard deviation
of $\Obs$ in equilibrium. In particular, $\sigma_0$ is a property 
of the system rather than of the Markov process. 
The integrated autocorrelation
time $\tau_{\Obs}$, on the other hand, often strongly depends
on the simulation algorithm.

The fact that the measured values of $\Obs$ are correlated 
leads to an increase of the variance 
of $\avg{\Obs}$ by the factor $2\tau_{\Obs}$ and thus
lowers the efficiency of the simulation. 
In practice one frequently chooses
to measure the observables only on a subsequence of states separated
by some fixed distance $\Delta t$ in simulation time. 
As long as $\Delta t$ is not much larger than 
$2\tau_{\Obs}$, the depletion of the measurements has
no or little influence on the
statistical error of $\avg{\Obs}$ and therefore 
helps reducing the computational load.

\section{Simulating the SU(3) gauge theory}

The theory of Markov chains developed in Section 2.2 
can be extended to non-discrete systems like
the pure SU(3) gauge theory on a finite lattice. 
Markov chains are sequences $U_1,U_2,\ldots,U_N$ of gauge-field
configurations in this case, which are generated according to some 
transition probability.
The latter must satisfy certain conditions analogous to those
listed in Section 2.2.1 for the discrete system.
The mathematics required at this point however tends to be 
quite heavy and no attempt will be made to
prove of the correctness of the procedure
(see \shortciteN{Tierney}, for example).

\subsection{Transition probability densities}

The equilibrium probability density 
to be simulated is
\begin{equation}
  p(U)={1\over\mathcal Z}\rme^{-\Sg(U)},
  \qquad
  {\mathcal Z}=\int\rmD[U]\,\rme^{-\Sg(U)},
\end{equation}
where the gauge action $\Sg(U)$ is assumed to be 
a bounded function of the gauge field $U$.
Markov processes in this theory are characterized
by a transition probability density $T(U\to U')$ 
that specifies the probability $\rmD[U']\,T(U\to U')$
for the next configuration to be in the volume element $\rmD[U']$
at $U'$ when the current configuration is $U$.

Note that transition probability densities 
may involve $\delta$-function and other singularities,
but the product $\rmD[U']\,T(U\to U')$ 
must be a well-defined measure on the field manifold for 
any given $U$. The obvious requirements are then

\begin{itemize}
\itemsep=1.5ex\it
\item[\rm 1.]{$T(U\to U')\geq0$ for all $U,U'$ and
          $\int\rmD[U']\,T(U\to U')=1$ for all $U$.}
\item[\rm 2.]{$\int\rmD[U]\,p(U)T(U\to U')=p(U')$ for all $U'$.}
\end{itemize}

\noindent
Further conditions need to be added, however,
in order to guarantee the aperiodicity, the ergodicity and thus 
the convergence of the Markov process. 
A sufficient but fairly strong condition is

\begin{itemize}
\itemsep=1.5ex\it
\item[\rm 3.]{Every gauge field $V$ has an open neighbourhood 
          ${\mathcal N}$ in field space\\ 
          such that $T(U\to U')\geq\eps$ for some $\eps>0$ and all 
          $U,U'\in{\mathcal N}$.}
\end{itemize}

\noindent
This property ensures that, in every step, the Markov process spreads out
in an open neighbourhood of the current field. Moreover,
using the compactness of the field manifold, it is possible
to show that the process will reach any region in field space
in a finite number of steps.

While properties 1--3 guarantee the asymptotic
correctness of the simulations,
the rigorous upper bounds on the exponential autocorrelation time
obtained in the course of the convergence proofs
tend to be astronomically large.
In practice, simulations of lattice QCD therefore remain
an empirical science to some extent, where 
one cannot claim, with absolute certainty, that 
the simulation results are statistically correct.

\subsection{Link-update algorithms}

If $T_1$ and $T_2$ are two transition probability densities satisfying
1 and 2, so does their composition
\begin{equation}
  T(U\to U')=\int\rmD[V]\,T_1(U\to V)T_2(V\to U').
\end{equation}
An update step according to the composed transition probability density
first obtains the intermediate field $V$ with probability
$\rmD[V]\,T_1(U\to V)$ and then generates $U'$ with probability
$\rmD[U']\,T_2(V\to U')$. Composition allows
simulation algorithms for fields to be constructed
from elementary transitions, where a single field variable is
changed at the time.

Link-update algorithms generate the next gauge field by updating 
the link variables one after another in some order. 
The Metropolis algorithm, for example, proceeds as follows
(the SU(3) notation is summarized in Section 2.3.6):

\begin{itemize}
\itemsep=1.5ex\it
\item[\rm(a)]{Select a link $(x,\mu)$ and choose $X\in\Lie$ randomly
           in the ball $\|X\|\leq\eps$,\\ with uniform distribution,
           where $\eps$ is some fixed positive number.}
\item[\rm(b)]{Accept $U'(x,\mu)=\rme^XU(x,\mu)$ as the new value
           of the link variable on\\ the selected link with
           probability $\Pacc=\min\{1,\rme^{\Sg(U)-\Sg(U')}\}$.}
\item[\rm(c)]{Leave the link variable unchanged if the new value
              proposed in step (b)\\ is not accepted.}
\end{itemize}

\noindent
It is not difficult to write down the transition probability density
corresponding to the steps (a)--(c) and to check that 
it satisfies conditions 1 and 2. Moreover, 
the complete update cycle, where each link is visited once,
satisfies condition 3 as well.

If the gauge action is local, 
the calculation of the action difference
$\Sg(U')-\Sg(U)$ in step (b) involves only the field variables residing 
in the vicinity of the selected link.
The computer time required per
link update is then quite small.
When all factors are taken into account, including the
autocorrelation times, a more efficient
link-update algorithm is however provided by the 
combination of the heatbath algorithm 
discussed below
with a number of micro\-canonical moves (Section 2.3.5).

\subsection{Heatbath algorithm}

As a function of the field variable $U(x,\mu)$ residing
on a given link $(x,\mu)$,  
the Wilson plaquette action is of the form
\begin{equation}
  \Sg(U)=-\Re\tr\{U(x,\mu)M(x,\mu)\}+\ldots,
\end{equation}
where $M(x,\mu)$ and the terms represented by the ellipsis
do not depend on $U(x,\mu)$.
Up to a constant factor 
involving the bare gauge coupling $g_0$,
the complex $3\times3$ matrix 
\begin{equation}
  M(x,\mu)={2\over g_0^2}\sum_{\nu\neq\mu}\left\{
  \begin{array}{l}
  \scriptstyle x+\hat{\nu}\hspace{1.65cm}x\hspace{0.5cm}x+\hat{\mu}\\
  \raise-0.35cm\hbox{\hspace{0.1cm}
  \includegraphics[clip,scale=0.40]{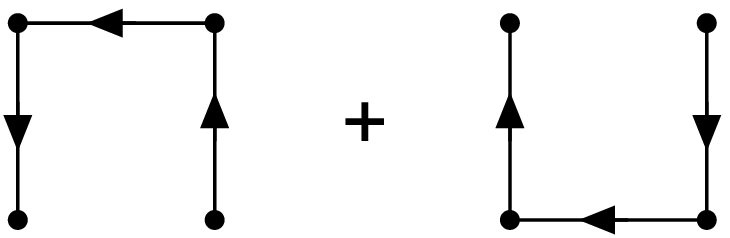}\hspace{0.2cm}}\\
  \scriptstyle\hspace{0.2cm}x\hspace{0.5cm}x+\hat{\mu}\hspace{0.6cm}x-\hat{\nu}
  \end{array}
  \right\}
  \label{StapleSum}
\end{equation} 
coincides with the ``staple sum'' of Wilson lines from $x+\hat{\mu}$ to $x$.
Other popular gauge actions, including the $\rmO(a^2)$-improved 
Symanzik action,
are of the same form except that $M(x,\mu)$ gets replaced by a sum of more
complicated Wilson lines.

The heatbath algorithm \shortcite{Creutz} is a link-update 
algorithm, where the new value 
$U'(x,\mu)$ of the field variable on the selected link
is chosen randomly with probability density proportional to 
$\exp\left(\Re\tr\{U'(x,\mu)M(x,\mu)\}\right)$.
In other words, the link variable is updated
according to its exact distribution in the presence of 
the other field variables.

While this algorithm fulfills conditions 1--3, 
it is difficult to implement in practice exactly as described here.
However, for gauge group $\SUtwo$
(the case considered by \shortciteN{Creutz}), the
situation is more favourable and there are highly efficient ways
to generate random link variables with the required
probability distribution \shortcite{FabriciusHaan,KennedyPendleton}.

\subsection{The Cabibbo--Marinari method}

The practical difficulties encountered when 
implementing the heatbath algorithm in the $\Group$ theory
can be bypassed as follows \shortcite{CabibboMarinari}.
Let $v\in\SUtwo$ be embedded in $\Group$ through
\begin{equation}
  v\to V=\left(
  \begin{array}{c@{\hspace{0.6em}}c@{\hspace{0.6em}}c}
   v_{11}& v_{12} & 0 \\[0.2ex]
   v_{21}& v_{22} & 0 \\[0.2ex]
   0& 0& 1 \\
  \end{array}
  \right).
\end{equation} 
A correct one-link update move, $U(x,\mu)\to U'(x,\mu)$, is
then obtained by choosing $v$ randomly with probability density
proportional to $\exp\left(\Re\tr\{VU(x,\mu)M(x,\mu)\}\right)$
and by setting $U'(x,\mu)=VU(x,\mu)$.
A moment of thought reveals that the distribution of $v$ is
of the same analytic form as the link distribution
in the $\SUtwo$ theory. The highly efficient methods developed for 
the latter can thus be used here too.

In order to treat all colour components of the link variables democratically,
different embeddings of $\SUtwo$ in $\SUthree$ should be used.
A popular choice is to perform $\SUtwo$ rotations 
in the $(1,2)$, $(2,3)$
and $(3,1)$ planes in colour space before proceeding 
to the next link. The ergodicity of the algorithm is then
again guaranteed.

\subsection{Microcanonical moves}

The link-update algorithms discussed so far 
tend to become inefficient when the lattice spacing $a$ is
reduced, a phenomenon known as ``critical slowing down''.
Typically the autocorrelation times of physical observables
increase approximately like $a^{-2}$.

Microcanonical algorithms are based on field transformations
that preserve the gauge action. Such transformations are
valid transitions, satisfying conditions 1 and 2, 
provided the field integration measure is preserved too.
The trajectories in field space generated by 
a microcanonical algorithm
do not involve random changes
of direction and are therefore quite different from 
the random walks performed by the other algorithms.
An acceleration of the simulation 
is then often achieved when microcanonical
moves are included in the update scheme.

A microcanonical link-update algorithm for the $\Group$ theory
is easily constructed following the steps taken in the 
case of the heatbath algorithm
\shortcite{CreutzOver,BrownWoch}.
The link variable $U(x,\mu)$ on the selected link
is again updated by applying 
Cabibbo--Marinari rotations, but the $\SUtwo$ matrix $v$
is now set to
\begin{equation}
  v={2w^2\over\tr\{w^{\dagger}w\}},
  \label{v-matrix}
\end{equation} 
where $w$ is a $2\times 2$ matrix implicitly defined by
\begin{equation}
  \Re\tr\{VU(x,\mu)M(x,\mu)\}=\tr\{vw^{\dagger}\}+\ldots
  \label{w-matrix}
\end{equation} 
and the requirement that $w$ is in $\SUtwo$ up to a real scale factor.
The existence and uniqueness of $w$ is implied by the 
reality and linearity properties of the expression on the 
left of eqn~(\ref{w-matrix}) (no update is performed if
$w$ is accidentally equal to zero).

The transformation $U(x,\mu)\to VU(x,\mu)$ defined in this way
preserves the gauge action since
$\tr\{vw^{\dagger}\}=\tr\{w^{\dagger}\}$, but
it is less obvious that it
also preserves the link integration measure,
because $w$ and therefore $v$ depend on $U(x,\mu)$,
i.e.~the transformation is non-linear.
It is straightforward to show, however, that
\begin{equation}
  \left.w\right|_{U(x,\mu)\to ZU(x,\mu)}=wz^{\dagger}
  \label{w-transform}
\end{equation}
for all $z\in\SUtwo$. Now if $f(U)$ is any integrable 
function of $U=U(x,\mu)$, the substitution $U\to ZU$ 
leads to the identity
\begin{equation}
  \int\rmd Uf(VU)=\int\rmd Uf(\tilde{V}U),
  \qquad
  \tilde{v}={2wz^{\dagger}w\over\tr\{w^{\dagger}w\}}.
\end{equation}
Since this equation holds for any $z$, it remains 
valid when integrated over $\SUtwo$. Using the invariance of the 
group integration measures under left- and right-multiplications,
one then deduces that
\begin{equation}
  \int\rmd Uf(VU)=\int\rmd z\int\rmd Uf(\tilde{V}U)
  =\int\rmd z\int\rmd Uf(Z^{\dagger}U)
  =\int\rmd Uf(U),
\end{equation}
which proves that the transformation $U\to VU$
preserves the link integration measure.

Microcanonical simulation algorithms are not ergodic and 
must therefore be combined with an ergodic one.
A recommended scheme consists in updating all link variables
once using the heatbath algorithm and subsequently
$n$ times using microcanonical moves.
This combination is more efficient than the pure Metropolis or
heatbath algorithm and it reportedly has an improved scaling
behaviour as a function of the lattice spacing 
if $n$ is scaled roughly like $a^{-1}$.

\subsection{Appendix: SU(3) notation}

The Lie algebra $\Lie$ of $\Group$ consists of all complex
$3\times3$ matrices $X$ that satisfy
\begin{equation}
  X^{\dagger}=-X\quad\hbox{and}\quad\tr\{X\}=0.
\end{equation}
With this convention, the Lie bracket $[X,Y]$ maps any 
pair $X,Y$ of $\Lie$ matrices to another element of $\Lie$.
Moreover, the exponential series
\begin{equation}
  \rme^X=1+\sum_{k=1}^{\infty}{X^k\over k!}
\end{equation}
converges to an element of $\Group$ (note the absence of 
factors of $i$ in these and the following formulae).

One can always choose a basis $T^a$, $a=1,\ldots,8$, of $\Lie$ such that
\begin{equation}
  \tr\{T^aT^b\}=-\frac{1}{2}\delta^{ab}.
\end{equation}
With respect to such a basis,
the elements $X\in\Lie$ are represented as
\begin{equation}
  X=\sum_{a=1}^8X^aT^a,\qquad
  X^a=-2\,\tr\{XT^a\}\in\rz.
\end{equation}
Moreover, the natural scalar product on $\Lie$ is given by
\begin{equation}
  (X,Y)=\sum_{a=1}^8X^aY^a=-2\,\tr\{XY\},
\end{equation}
the associated matrix norm being $\|X\|=(X,X)^{1/2}$.
If not specified otherwise, the Einstein summation convention is used
for group indices.

\section{The Hybrid Monte Carlo (HMC) algorithm}

The inclusion of the sea quarks in the simulations is difficult,
because the quark determinants in the QCD functional integral
depend non-locally on the gauge field. In particular, one-link
update algorithms would require a computational effort
proportional to the square of the lattice volume and 
are therefore not practical.

The HMC algorithm \shortcite{HMC} 
updates all link variables at once and has a much 
better scaling behaviour with respect to the lattice volume.
It will here be explained in general terms for an unspecified 
(possibly non-local) action $S(U)$, which is assumed
to be real and differentiable.

\subsection{Molecular dynamics}

As an intermediate device,
the HMC algorithm requires an $\Lie$-valued field 
\begin{equation}
  \pi(x,\mu)=\pi^a(x,\mu)T^a,
  \qquad
  \pi^a(x,\mu)\in\rz,
\end{equation}
to be added to the theory (the $\Group$ notation is as in Section 2.3.6).
The new field is interpreted as the canonical momentum
of the gauge field,  the associated Hamilton function being
\begin{equation}
  H(\pi,U)=\frac{1}{2}(\pi,\pi)+S(U),\qquad
  (\pi,\pi)=\sum_{x,\mu}\pi^a(x,\mu)\pi^a(x,\mu).
\end{equation}
Evidently, since
\begin{equation}
  \int\rmD[U]\,\Obs(U)\kern0.5pt\rme^{-S(U)}=
  \hbox{constant}\times\int\rmD[\pi]\rmD[U]\,\Obs(U)\,\rme^{-H(\pi,U)},  
  \label{ClassicalSystem}
\end{equation}
the addition of the momentum field does not affect the 
physics content of the theory. 

In the form (\ref{ClassicalSystem}), 
the theory is reminiscent of the classical statistical
systems that describe a gas of molecules.
Hamilton's equations\footnote{
The force $F(x,\mu)$ informally coincides with 
$\partial S(U)/\partial U(x,\mu)$.
Derivatives with respect to the link variables however
need to be properly defined. According to 
eqn~(\ref{MDequation1}), the force field is obtained
by substituting 
$U(x,\mu)\to\exp\{\omega^a(x,\mu)T^a\}U(x,\mu)$, 
differentiating with respect to the real variables $\omega^a(x,\mu)$
and setting $\omega^a(x,\mu)=0$ at the end of the calculation.},
\begin{eqnarray}
   \dot{\pi}(x,\mu)&=&-F(x,\mu),\qquad
   F^a(x,\mu)=
   \left.{\partial S(\rme^{\omega}U)\over\partial\omega^a(x,\mu)}
   \right|_{\omega=0},
   \label{MDequation1}
   \\[1.0ex]
   \dot{U}(x,\mu)&=&\pi(x,\mu)U(x,\mu),
   \label{MDequation2}
\end{eqnarray}
are therefore often referred to as the ``molecular-dynamics equations''.
As usual, the dot on the left of these equations 
implies a differentiation with respect to time $t$, which
is here a fictitious time unrelated to the time coordinate of space-time. 
The solutions of the molecular-dynamics equations, 
$\pi_t(x,\mu)$ and
$U_t(x,\mu)$, are uniquely determined by the initial values of 
the fields at $t=0$. They may be visualized as trajectories
in field space (or, more precisely, in phase space) parameterized
by the time $t$.

\subsection{The HMC strategy}

The basic idea underlying the HMC algorithm is to pass from the
original theory to the classical system (\ref{ClassicalSystem}) and to
evolve the fields by integrating the molecular-dynamics equations. 
Explicitly, the steps leading from the current gauge field $U(x,\mu)$ to 
the next field $U'(x,\mu)$ are the following:

\begin{itemize}
\itemsep=1.5ex\it
\item[\rm(a)]{A momentum field $\pi$ is generated randomly
           with probability\\ density proportional to 
           $\exp\bigl\{-\frac{1}{2}(\pi,\pi)\bigr\}$.}
\item[\rm(b)]{The molecular-dynamics equations are integrated from 
           time\\ $t=0$ to some later time $t=\tau$, taking 
           $\pi$ and $U$ as the initial\\ values
           of the fields.}
\item[\rm(c)]{The new gauge field $U'$ is set to the field 
           $U_{\tau}$ obtained  at time\\ $t=\tau$
           through the molecular-dynamics evolution.}
\end{itemize}

\noindent
If $\tau$ is set to a fixed value, as is usually done,
the transition probability density corresponding to the 
steps (a)--(c) is given by
\begin{equation}
  T(U\to U')={1\over{\mathcal Z}_{\pi}}
  \int\rmD[\pi]\,\rme^{-\frac{1}{2}(\pi,\pi)}
  \prod_{x,\mu}
  \delta(U'(x,\mu),U_{\tau}(x,\mu)),
  \label{HMCtransition1}
\end{equation}
where the partition function ${\mathcal Z}_{\pi}$ of the momentum
field ensures the correct normalization and
the Dirac $\delta$-function is the one 
appropriate to the gauge-field integration measure.

It is trivial to check that 
the transition probability density (\ref{HMCtransition1})
satisfies the first of the three conditions 
listed in Section 2.3.1.
The second condition is also fulfilled, but some work is required to 
show this.
An important point to note is that the 
molecular-dynamics equations are invariant
under time reversal $\pi_t,U_t\to -\pi_{\tau-t},U_{\tau-t}$.
The molecular-dynamics evolution $\pi_0,U_0\to\pi_{\tau},U_{\tau}$ 
therefore defines an invertible 
transformation of phase space.
Moreover, it preserves the Hamilton function and,
by Liouville's theorem, also the phase space integration measure.
It follows from these remarks that
\begin{eqnarray}
  \lefteqn{\int\rmD[U]\,\rme^{-S(U)}
  T(U\to U')
  ={1\over{\mathcal Z}_{\pi}}
  \int\rmD[\pi]\rmD[U]\,\rme^{-H(\pi,U)}
  \prod_{x,\mu}
  \delta(U'(x,\mu),U_{\tau}(x,\mu))}
  \nonumber\\[1.5ex]
  &&\quad={1\over{\mathcal Z}_{\pi}}
  \int\rmD[\pi_{\tau}]\rmD[U_{\tau}]\,\rme^{-H(\pi_{\tau},U_{\tau})}
  \prod_{x,\mu}
  \delta(U'(x,\mu),U_{\tau}(x,\mu))=
  \rme^{-S(U')},\hspace{1.0cm}
\end{eqnarray}
where the second equation is obtained by performing a
change of variables from $\pi,U=\pi_0,U_0$ to
$\pi_{\tau},U_{\tau}$. Condition 2 is thus satisfied too.

For sufficiently small $\tau$,
the ergodicity of the algorithm (condition 3) can be 
proved as well.
The proof is based on an expansion of 
$\pi_{\tau},U_{\tau}$ in powers of $\tau$,
from which one infers that the range of $U_{\tau}$ 
includes an open neighbourhood of $U=U_0$
when the initial momentum $\pi=\pi_0$ varies
over a neighbourhood of the origin.
In view of the non-linear nature of the theory,
HMC simulations of lattice QCD are however expected to be
ergodic at any (non-zero) value of $\tau$, even if
one is unable to show this.
Ergodicity can, in any case, always 
be rigorously ensured by
choosing $\tau\in[0,\tau_{\rm max}]$ randomly
from one update step to the next.

\subsection{Numerical integration of the molecular-dynamics equations}

In practice, the molecular-dynamics equations cannot be integrated exactly
and one must resort to 
some numerical integration method. As explained in 
Section 2.4.4, the integration error can be compensated by including
an acceptance-rejection step in the HMC algorithm so that 
the correctness of the simulation is not compromised.

\begin{figure}[t]
\begin{center}
  \includegraphics[clip,scale=0.65]{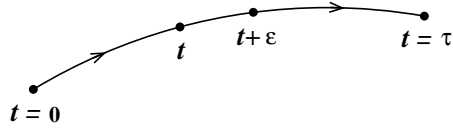}
\end{center}
\caption{
The numerical integration of the molecular-dynamics equations
proceeds in time steps of size $\eps$. In each step,
the fields at time $t+\eps$ are computed from the fields at time $t$ and 
possibly those at earlier times (if a higher-order scheme is used).
The integration rule and the step size $\eps$ should evidently be such that
the calculated fields at time $t=n\eps$, $n=0,1,\ldots,N_0$, closely
follow the exact trajectory in phase space.
}
\label{Trajectory}
\end{figure}

The numerical integration 
proceeds by dividing the time interval $[0,\tau]$
in $N_0$ steps of size $\eps$ and by applying a discrete 
integration rule
that gives the correct result in the limit $\eps\to0$
(see Fig.~\ref{Trajectory}).
Considering the Taylor expansions
\begin{eqnarray}
  \pi_{t+\eps}&=&
  \pi_t-\eps F|_{U=U_t}+\rmO(\eps^2),
  \\[1.7ex]
  U_{t+\eps}&=&
  U_t+\eps\pi_tU_t+\rmO(\eps^2),
\end{eqnarray}
it is clear that
acceptable integration schemes 
can be built from the elementary operations
\begin{eqnarray}
  \Imom(\eps):\;&&
  \pi,U\to\pi-\eps F,U,
  \label{MomStep}
  \\[1.7ex]
  \Ifld(\eps):\;&&
  \pi,U\to\pi,\rme^{\eps\pi}U.
  \label{FldStep}
\end{eqnarray}
The combination $\Imom(\frac{1}{2}\eps)\Ifld(\eps)\Imom(\frac{1}{2}\eps)$,
for example, takes $\pi_t,U_t$ to $\pi_{t+\eps},U_{t+\eps}$
up an error of order $\eps^3$. The complete integration
from time $t=0$ to time $t=\tau$ then amounts to applying the product
\begin{equation}
  \Jint_0(\eps,N_0)=\left\{
  \Imom(\frac{1}{2}\eps)\Ifld(\eps)\Imom(\frac{1}{2}\eps)
  \right\}^{N_0},
  \qquad
  \eps={\tau\over N_0},
  \label{LeapFrog}
\end{equation}
to the initial fields.

The ``leap-frog integrator'' (\ref{LeapFrog}) is 
remarkably simple and has a number of good properties.
In particular, the integration is reversible,
\begin{equation}
  \Jint_0(-\eps,N_0)\Jint_0(\eps,N_0)=1,
\end{equation}
and $\Jint_0(\eps,N_0)$ is therefore an invertible mapping of 
phase space.
Starting from the elementary integration steps (\ref{MomStep})
and (\ref{FldStep}),
it is also trivial to show that the integrator preserves
the field integration measure $\rmD[\pi]\rmD[U]$.
Through the numerical integration, these important properties
of the molecular-dynamics evolution are thus not lost.

The leap-frog integrator is widely used
and appreciated for its simplicity,
but there are many other integration schemes that can 
be employed \shortcite{LeimkuhlerReich,Hairer}.
In particular, the so-called symplectic integrators all have
the good properties mentioned above.

\subsection{Acceptance-rejection step}

For a fixed step size $\eps$,
the numerical integration of the molecular-dynamics equations
normally does not preserve the Hamilton function. In particular,
the difference
\begin{equation}
  \Delta H(\pi,U)=\left\{H(\pi_{\tau},U_{\tau})-H(\pi_0,U_0)
  \right\}_{\pi_0=\pi,U_0=U}
  \label{DeltaH}
\end{equation}
does not vanish in general.
The HMC algorithm (as defined through steps (a)--(c) in Section 2.4.2)
consequently
violates condition 2 if the integration is performed numerically.
It is possible to correct for this deficit by replacing
step (c) through

\begin{itemize}
\itemsep=1.5ex\it
\item[\rm(${\rm c}'$)]{The new gauge field $U'$ is set to the field 
           $U_{\tau}$ obtained through the\\
           integration of the molecular-dynamics
           equations with probability
\begin{equation}
  \Pacc(\pi,U)=\min\bigl\{1,\rme^{-\Delta H(\pi,U)}\bigr\}.
\end{equation}
           Otherwise, i.e.~if the proposed field is rejected,
           $U'$ is set to $U$.}
\end{itemize}

\noindent
The transition
probability density of the modified algorithm,
\begin{eqnarray}
  T(U\to U')&=&{1\over{\mathcal Z}_{\pi}}
  \int\rmD[\pi]\,\rme^{-\frac{1}{2}(\pi,\pi)}
  \Bigl\{\Pacc(\pi,U)\prod_{x,\mu}\delta(U'(x,\mu),U_{\tau}(x,\mu))
  \nonumber\\[1.5ex]
  &&+(1-\Pacc(\pi,U))
  \prod_{x,\mu}\delta(U'(x,\mu),U(x,\mu))\Bigr\},
  \label{HMCtransition2}
\end{eqnarray}
can then again be shown to have the required properties, 
for any value of the integration step size $\eps$, provided
the integrator is reversible and measure-preserving.

The adjustable parameters of the HMC algorithm are then
the trajectory length $\tau$, the integration step size $\eps$
and further parameters of the integration scheme (if any).
Evidently, the simulation will be inefficient if
the average acceptance rate is low, i.e.~if the numerical
integration is not very accurate.
The acceptance rate must however be balanced against the
computer time required for the simulation, which
grows roughly linearly with the step number $N_0=\tau/\eps$.
In the case of the leap-frog integrator, for example,
$\langle\Pacc\rangle=1-\rmO(\eps^2)$ and
the integration step size is then usually tuned so that
acceptance rates of $70-80$ percent are achieved.

It is more difficult to give a recommendation on the value of 
$\tau$. Traditionally, the trajectory length is set to $1$, but the
choice of $\tau$ can have an influence on 
the auto\-cor\-re\-lation times and
the stability of the numerical integration.
Some empirical studies are therefore
required in order to determine the optimal value of $\tau$ in 
a given case.

\section{Application to two-flavour QCD}

In QCD with a doublet of mass-degenerate sea quarks, the 
probability density to be simulated is
\begin{equation}
  p(U)={1\over\mathcal Z}\rme^{-S(U)},
  \qquad
  S(U)=\Sg(U)-\ln\left|\det D(U)\right|^2,
  \label{TwoFlavourDistribution}
\end{equation}
where $D(U)$ denotes the lattice Dirac operator
in presence of the gauge field $U$ (cf.~Section 2.1.1). 
The HMC algorithm can in principle be used to simulate this 
distribution, but a straightforward application of the algorithm
is not possible in practice, because
the calculation of the quark determinant
and of the force deriving from it would require 
an unreasonable amount of computer time.

\subsection{Pseudo-fermion fields}

This difficulty can fortunately be overcome using the
pseudo-fermion representation
\begin{eqnarray}
  &&\left|\det D(U)\right|^2=
  \hbox{constant}\times\int\rmD[\phi]\,\rme^{-\Spf(U,\phi)},
  \\[1.8ex]
  &&\Spf(U,\phi)=\bigl(D(U)^{-1}\phi,D(U)^{-1}\phi\bigr),
  \qquad
  \rmD[\phi]=\prod_{x,A,\alpha}
  \rmd\phi_{A\alpha}(x)\rmd\phi_{A\alpha}(x)^{\ast},\hspace{0.5cm}
  \label{PseudoFermionAction}
\end{eqnarray}
of the quark determinant. The auxiliary field $\phi(x)$ introduced
here carries a Dirac index $A$ 
and a colour index $\alpha$, like a quark
field, but its components are complex numbers 
rather than being elements
of a Grassmann algebra. For the scalar product 
in eqn~(\ref{PseudoFermionAction}) one can take the obvious one
for such fields, with any convenient normalization. 
Step (a) of the HMC algorithm is then replaced by

\begin{itemize}
\itemsep=1.5ex\it
\item[\rm(${\rm a}'$)]{A momentum field $\pi$ and a pseudo-fermion field
           $\phi$ are generated randomly\\ with probability density 
           proportional to 
           $\exp\bigl\{-\frac{1}{2}(\pi,\pi)-\Spf(U,\phi)\bigr\}$.}
\end{itemize}

\noindent
Note that the pseudo-fermion action is quadratic in $\phi$. The field 
can therefore be easily generated
following the lines of Section 2.1.4.
Once this is done, the algorithm proceeds as before,
where the Hamilton function to be used in
steps (b) and (${\rm c}'$) is
\begin{equation}
  H(\pi,U)=\frac{1}{2}(\pi,\pi)+\Sg(U)+\Spf(U,\phi).
  \label{TwoFlavourHamilton}
\end{equation}
Only the momentum $\pi$ and
the gauge field $U$ are evolved by the molecular-dynamics
equations. The pseudo-fermion field $\phi$ remains unchanged and
thus plays a spectator r\^ole at this point.

The steps (${\rm a}'$), (b)
and (${\rm c}'$) implement the transition probability density
\begin{eqnarray}
  &&T(U\to U')={1\over{\mathcal Z}_{\pi}{\mathcal Z}_{\rm pf}(U)}
  \int\rmD[\pi]\rmD[\phi]\,\rme^{-\frac{1}{2}(\pi,\pi)-\Spf(U,\phi)}
  \nonumber\\[1.5ex]
  &&\quad\times\Bigl\{\Pacc(\pi,U)\prod_{x,\mu}\delta(U'(x,\mu),U_{\tau}(x,\mu))
  +(1-\Pacc(\pi,U))
  \prod_{x,\mu}\delta(U'(x,\mu),U(x,\mu))\Bigr\},\nonumber\\
  \label{HMCtransition3}
\end{eqnarray}
where ${\mathcal Z}_{\rm pf}(U)$ is the partition function of 
the pseudo-fermion field $\phi$ in presence of the gauge field $U$.
For simplicity, the dependence of $U_{\tau}$ and $\Pacc$ on
$\phi$ has been suppressed.
Starting from this formula, it is then not difficult to show that 
the algorithm correctly simulates the distribution 
(\ref{TwoFlavourDistribution}).

\begin{figure}[t]
\begin{center}
  \includegraphics[clip,scale=0.55]{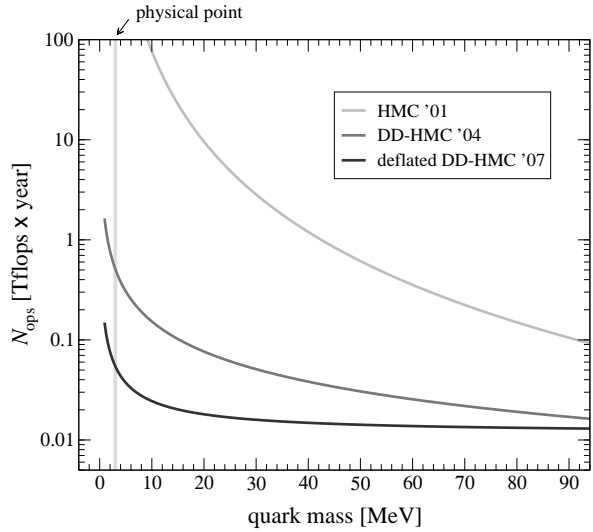}
\end{center}
\caption{
Number of floating-point operations required for the generation of
100 statistically independent
gauge-field configurations in $\rmO(a)$-improved
two-flavour QCD on a $64\times32^3$ lattice
with spacing $a=0.08$ fm. The top curve \protect\shortcite{UkawaBerlin}
represents the status reported at the memorable Berlin lattice conference 
in 2001, the middle one
was obtained a few years later,
using the so-called domain-decomposed HMC algorithm 
\protect\shortcite{DDHMC,TwoFlavourI},
and the lowest curve shows the performance of
a recently developed
deflated version of the latter
\protect\shortcite{DFLHMC}.
}
\label{Berlin}
\end{figure}

\subsection{Performance of the HMC algorithm}

The force $F$ that drives the molecular-dynamics evolution
in step (b) of the algorithm 
has two parts, $F_0$ and $F_1$, the first deriving from the gauge
action and the other from the pseudo-fermion action
in the Hamilton function (\ref{TwoFlavourHamilton}).
In the case of the Wilson theory, for example, the forces are
\begin{eqnarray}
  F_0^a(x,\mu)&=&-\Re\tr\{T^aU(x,\mu)M(x,\mu)\},
  \\[1.8ex]
  F_1^a(x,\mu)&=&-2\,\Re\bigl(
  \dirac{5}D^{-1}\dirac{5}\psi,\delta^a_{x,\mu}D\psi\bigr),  
  \qquad \psi=D^{-1}\phi,
\end{eqnarray}
where $M(x,\mu)$ is the staple sum (\ref{StapleSum}) previously
encountered in the pure gauge theory and 
\begin{eqnarray}
  \lefteqn{(\delta^a_{x,\mu}D\psi)(y)=}
  \nonumber\\[1.8ex]
  &&\quad
  \delta_{x+\hat{\mu},y}
  \frac{1}{2}(1+\dirac{\mu})U(x,\mu)^{-1}T^a\psi(x)
  -\delta_{x,y}
  \frac{1}{2}(1-\dirac{\mu})T^aU(x,\mu)\psi(x+\hat{\mu}).
  \qquad
\end{eqnarray}
Note that the computation of the pseudo-fermion force $F_1$ requires
the Dirac equation to be solved two times.
The by far largest fraction of the 
computer time is then usually spent in this part of 
the simulation program, particularly so at
small sea-quark masses where the Dirac operator becomes increasingly
ill-conditioned.

Traditionally the performance of QCD simulation algorithms is
measured by counting the number of floating-point operations
required for the generation of a sample of statistically independent
gauge-field configurations. Such performance estimations
are quite primitive and tend to be subjective to some extent,
because the term ``statistically independent''
is only loosely defined in this context.
Moreover, the specific capabilities of the computers used for the 
simulation are not taken into account. The performance estimates
plotted in Fig.~\ref{Berlin} nevertheless clearly 
show that the simulations have become significantly faster
since the beginning of the decade. 
In the following sections, some of the now widely used
acceleration techniques are briefly described.

The curves shown in Fig.~\ref{Berlin}
refer to a particular choice of the lattice 
action and the lattice parameters. 
If the lattice
volume $V$ is increased at fixed
lattice spacing and quark mass, and
if the leap-frog integrator is used, the numerical effort
required for the simulations is known to 
scale like $V^{5/4}$.
As a function of the lattice spacing $a$, the 
scaling behaviour of the HMC algorithm is more difficult to determine
and is currently a debated issue, but there is little doubt
that the required computer time increases at least like $a^{-7}$.

\subsection{Multiple time-step integration}

An acceleration of the HMC algorithm can often be achieved
using adapted integration step sizes for different
parts of the force $F$ in the molecular-dynamics
equations \shortcite{SextonWeingarten}.
The quark force $F_1$, for example, tends to be
significantly smaller than the gauge force $F_0$ and
can therefore be integrated with a larger step size 
than the latter.

\begin{figure}[t]
\begin{center}
  \includegraphics[clip,scale=0.50]{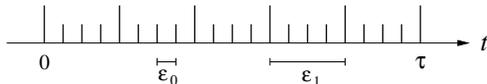}
\end{center}
\caption{
Multiple time-step integration schemes divide the integration 
range $[0,\tau]$ in a hierarchy of intervals of increasing
sizes $\eps_0,\eps_1,\ldots$ such that $\eps_{k+1}$ is an
integer multiple of $\eps_k$.
}
\label{Multistep}
\end{figure}

If the integration step sizes for 
the forces $F_0$ and $F_1$ are taken to be
\begin{equation}
  \eps_0={\tau\over N_0N_1},
  \qquad
  \eps_1={\tau\over N_1},
\end{equation}
where $N_0,N_1$ are some positive integers, the leap-frog integrators
\begin{eqnarray}
  \Jint_0(\eps_0,N_0)&=&\left\{\Iint_0(\frac{1}{2}\eps_0)\Ifld(\eps_0)
  \Iint_0(\frac{1}{2}\eps_0)\right\}^{N_0},
  \\[1.8ex]
  \Jint_1(\eps_1,N_1)&=&\left\{\Iint_1(\frac{1}{2}\eps_1)\Jint_0(\eps_0,N_0)
  \Iint_1(\frac{1}{2}\eps_1)\right\}^{N_1},
\end{eqnarray}
integrate the molecular-dynamics equations from time $t$ to $t+\eps_1$ 
and $t+\tau$, respectively
(see Fig.~\ref{Multistep}). In these equations, 
\begin{equation}
  \Iint_k(\eps):\;
  \pi,U\to\pi-\eps F_k,U
\end{equation}
are the elementary
integration steps involving the force $F_k$. In particular, the application
of $\Jint_0(\eps_0,N_0)$ consumes relatively little computer time, because
a computation of the quark force is not required. 

The hierarchical integration is profitable if 
the step size $\eps_1$ can be set to a value larger 
than $\eps_0$ without compromising the 
accuracy of the numerical integration too much.
An acceleration of the 
simulation by a factor approximately equal to $N_0$ is then 
achieved.

\subsection{Frequency splitting of the quark determinant}

The factorization
\begin{equation}
  \left|\det D\right|^2
  =\det\{\DdagD+\mu^2\}\times
  \det\biggl\{{\DdagD\over\DdagD+\mu^2}\biggr\}
  \label{tmFactorization}
\end{equation}
of the quark determinant
separates the contribution of the eigenvalues of
$\DdagD$ larger than $\mu^2$ from the contribution of 
the lower ones. 
In the HMC algorithm, the two factors may 
be represented by two pseudo-fermion fields, $\phi_1$ and $\phi_2$,
with action
\begin{equation}
  \Spf(U,\phi)=
  \bigl(\phi_1,(\DdagD+\mu^2)^{-1}\phi_1\bigr)+
  \bigl(\phi_2,\phi_2+\mu^2(\DdagD)^{-1}\phi_2\bigr).
\end{equation}
The quark force accordingly splits into two forces,
$F_1$ and $F_2$, where the first is nearly insensitive to the 
quark mass while the second involves the inverse of $\DdagD$ 
and is, in this respect, similar to the force derived from
the full quark determinant.

\begin{figure}[t]
\begin{center}
  \includegraphics[clip,scale=0.27]{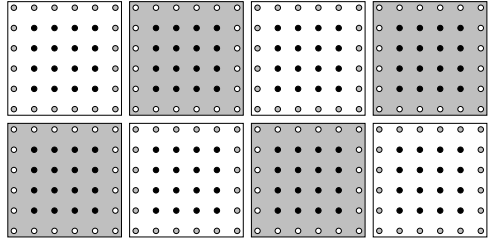}
\end{center}
\caption{
Divisions of the lattice into non-overlapping blocks of lattice points
can be chessboard-coloured if there is an even number of blocks
along each coordinate axis. The union of the sets of points contained 
in the black blocks is denoted by $\om$ and its complement (the points
in the white blocks) by $\oms$. Their exterior boundaries,  
$\dom$ and $\doms$, consist of all points in the other set
with minimal distance
from the surfaces separating the blocks
(open points).
}
\label{BlkGrid2}
\end{figure}

When such determinant factorizations were first considered
\shortcite{Hasenbusch,HasenbuschJansen},
the main effect appeared to be that the fluctuations of 
the quark force along the molecular-dynamics trajectories 
were reduced. The integration step size required for
a given acceptance rate could consequently be
increased by a factor 2 or so. 
\shortciteN{UrbachEtAl} later noted that $F_1$ 
is the far dominant
contribution to the quark force at small $\mu$.
Using a multiple time-step integrator,
and after some tuning of $\mu$,
an important acceleration of the simulation was
then achieved.

Another factorization of the quark determinant,
leading to the DD-HMC algorithm \shortcite{DDHMC},
is obtained starting from a domain decomposition of 
the lattice like the ones previously considered in 
Section 1.3.4. The r\^ole of the scale $\mu$ 
that separates the high modes of the Dirac operator from low modes
is here played by the block sizes,
which are usually
chosen to be in the range from $0.5$ to 1 fm or so.

For the associated factorization of the quark determinant to work
out, one needs to assume that the lattice Dirac operator has
only nearest-neighbour hopping terms, as in the case of 
the (improved) Wilson--Dirac operator, and that 
the block division of the lattice 
is chessboard-colourable (see Fig.~\ref{BlkGrid2}).
With respect to the subset $\om$ of points contained
in the black blocks and its complement $\oms$,
the Dirac operator then naturally decomposes into four parts,
\begin{equation}
  D=\Dom+\Doms+\Ddom+\Ddoms,
  \label{DDecomposition}
\end{equation}
where $\Dom$ includes all (diagonal and hopping) terms 
acting inside $\om$ and $\Doms$ those acting in $\oms$.
The hopping terms from the white to the black blocks
and the ones going in the opposite direction are included in
$\Ddom$ and $\Ddoms$, respectively.

Using the fact that $\om$ and $\oms$ are disjoint sets of lattice 
points, the factorization
\begin{equation}
  \det D=\det\Dom\det\Doms
  \det\{1-\Dom^{-1}\Ddom\Doms^{-1}\Ddoms\},
  \label{DDfactorization}
\end{equation}
may now be derived, where the first two factors
further factorize into the determinants of the 
block Dirac operators.
The factorization corresponds to a splitting of 
the quark force into an easy high-mode part 
(the block determinants) and 
an ``expensive'' but much smaller low-mode part.
No fine-tuning is required in this case and
an acceleration of the simulation 
is again achieved using a multiple time-step
integrator.
Moreover, the 
domain decomposition is, as usual, favourable
for the parallel processing of large lattices\footnote{
An efficient ISO C program implementing the DD-HMC algorithm
for two-flavour QCD can be downloaded from
{\tt http://cern.ch/luscher/DD\-HMC/index.html}
under the GNU Public License (GPL).
Many of the acceleration
techniques discussed here are included
in this package.}.

\subsection{Chronological inversion and deflation acceleration}

In the course of the integration of the molecular-dynamics 
equations, the Dirac equation must be solved many times.
The exact details depend on which integrator and acceleration 
techniques are used, but the equations to be solved 
are usually of the form
\begin{equation}
  D\psi=\phi,\qquad D\chi=\dirac{5}\psi,
  \label{DiracEqns}
\end{equation}
where the source field $\phi$ does not depend on the 
integration time $t$.

For small integration step sizes, the 
gauge field changes only little from one step to the next
and the solutions $\psi_t$ and $\chi_t$ obtained at time $t$
therefore tend to evolve smoothly.
One may thus
attempt to predict them from the previous solutions through
a polynomial extrapolation in $t$, for example
\shortcite{Chrono}.
In general, the predicted solutions of the Dirac
equation are not sufficiently accurate, but 
this deficit can easily be removed 
through iterative improvement (Section 1.1.2).
The total number of iterations performed 
by the solver program is then often significantly reduced.

\begin{figure}[t]
\begin{center}
  \includegraphics[clip,scale=0.38]{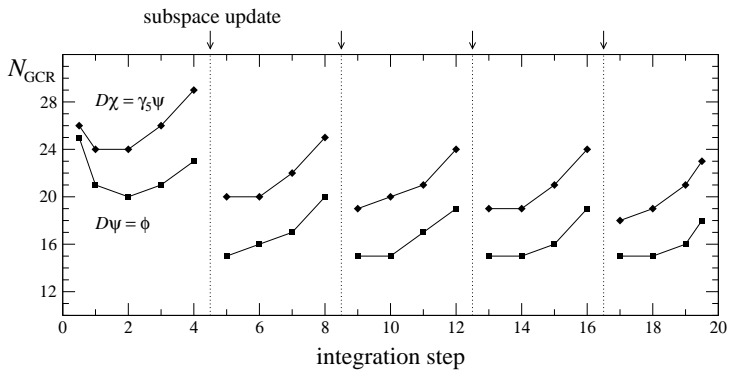}
\end{center}
\caption{
History of the GCR solver iteration numbers required for the solution
of the Dirac equation along a molecular-dynamics trajectory, showing the
effect of the refreshing of the deflation subspace. The data points 
were obtained at a sea-quark mass of $26$ MeV
on a $64\times32^3$ lattice with spacing $a=0.08$ fm. 
On other lattices, the situation 
is practically the same.
}
\label{GCRhistory}
\end{figure}

An important further reduction of the solver iteration numbers 
can be achieved using the low-mode deflation method described
in Section 1.3 \shortcite{DFLHMC}.
Since the gauge field
changes along a molecular-dynamics trajectory, the 
(domain-decomposed) deflation subspace
must be refreshed from time to time in order to preserve
its efficiency. An example illustrating this
process is shown in Fig.~\ref{GCRhistory}. 
The subspace generation at the beginning of the trajectory
and the periodic refreshing of the subspace requires
some computer time,
but this overhead is largely compensated
by the fact that the solutions of the 
Dirac equations (\ref{DiracEqns}) are obtained
much more rapidly than without deflation, 
particularly so at small quark masses (see Fig.~\ref{Berlin}).

\subsection{Improved integrators}

The use of integration schemes other than the leap-frog integrator 
can be profitable if fewer evaluations of the quark force
are required for a given integration accuracy.
Higher-order schemes, for example, where the one-step error is reduced 
to $\rmO(\eps^5)$, 
are not difficult to construct \shortcite{LeimkuhlerReich,Hairer},
but so far did not prove to be significantly faster than
the leap-frog integrator.

Another possibility is to look for $\rmO(\eps^3)$ 
integrators that minimize the integration error according to
some criterion
(\shortciteANP{OmelyanEtAlI},~\citeyearNP{OmelyanEtAlI,OmelyanEtAlII};
\shortciteANP{TakaishiForcrand},~\citeyearNP{TakaishiForcrand};
\shortciteANP{ClarkKennedySilva},~\citeyearNP{ClarkKennedySilva}).
An integrator of this kind is,
in the case of a single time-step integration, given by
\begin{equation}
  \widetilde{\Jint}_0(\eps,N_0)=
  \left\{\Imom(\frac{1}{2}\epst)\Ifld(\frac{1}{2}\eps)
  \Imom(\eps-\epst)\Ifld(\frac{1}{2}\eps)\Imom(\frac{1}{2}\epst)
  \right\}^{N_0},
  \label{OmelyanIntegrator}
\end{equation}
where $\epst\propto\eps$ is a tunable parameter.
Experience suggests that the optimal values of $\epst/\eps$
are in the range $0.3-0.5$, depending a bit on the chosen
accuracy criterion.
Although the quark force needs to be computed two times per 
integration step, this integrator achieves 
a net acceleration by a factor $1.5$ or so with respect to
the leap-frog integrator, because fewer integration steps
are required.

\section{Inclusion of the strange quark}

Unlike the light quarks, the strange and the heavy quarks
do not pair up in approximately mass-degenerate doublets.
For various technical reasons, 
single quarks are more difficult to include in the 
simulations than pairs and a special treatment is
required.

\subsection{Strange-quark determinant}

One of the issues that needs to be addressed is the fact that
the determinant
\begin{equation}
   \det\Ds=\pm\left|\det\Ds\right|
\end{equation}
of the strange-quark Dirac operator $\Ds$ 
may not have the same sign for all gauge-field configurations.
If chiral symmetry is exactly preserved on the lattice, the 
determinant is guaranteed to be positive, but sign changes
from one configuration to another
are not excluded in the case of the (improved) Wilson--Dirac operator,
for example.
The presence of positive
and negative contributions to the QCD partition function 
potentially ruins the foundations on which numerical simulations
are based. In particular, importance sampling ceases to have a clear
meaning.

In all current simulations of QCD that include the strange quark,
the regions in field space,
where the strange-quark determinant is negative, 
are assumed (or can be shown) to have
a totally negligible weight in the functional integral.
The operator in the determinant may then be replaced by
the non-negative hermitian operator
\begin{equation}
   |\Qs|=\bigl(\Qs^2\bigr)^{1/2},
   \qquad
   \Qs=\dirac{5}\Ds,
\end{equation} 
without affecting the simulation results. As explained
in the following sections, the so modified theory can
again be simulated using the HMC algorithm.

\begin{figure}[t]
\begin{center}
  \includegraphics[clip,scale=0.48]{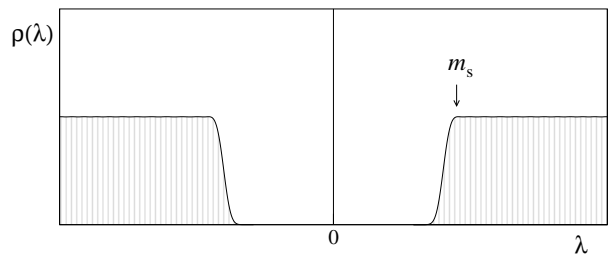}
\end{center}
\caption{
Typical shape of the density $\rho(\lambda)$ of the
eigenvalues $\lambda$ of
the hermitian Wilson--Dirac operator $\Qs$ 
near the origin.
At lattice spacings
$a\leq 0.1$ fm, and if the strange-quark mass $\ms$ 
is greater or equal to its physical value,
the distribution of the eigenvalues with the smallest
magnitude extends only slightly inside 
the spectral gap of the density 
in the continuum limit 
\protect\shortcite{Stability}. 
}
\label{SpectralGap}
\end{figure}

In the Wilson theory, the justification of this procedure
rests on the observation
that the physical strange quark is relatively heavy
and that the hermitian Dirac operator $\Qs$ consequently
tends to have a solid spectral gap in presence
of the gauge fields that dominate the QCD functional integral
(see Fig.~\ref{SpectralGap}). Since the gap tends to widen
when the strange-quark mass is increased,
the sign of the determinant of $\Qs$ cannot change
and is therefore
the same as the one at large masses, where it can be
proved to be positive.

\subsection{Pseudo-fermion representation}

Although well defined, the operator $|\Qs|$ is not directly accessible and
one is forced to use approximations whenever its action
on a quark field needs to be computed. In the present context,
an approximation of its inverse is required,
i.e.~a tractable function $R$ of $\Qs^2$ such that 
$|\Qs|R\simeq\hbox{constant}$. An exact representation
of the strange-quark determinant can then be given using
two pseudo-fermion fields,
\begin{equation}
   \det|\Qs|=
   \hbox{constant}\times
   \int\rmD[\phi_1]\rmD[\phi_2]\,\rme^{-\Spfs(U,\phi_1,\phi_2)}
\end{equation}
where only the second term in the pseudo-fermion action
\begin{equation}
   \Spfs=(\phi_1,(|\Qs|R)^{-1}\phi_1)+
         (\phi_2,R\phi_2)
   \label{SquarkAction}
\end{equation}
is included in the molecular-dynamics Hamilton function. The 
first term
is nearly independent of the gauge field 
and can be taken into account in the 
acceptance-rejection step at the end of the molecular-dynamics
evolution.

So far not many different approximations $R$ were considered.
The PHMC algorithm, for example,
employs a polynomial approximation
(\shortciteANP{PHMCI},~\citeyearNP{PHMCI,PHMCII}),
while the RHMC algorithm 
is based on a rational approximation
\shortcite{RHMCI,RHMCII}. There is also a complex version
of the PHMC algorithm, where one starts from an approximation 
of $\Ds^{-1}$ by a
polynomial in $\Ds$
\shortcite{PHMCIII,PHMCIV}.
None of these algorithms is completely trivial to implement
or obviously preferable in view of its simplicity, accessibility to acceleration
techniques or speed.

In the following, the RHMC algorithm is worked out 
as a representative case.
The steps to be taken in the
PHMC algorithm are similar,
the main differences being
issues of approximation accuracy and numerical stability.

\subsection{Optimal rational approximation}

The operator norm of the difference of two functions of $\Qs^2$
is bounded from above by the maximal absolute deviation of the functions
in the range covered by the eigenvalues of $\Qs^2$. 
For the construction of optimal approximations of $|\Qs|^{-1}$,
some information on the spectral range of $\Qs^2$ 
is therefore required as input.

Since $\Qs^2$ has a solid spectral gap,
one can choose some fixed numbers
$M$ and $\eps>0$ such
that the spectrum of $\Qs^2$ is fully contained in the interval
$[\eps M^2,M^2]$ with high probability (i.e.~for most
configurations in a representative ensemble of gauge fields).
Now let
\begin{equation}
   R(y)=A{(y+a_1)(y+a_3)\ldots(y+a_{2n-1})\over
          (y+a_2)(y+a_4)\ldots(y+a_{2n})},
   \label{Rofy}
\end{equation}
be a rational function of degree $[n,n]$ in $y$,
which approximates the function $1/\sqrt{y}$ in the range
$[\eps,1]$. 
The operator $R(\Qs^2/M^2)$
then approximates $M/|\Qs|$ up to a relative error
(in the operator norm) less than or equal to
\begin{equation}
   \delta=\max_{\eps\leq y\leq1}\left|
   1-\sqrt{y}R(y)\right|
   \label{Rerror}
\end{equation}
if the spectrum of $\Qs^2$ is fully contained in $[\eps M^2,M^2]$ and 
a somewhat larger error if the 
spectrum extends slightly beyond the limits of this interval.

For a specified degree $n$, one would evidently like the coefficients
$a_r$ in eqn~(\ref{Rofy}) to be such that
the error $\delta$ is minimized. 
It seems unlikely that the optimal coefficients can be 
worked out analytically,
but the mathematician Zolotarev was able to do that
a long time ago by relating the optimization problem
to the theory of elliptic functions
(see \shortciteN{Achieser}, for example; the coefficients
are given explicitly in Section 2.6.5)\footnote{Zolotarev
obtained two different optimal rational approximations $R(y)$
to the function $1/\sqrt{y}$, one of degree $[n,n]$ and
the other of degree $[n-1,n]$. Both are used in lattice
QCD and are commonly referred to as the Zolotarev rational
approximation.}.

The optimal rational approximation has a number of 
remarkable properties. One of them is that the error $\delta$ is
a very rapidly decreasing function of the degree $n$. For 
$\eps=10^{-5}$, for example, the approximation error is
$5\times10^{-4}$ if $n=6$ and decreases to 
$1\times10^{-7}$ and $8\times10^{-15}$ if $n=12$ and $n=24$, respectively.
The fact that the minimizing coefficients satisfy
\begin{equation}
   a_1>a_2>\ldots>a_{2n}>0,
   \qquad A>0,
\end{equation}
is also very important. In particular,
the residues in the expansion in partial fractions,
\begin{equation}
   R(y)=A\left\{
   1+{r_2\over y+a_2}+\ldots+{r_{2n}\over y+a_{2n}}\right\},
\end{equation}
are all positive.
The expansion is therefore well suited for the numerical evaluation
of the action of the operator $R(\Qs^2/M^2)$ on the pseudo-fermion field 
$\phi_2$.
Note that this calculation essentially amounts to solving the equations
\begin{equation}
   \left(\Qs^2+\mu^2_{2k}\right)\psi_k=M^2\phi_2,
   \qquad
   \mu_r=M\sqrt{a_r},
   \label{MultiMassEqns}
\end{equation}
for $k=1,\ldots,n$. 
Since the right-hand sides of these equations are the same,
it is possible to solve all equations simultaneously
using a so-called multi-mass solver \shortcite{MultiMass}.
The total computational effort is then not very much larger than
what would be required for the sequential solution of the 
equations that are the most difficult to solve
(the ones at the largest values of $k$).

\subsection{The RHMC algorithm}

Having specified the operator $R$,
the inclusion of the strange quark in the HMC algorithm 
is now straightforward. The algorithm proceeds according to 
the steps (${\rm a}'$), (b) and (${\rm c}'$), as before, and 
one merely has to add the contribution of the strange-quark 
pseudo-fermion fields. Some specific remarks
on what exactly needs to be done may nevertheless be useful.

\subsubsection{(a) Pseudo-fermion generation}
The random generation of the strange-quark pseudo-fermion fields 
in the first step
requires two operators $B$ and $C$ to be found such that
\begin{equation}
   BB^{\dagger}=|\Qs|R,
   \quad\hbox{and}\quad
   CC^{\dagger}=R^{-1}
\end{equation}
(cf.~Section~2.1.4). For $C$ one can take the rational
function
\begin{equation}
   C=A^{-1/2}{(\Qs+i\mu_2)\ldots(\Qs+i\mu_{2n})
              \over
              (\Qs+i\mu_1)\ldots(\Qs+i\mu_{2n-1})},
\end{equation}
but there is no similarly simple choice for the other operator.
However, since
\begin{equation}
   Z={\Qs^2R^2\over M^2}-1
\end{equation}
is of order $\delta$, the series
\begin{equation}
   B=\sqrt{M}(1+Z)^{1/4}=\sqrt{M}\left\{1+\frac{1}{4}Z-\frac{3}{32}Z^2+\ldots
   \right\}
\end{equation}
converges rapidly and may be truncated after the 
first few terms. The strange-quark pseudo-fermion fields
can thus be generated with a computational effort equivalent to 
the one required for a few applications of the operator 
$R$ to a given quark field.

\subsubsection{(b) Strange-quark force}
As already mentioned, only the second term of the pseudo-fermion
action
(\ref{SquarkAction}) is included in the molecular-dynamics
Hamilton function. The computation of the associated force, 
\begin{equation}
   F^a_{\rm s}(x,\mu)=-{2A\over M^2}\sum_{k=1}^n
   r_{2k}\Re\bigl(\psi_k,\Qs\dirac{5}\delta^a_{x,\mu}\Ds\psi_k\bigr),
\end{equation}
requires the $n$ linear systems (\ref{MultiMassEqns}) to be solved.
The computer time needed for this calculation is therefore essentially
the same as for one application of the operator $R$.

\subsubsection{(c) Acceptance step}
The acceptance probability is calculated as usual
except for the fact that the change
of the first term of the pseudo-fermion action (\ref{SquarkAction}),
\begin{equation}
   \left.(\phi_1,(|\Qs|R)^{-1}\phi_1)\right|_{U=U_{\tau}}
   -
   \left.(\phi_1,(|\Qs|R)^{-1}\phi_1)\right|_{U=U_0},
   \label{pfActionDifference}
\end{equation}
must be added to the difference (\ref{DeltaH})
of the molecular-dynamics Hamilton function.
Note that the action difference (\ref{pfActionDifference}) can be computed
by expanding $(|\Qs|R)^{-1}\phi_1$ in powers of $Z$.

\subsection{Appendix: Coefficients of the optimal rational approximation}

The analytic expressions for the 
coefficients of the rational function (\ref{Rofy})
that minimizes the approximation error (\ref{Rerror})
involve the Jacobi elliptic functions $\sn(u,k)$, $\cn(u,k)$
and the complete elliptic integral $K(k)$ 
(see \shortciteN{Abramowitz},
for example, for the definition of these functions). 
Explicitly, they are given by
\begin{equation}
  a_r={\cn^2(rv,k)\over\sn^2(rv,k)},
  \quad
  r=1,2,\ldots,2n,
\end{equation}
where 
\begin{equation}
  k=\sqrt{1-\eps},\qquad
  v={K(k)\over2n+1}.
\end{equation}
The formulae for the amplitude $A$ and the error $\delta$,
\begin{eqnarray}
   A&=&{2\over1+\sqrt{1-d^2}}\,{c_1c_3\ldots c_{2n-1}\over
                          c_2c_4\ldots c_{2n}},
   \\[2.5ex] 
   \delta&=&{d^2\over\left(1+\sqrt{1-d^2}\right)^2},
\end{eqnarray}
involve the coefficients 
\begin{eqnarray}
   c_r&=&\sn^2(rv,k),
   \quad
   r=1,2,\ldots,2n,
   \\[2.5ex] 
   d&=&k^{2n+1}\left(c_1c_3\ldots c_{2n-1}\right)^2.
\end{eqnarray}
All these expressions are free of singularities and can be
programmed straightforwardly, using the well-known methods
for the numerical evaluation of the Jacobi elliptic functions.
An ISO C program that calculates the coefficients $A$, 
$a_1,\ldots,a_{2n}$ and the error
$\delta$ to machine precision can be downloaded from
{\tt http://cern.ch/luscher/}.

\chapter{Variance reduction methods}

Lattice QCD simulations produce ensembles
$\{U_1,\ldots,U_N\}$ of gauge fields, which are
representative of the functional integral at the specified
gauge coupling and sea-quark masses.
In this chapter it is
taken for granted that autocorrelation effects can be
safely neglected, i.e.~that the separation in 
simulation time of subsequent field configurations 
is sufficiently large for this to be the case.
As discussed in Section 2.2.4,
the expectation value of any (real or complex) observable $\Obs(U)$ 
may then be calculated through
\begin{equation}
  \langle\Obs\rangle={1\over N}\sum_{k=1}^N\Obs(U_k)+\rmO(N^{-1/2}),
\end{equation}
where the statistical error is, to leading order in $1/N$, 
given by
\begin{equation}
  \sigma(\Obs)={\sigma_0(\Obs)\over N^{1/2}},
  \qquad
  \sigma_0(\Obs)=\langle|\Obs-\langle\Obs\rangle|^2\rangle^{1/2}.
\end{equation}
In practice, the error is estimated from the variance 
of the ``measured'' values $\Obs(U_k)$, which is a correct
procedure up to subleading terms.

For a given observable $\Obs$, 
another observable $\Obs'$ satisfying
\begin{equation}
  \langle\Obs'\rangle=\langle\Obs\rangle,
  \qquad
  \sigma_0(\Obs')\ll\sigma_0(\Obs),
\end{equation}
can sometimes be found.
The desired expectation value is then obtained with 
a much smaller statistical error if $\Obs$ is replaced by $\Obs'$.
Most variance-reduction methods are based on this simple observation.
The construction of effective 
alternative observables is non-trivial, however, 
and may involve
auxiliary stochastic variables and transformations of the functional
integral.

The discussion in this chapter is often of a general nature
and applies to most forms of lattice QCD, but if not specified
otherwise,
the Wilson formulation will be assumed, 
with or without O($a$)-improvement and with two or more flavours
of sea quarks.

\section{Hadron propagators}

The calculation of the properties of the light mesons and baryons 
is a central goal in lattice QCD. Finding good
variance-reduction methods proves to be difficult in this field,
but some important progress has nevertheless been made.
In this section, the aim is to shed some light on the problem
by discussing the statistical variance of hadron 
propagators.

\subsection{The pion propagator}

\begin{figure}[t]
\begin{center}
  \includegraphics[clip,scale=0.5]{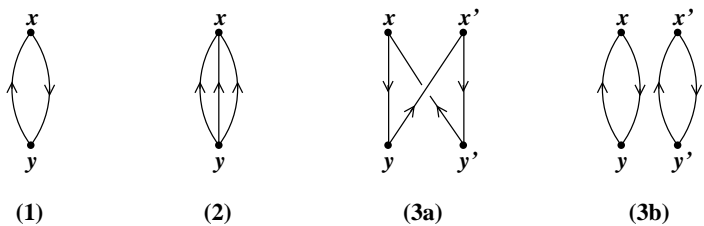}
\end{center}
\caption{
Quark-line diagrams contributing to the pion propagator 
(1), the nucleon propagator (2) and the pion four-point function 
(3a and 3b).
Directed lines from $y$ to $x$ stand for the 
light-quark propagator $S(x,y)$.
At the vertices, the spinor indices
are contracted with the appropriate colour tensor and Dirac matrix
($\dirac{5}$ in the case of the pion-field vertices).
}
\label{QuarkLineDiagrams}
\end{figure}

Once the sea-quark fields are integrated out,
the correlation functions of local fields like the
isospin pseudo-scalar density
\begin{equation}
   P^a=\psibar\frac{1}{2}\tau^a\dirac{5}\psi,
   \qquad
   \psi=\left(\begin{array}{c}
              u\\[0.0ex]
              d
              \end{array}\right),  
   \qquad
   \hbox{$\tau^a$: Pauli matrices},
\end{equation}
become expectation values of quark-line diagrams.
The diagrams are products of quark propagators,
Dirac matrices and invariant colour tensors,
with all Dirac and colour indices properly contracted
(see Fig.~\ref{QuarkLineDiagrams}).

For any given gauge field,
the light-quark
propagator is determined through the field equation
\begin{equation}
   DS(x,y)=\delta_{xy}
\end{equation}
and the chosen boundary conditions. The spinor
indices have been suppressed in this equation, but one should
keep in mind that $S(x,y)$ is, for fixed $x$ and $y$, 
a complex $12\times12$ matrix in spinor space. 
In view
of the $\dirac{5}$-hermiticity of the Dirac operator,
forward and backward propagators are related by
\begin{equation}
   \dirac{5}S(y,x)\dirac{5}=S(x,y)^{\dagger}, 
\end{equation}
where the hermitian conjugation refers to spinor space only.

In lattice QCD, the pion mass $\mpi$ is usually determined
from the exponential decay of the ``pion propagator''
\begin{equation}
  \langle P^a(x)P^b(y)\rangle=-\frac{1}{2}\delta^{ab}
  \langle\tr\{S(x,y)S(x,y)^{\dagger}\}\rangle
\end{equation}
at large distances $|x-y|$. It is advantageous for this 
calculation to pass to the zero-momentum component
of the propagator, 
\begin{eqnarray}
  \gpi(x_0-y_0)&=&\langle\Obs_{\pi}(x_0,y)\rangle,
  \\[1.8ex]
  \Obs_{\pi}(x_0,y)&=&\sum_{\vec{x}}
  \tr\{S(x,y)S(x,y)^{\dagger}\},
\end{eqnarray}
whose asymptotic form at large times,
\begin{equation}
  \gpi(t)\mathrel{\mathop\propto_{t\to\infty}}
  \rme^{-\mpi t}+\rmO(\rme^{-3\mpi t}),
\end{equation}
is dominated by a single exponential with exponent
equal to the pion mass.

\subsection{Statistical error estimation}

For a given gauge field and source point $y$, 
the calculation of the propagator $S(x,y)$ 
allows the observable $\Obs_{\pi}(x_0,y)$ to be evaluated
at all times $x_0$. The ensemble average of 
$\Obs_{\pi}(x_0,y)$ then provides a stochastic 
estimate of the zero-momentum pion propagator,
from which one may be able to extract the pion mass.

The statistical error of the average of $\Obs_{\pi}(x_0,y)$ is
proportional to the square root of the a priori variance
\begin{equation}
  \sigma_0(\Obs_{\pi})^2=\langle\Obs_{\pi}(x_0,y)^2\rangle-
                     \langle\Obs_{\pi}(x_0,y)\rangle^2,
  \label{PionPropVariance}
\end{equation}
which may symbolically be written in the form
\begin{equation}
  \sigma_0(\Obs_{\pi})^2=\sum_{\vec{x},\vec{x}'}\biggl\{
  \raise-0.80cm\hbox{\hspace{0.1cm}%
  \includegraphics[clip,scale=0.40]{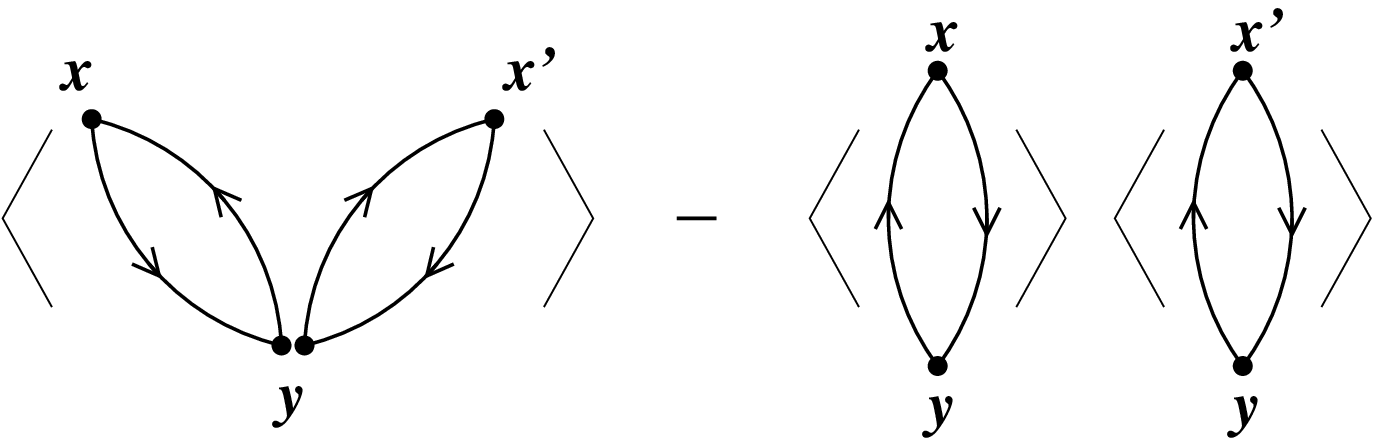}\hspace{0.1cm}}\biggr\}
  _{x_0'=x_0}.
\end{equation}
An important point to note here is that 
the diagram in the first term 
also contributes to the $\pi\pi$ propagator
$\langle P^a(x)P^b(x')P^a(y)P^b(y)\rangle$ 
(see Fig.~\ref{QuarkLineDiagrams}).
Moreover, the second term
is just the square of the pion propagator. 
The variance is therefore expected to decay like 
$\rme^{-2\mpi|x_0-y_0|}$ at large time separations,
which implies that the pion propagator is obtained
with a nearly time-independent relative statistical error.

Lattice QCD simulations tend to confirm this and they also
show that the quark propagator typically falls off like
\begin{equation}
  \tr\{S(x,y)S(x,y)^{\dagger}\}^{1/2}
  \propto
  \rme^{-{1\over2}\mpi|x-y|}
  \label{PropDecay}
\end{equation}
at large distances,
for every gauge field
in a representative ensemble of fields.
There is actually little room for a different
behaviour of the propagator, since both the mean and
width of the distribution of $\tr\{S(x,y)S(x,y)^{\dagger}\}$ 
decay exponentially with the same exponent.

\subsection{Baryons and the exponential SNR problem}

The diagrams contributing to the nucleon
two-point function involve 3 quark propagators. 
At zero spatial momentum and for every
gauge-field configuration,
the associated observable $\Obs_N(x_0,y)$ thus falls off roughly like
$\rme^{-{3\over2}\mpi|x_0-y_0|}$ at large time separations.
The nucleon propagator, however, decays much more rapidly,
since the exponent (the nucleon mass $\mN$) is 
significantly larger than ${3\over2}\mpi$.

In the sum over all gauge fields, there must therefore be 
important cancellations
among the measured values of $\Obs_N(x_0,y)$.
More precisely, after averaging over $N$ configurations,
the nucleon propagator is obtained with 
a signal-to-noise ratio (SNR) 
proportional to 
$\sqrt{N}\rme^{-(\mN-{3\over2}\mpi)|x_0-y_0|}$.
The number of measurements required for a specified statistical
accuracy at a given time separation thus scales like
\begin{equation}
  N\propto\rme^{(2\mN-3\mpi)|x_0-y_0|}.
\end{equation}
Note that the exponent $2\mN-3\mpi$ is, in all cases of interest, 
not small and as large as $7\,\fm^{-1}$ at the physical point.
The calculated values of the nucleon propagator
therefore tend to rapidly disappear in the statistical noise
at large time separations $|x_0-y_0|$. 

Computations of other hadron propagators are similarly affected
by an exponential loss of significance, the 
only exception being the propagators of the stable
pseudo-scalar mesons.
The masses and decay constants of the latter can thus
be determined far more easily
than those of the vector mesons and the baryons.

\section{Using random sources}

As already noted in Section 2.1.3, the link variables
in distant regions of a large lattice
are practically decoupled.
Hadron propagators calculated at widely separated
source points therefore tend to be sampled independently
and their average consequently has smaller statistical fluctuations
than the propagator at a single source point. 
Averaging over as many source points as possible is thus desirable, but
tends to be expensive in terms of computer time, because
the quark propagators must be recomputed at each point.

The random source method performs the sum
over all or a selected set of source points
stochastically \shortcite{MichaelPeisa}.
In many cases, a significant acceleration of the computation
can be achieved in this way. The idea of introducing random sources
is also quite interesting from a purely theoretical point
of view, because it extends the notion of an observable
to stochastic observables (i.e.~functions of the gauge field that
depend on random auxiliary variables).

\subsection{Gaussian random fields}

Random sources may be thought of as a set of
additional fields that are decoupled from the dynamical fields
and therefore do not change the physics content of the theory.
In the case considered here, the added fields are a multiplet
\begin{equation}
  \eta_i(\vec{x}),\quad i=1,\ldots,\Nsrc,
\end{equation}
of pseudo-fermion fields on the fixed-time spatial lattice.
Their action is taken to be
\begin{equation}
  \Ssrc(\eta)=\sum_{i=1}^{\Nsrc}(\eta_i,\eta_i),
\end{equation}
where the scalar product is the obvious one for such fields.
For each gauge-field configuration in a representative ensemble
of fields, the source fields are chosen randomly 
with probability density proportional to $\rme^{-\Ssrc(\eta)}$.
Evidently,
the set of fields obtained in this way is a
representative ensemble
for the joint probability density of the gauge field and the source fields.

One may now consider observables 
$\Obs(U,\eta)$ that depend on both the gauge field and the sources.
When the latter are integrated out, such observables reduce to
ordinary (non-stochastic) observables. Noting
\begin{equation}
  \langle\eta_i(\vec{x})\eta_j(\vec{y})^{\dagger}\rangle_{\src}=
  \delta_{ij}\delta_{\vec{x}\vec{y}},
\end{equation}
the integral over the source fields
is easily worked out, using Wick's theorem, 
if $\Obs(U,\eta)$ is a polynomial in the source fields.
In the case of the observable
\begin{equation}
  \Obs={1\over\Nsrc}\sum_{i=1}^{\Nsrc}
  \sum_{\vec{x},\vec{y}}\eta_i(\vec{x})^{\dagger}
  S(x,y)|_{x_0=y_0}\eta_i(\vec{y}),
  \label{TraceObs}
\end{equation}
for example, the calculation yields
\begin{equation}
  \langle\Obs\rangle_{\src}=\sum_{\vec{x}}
  \tr\{S(x,x)\}.
  \label{TraceSum}
\end{equation}
The random sources thus allow
the trace (\ref{TraceSum}) to be estimated
stochastically.

\subsection{The pion propagator revisited}

For any fixed time $y_0$, the spinor fields
\begin{equation}
  \phi_i(x,y_0)=\sum_{\vec{y}}S(x,y)\eta_i(\vec{y}),
  \qquad i=1,\ldots,\Nsrc,
\end{equation}
can be computed
by solving the Dirac equations $D\phi_i(x,y_0)=\delta_{x_0y_0}\eta_i(\vec{x})$.
It is then straightforward to show that the 
observable
\begin{equation}
  \Obshat_{\pi}(x_0,y_0)={1\over\Nsrc\Vol}\sum_{i,\vec{x}}
  \left|\phi_i(x,y_0)\right|^2,
  \qquad\hbox{$\Vol$: spatial lattice volume},
\end{equation}
has the same expectation value as $\Obs_{\pi}(x_0,y)$ and 
may therefore be used in place of the latter in a 
calculation of the pion propagator.
Note that through the
average over the source fields,
\begin{equation}
  \langle\Obshat_{\pi}(x_0,y_0)\rangle_{\src}=
  {1\over\Vol}\sum_{\vec{y}}
  \Obs_{\pi}(x_0,y),
  \label{VolAverage}
\end{equation}
one effectively 
sums over all source points at time $y_0$.

Whether the new observable is 
any better than the old one depends on whether it has
smaller statistical fluctuations or not.
A short calculation,
using Wick's theorem, shows that the associated a priori
variance is given by
\begin{equation}
  \sigma_0(\Obshat_{\pi})^2=
  \sigma_0(\Obsbar_{\pi})^2+{1\over\Nsrc\Vol^2}
  \sum_{\vec{x},\vec{x}'}
  \sum_{\vec{y},\vec{y}'}\Biggl\{
  \raise-0.76cm\hbox{\hspace{0.15cm}%
  \includegraphics[clip,scale=0.38]{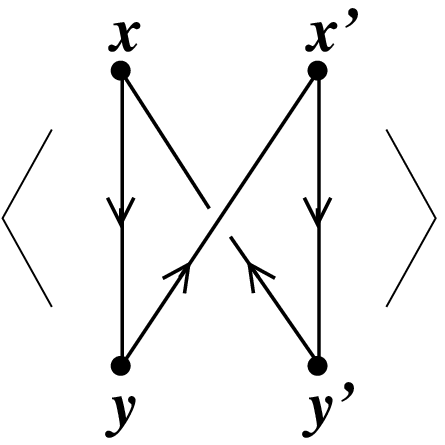}\hspace{0.1cm}}\Biggr\}
  _{x_0'=x_0,y_0'=y_0},
\end{equation}
where $\Obsbar_{\pi}(x_0,y_0)$ denotes the volume-averaged 
observable (\ref{VolAverage}).
Recalling the exponential decay (\ref{PropDecay}) of the quark propagator,
the second term is readily estimated to be of order
$(\Nsrc\mpi^3\Vol)^{-1}$ at large volumes. Since the 
first term scales like $(\mpi^3\Vol)^{-1}$, 
it tends to be the dominant contribution to the variance
already for moderately large numbers of 
source fields (setting $\Nsrc=12$, for example,
is often sufficient).

With respect to a calculation of the pion propagator
at a single source point, the random source method described here 
thus achieves a reduction of 
the statistical error by a factor proportional to $(\mpi^3\Vol)^{-1/2}$
for approximately the same computational cost.
The use of random sources is therefore recommended in this case, 
particularly so on large lattices.

\subsection{Further applications}

Random sources are an interesting and useful tool.
Here only one particular application and variant of the method
was discussed. Source fields on different subsets of points
can be considered as well as random fields
taking values in a group or, more generally,
fields with any non-gaussian distribution. 
The method combines well
with low-mode averaging techniques
\shortcite{NeffEtAl,LowModeGiusti,LowModeDeGrand,BaliEtAl,AllToAllTrinity}
and, to mention just one further example, 
it also played a central r\^ole in a recent
calculation of the spectral density
of the hermitian Wilson--Dirac operator
\shortcite{GiustiLuscherSigma}.

In the case of the vector-meson and baryon propagators, the 
application of random source methods is complicated by the 
fact that the index contractions at the vertices of the
quark-line diagrams couple different components of the quark
propagators. Different kinds of random sources
(one for each component, for example) then need to be introduced
to be able to write down a correct random-source representation
of the hadron propagator. 
The variance of such observables typically involves many diagrams,
among them often also disconnected ones.
As a consequence, the use of random sources for these propagators 
is not obviously profitable, at least as long as the 
spatial extent of the lattices
considered is not significantly larger than 2 or 3 fm.

Random source methods should preferably be applied only after
a careful analysis of the variance of the proposed 
stochastic observables. Such an 
analysis can be very helpful in deciding which 
kind of random fields to choose and 
how exactly the stochastic observable must be constructed
in order to achieve a good scaling of the statistical
error with the lattice volume.

\section{Multilevel simulations}

Random source methods can lead to an important reduction of 
statistical errors, but are unable to overcome
the exponential SNR problem encountered in the 
case of the nucleon propagator, for example.
In a multilevel simulation, the improved observables
are constructed through a stochastic process, i.e.~through
a secondary or nested simulation. Exploiting 
the locality of the theory, an exponential reduction
of the statistical error can then be achieved in 
certain cases.

So far multilevel simulations have been limited to 
bosonic field theories, essentially because 
manifest locality is 
lost when the fermion fields are integrated out.
Whether this limitation is a transient one is unclear
at present, but it is certainly worth explaining the
idea here and to show its impressive potential.

\subsection{Statistical fluctuations of Wilson loops}

In the following, a multilevel algorithm 
for the computation of the expectation values of 
Wilson loops in the pure
$\Group$ lattice gauge theory will be described.
The lattice action is assumed to be the Wilson plaquette action.

Similarly to the hadron propagators,
Wilson loop expectation values suffer from an exponential
SNR problem. Let $\Loop$ be a $T\times R$ rectangular loop
in the $(x_0,x_1)$-plane,
$U(\Loop)$ the ordered product of the link variables around
the loop and $\Wloop=\tr\{U(\Loop)\}$ its trace.
For large loops, the expectation value of $\Wloop$
satisfies the area law
\begin{equation}
  \langle\Wloop\rangle\sim\rme^{-\sigma A},
  \qquad
  A=TR,
  \label{AreaLaw}
\end{equation}
where $\sigma\simeq1\,\GeV/\fm$.
Since $\Wloop$ is a number of order 1 for every 
gauge-field configuration, the a priori variance $\sigma_0(\Wloop)$
is practically equal to $\langle|\Wloop|^2\rangle$ and thus of order 1 too.
One therefore needs to generate ensembles of at least
\begin{equation}
  N=\langle\Wloop\rangle^{-2}\sim\rme^{2\sigma A}
  \label{WloopEnsembleSize}
\end{equation}
configurations to be able to 
calculate the expectation value $\langle\Wloop\rangle$ 
to a useful precision.

For loop areas $A$ equal to 1, 2 and 4 $\fm^2$, for example, the 
minimal ensemble sizes are thus estimated to be
$2\times10^4$, $6\times10^8$ and $4\times10^{17}$,
respectively. These figures may not be exactly right, because
the Wilson loop is averaged over all possible translations 
in practice and since there are
important subleading corrections to the area law
(\ref{AreaLaw}). However, for large areas $A$, 
the minimal ensemble size
is essentially determined by the rapidly growing
exponential factor (\ref{WloopEnsembleSize}).

\subsection{Factorization and sublattice expectation values}

A multilevel algorithm invented many years ago
is the ``multihit method'' 
\shortcite{MultiHit}.
In this case, a reduction of the 
statistical error by an exponential factor with exponent
proportional to $T$ is achieved by replacing
the time-like link variables
along the Wilson loop through
stochastic estimates of their average values in presence
of the other link variables.
The method thus exploits the fact that the Wilson loop
factorizes into a product of link variables.

\begin{figure}[t]
\begin{center}
  \includegraphics[clip,scale=0.4]{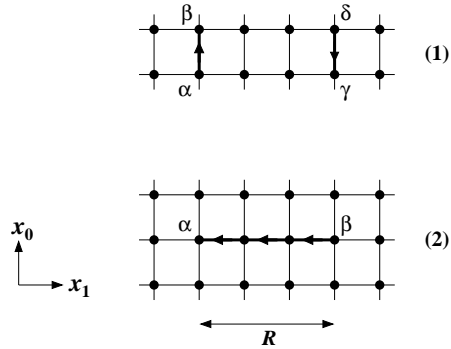}
\end{center}
\caption{
A $T\times R$ Wilson loop in the $(x_0,x_1)$-plane with lower-left
corner at $x=0$ can be factorized into the product 
(\ref{WloopFactorization}) of two-link and line operators. 
The $\Group$ indices of the two-link operator (\ref{TwoLinkOp}) are assigned
as shown in figure (1).
The line operators (\ref{LineOp}) coincide with the spatial 
lines of the Wilson loop and have only two indices, 
as indicated in figure (2).
}
\label{TwoLink}
\end{figure}

If the loop is instead factorized into
a product of two-link operators, an
exponential error reduction is obtained
with an exponent proportional to the area $A$
\shortcite{Multilevel}.
A two-link operator is an object with four $\Group$ indices,
\begin{equation}
  \twolink(x_0)_{\alpha\beta\gamma\delta}
  =U(x,0)_{\alpha\beta}^{\ast}U(x+R\hat{1},0)_{\gamma\delta},
  \label{TwoLinkOp}
\end{equation}
residing at time $x_0$ and $\vec{x}=0$ (see Fig.~\ref{TwoLink}).
Two-link operators
at adjacent times can be multiplied and their product from time
0 to $T-a$ yields the tensor product of the time-like lines
of the Wilson loop. The latter is then given by 
\begin{equation}
  \Wloop=\tr\{\lineop(0)\twolink(0)\twolink(a)\ldots\twolink(T-a)\lineop(T)\},
  \label{WloopFactorization}
\end{equation}
where the product
\begin{equation}
  \lineop(x_0)_{\alpha\beta}
  =\left\{U(x,1)U(x+a\hat{1},1)\ldots U(x+(R-a)\hat{1},1)
  \right\}_{\alpha\beta}
  \label{LineOp}
\end{equation}
denotes the spatial Wilson line at time $x_0$.

\begin{figure}[t]
\begin{center}
  \includegraphics[clip,scale=0.40]{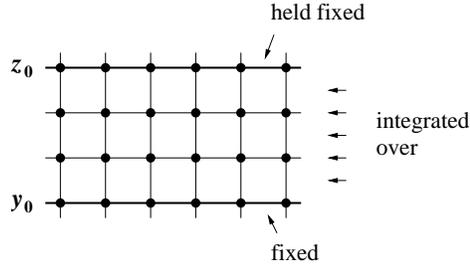}
\end{center}
\caption{Factorized representations
of the Wilson loop expectation value 
such as (\ref{FactorizedExpectation}) are based on
a division of the lattice into sublattices separated by equal-time
hyperplanes. The sublattice expectation values $[\ldots]$ in these
expressions involve an integration
over the field variables residing on the links between the hyperplanes, all
other link variables being held fixed.
}
\label{Sublattice}
\end{figure}

Consider now the sublattice bounded by
the equal-time hyperplanes at some time $y_0$ and some later time $z_0$
(see Fig.~\ref{Sublattice}).
The Wilson action couples the link variables inside the sublattice
(the ``interior'' link variables) 
to themselves and the spatial fields on the boundaries, but there
is no interaction with the field variables elsewhere on the lattice.
If $\Obs$ is any function of the interior link variables, its
sublattice expectation value is defined by
\begin{equation}
  [\Obs]=
  {1\over{\mathcal Z}_{\inslice}}\int\rmD[U]_{\inslice}\,
  \Obs(U)\kern1pt\rme^{-S(U)},
  \label{SublatExpectation}
\end{equation}
where one integrates over the interior variables. Note that
the part of the action that does not depend on the latter
drops out in the expectation value, which is thus a function of the spatial
link variables at time $y_0$ and $z_0$ only.

If the lattice is decomposed into non-overlapping sublattices of this
kind, the functional integral divides into an integral
over the interior variables and an integral over the boundary
fields. For even $T/a$, for example,
the Wilson loop expectation value may be rewritten in the form
\begin{equation}
  \Wloop=\langle\tr\{\lineop(0)[\twolink(0)\twolink(a)]
  [\twolink(2a)\twolink(3a)]\ldots\lineop(T)\}\rangle.
  \label{FactorizedExpectation}
\end{equation}
The outer expectation value in this expression is the usual one
involving an integration over all field variables.
However, since the observable depends only on the spatial fields
at time $x_0=0,2a,4a,\ldots,T$, the integral over the interior 
field variables yields the product of the sublattice partition
functions. This factor exactly cancels the product of the 
normalization factors of the sublattice
expectation values.

The sublattice expectation value of a product of two-link operators
is a correlation function of two segments of Wilson lines in 
presence of the boundary fields. A single segment transforms
non-trivially under the center symmetry of the theory,
where all time-like link variables at a given time are
multiplied by an element of the center of $\Group$.
Barring spontaneous symmetry breaking, 
its sublattice expectation value therefore vanishes
and the correlation function of two segments is consequently
expected to go to zero at large separations $R$.
Experience actually suggests that
\begin{equation}
  [\twolink(y_0)\twolink(y_0+a)\ldots\twolink(z_0-a)]
  \sim\rme^{-\sigma (z_0-y_0)R}
  \label{TwolinkCorrelation}
\end{equation}
if the time difference $z_0-y_0$ is larger than, say,
$0.5$ fm or so.
Recalling eqn~(\ref{FactorizedExpectation}),
the rapid decay of the Wilson loop expectation value
at large times $T$ (and the area law if eqn~(\ref{TwolinkCorrelation})
holds) is thus seen to arise from a product of small factors.

\subsection{Multilevel update scheme}

The foregoing suggests that the expectation value of 
the Wilson loop may be accurately calculated, at any value of $T$,
starting from a factorized representation 
like (\ref{FactorizedExpectation}).
An algorithm that implements the idea for this particular
factorization proceeds in cycles consisting
of the following steps:

\begin{itemize}
\itemsep=1.5ex\it
\item[\rm(a)]{Update the gauge field $N_0$ times using
              a combination of the heatbath\\ and microcanonical
              link-update algorithms.}
\item[\rm(b)]{Estimate the expectation values 
              $[\twolink(x_0)\twolink(x_0+a)]$ by
              updating the\\ field inside the associated
              sublattices $N_1$ times and by averaging the\\
              product of the two-link operators over
              the generated configurations.}
\item[\rm(c)]{Compute the trace
              $\tr\{\lineop(0)[\twolink(0)\twolink(a)]
              [\twolink(2a)\twolink(3a)]\ldots\lineop(T)\}$
              using the\\ estimates of the sublattice 
              expectation values obtained in step (b).}
\end{itemize}

\noindent 
Each cycle thus yields an estimate of the trace
$\tr\{\lineop(0)[\twolink(0)\twolink(a)]\ldots\lineop(T)\}$.
A moment of thought shows that
these estimates are averages of the Wilson loop over a particular set field
configurations generated through a valid simulation algorithm. Their
average over many cycles therefore coincides with the expectation value of the
Wilson loop up to statistical errors. 

As explained in Section 3.3.2, the trace 
$\tr\{\lineop(0)[\twolink(0)\twolink(a)]\ldots\lineop(T)\}$ 
tends to decay exponentially, for every field configuration in
a representative ensemble of fields. The statistical error
of the stochastic estimates of the trace 
produced by the multilevel algorithm (a)--(c) 
is therefore guaranteed to 
fall off exponentially as well. With respect to 
a straightforward computation of the Wilson loop expectation
value, the sublattice averaging thus achieves 
an exponential error reduction.

\begin{figure}[t]
\begin{center}
  \includegraphics[clip,scale=0.65]{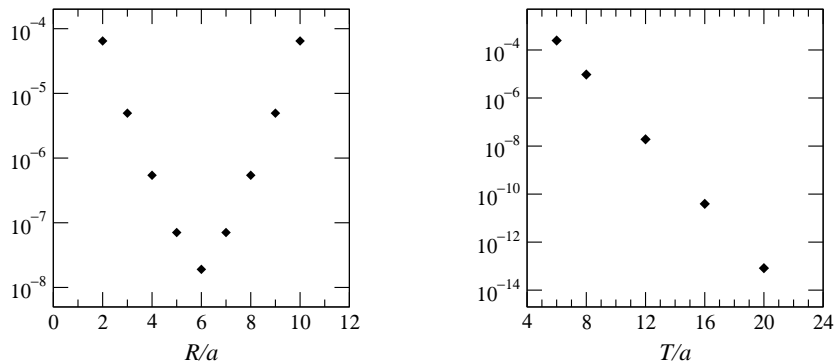}
\end{center}
\caption{Values of the correlation function of the Polyakov loop on a 
$(T/a)\times12^3$ lattice with spacing $a\simeq0.17$ fm
and periodic boundary conditions,
plotted as a function of the distance $R$ at
$T=12a$ (left figure) and as a function of $T$ at $R=6a$ (right figure).
}
\label{PPexample}
\end{figure}

For illustration, some results obtained in the course
of an early application
of the multilevel algorithm are shown in Fig.~\ref{PPexample}. 
In order to simplify the 
situation a little bit, the correlation function of 
two Polyakov loops (i.e.~Wilson lines that wrap around
the lattice in the time direction) was considered in this study.
The line operators $\lineop(x_0)$ are then not needed and 
the observable coincides with the trace of a product of $T/a$
two-link operators, $T$ being the time-like extent of the lattice.
As is evident from the data plotted in the figure, the
multilevel algorithm allows the exponential decay of
the correlation function to be followed over many orders of 
magnitude, which would not be possible (or only with 
an astronomical computer budget) using the standard simulation
techniques.

\subsection{Remarks and further developments}

Multilevel algorithms of the kind described here 
have been employed in studies of the string behaviour
\shortcite{BosonicString,KratochvilaForcrand,PepeWiese},
the glueball spectrum 
(\shortciteANP{HarveyTwoLevelI},~\citeyearNP{HarveyTwoLevelI,HarveyTwoLevelII})
and the bulk viscosity
\shortcite{HarveyViscosity}
in pure gauge theories.
Another, closely related multilevel algorithm has recently 
been proposed
for the calculation of the energies of the lightest states
with a specified (non-trivial) transformation behaviour under 
the exact symmetries of the theory
\shortcite{DellaMorteGiusti}.
In all these cases,
an important and sometimes impressive acceleration of the simulation
was achieved.

The exponential SNR problem is, however, often not completely
eliminated.
Computations of Wilson loop expectation values, for example,
require sublattice expectation values of products of 
two-link operators to be determined to some accuracy.
In view of eqn~(\ref{TwolinkCorrelation}), 
the number of sublattice updates 
that need to be performed in this part of the calculation
is therefore expected to scale roughly like $\rme^{2\sigma(z_0-y_0)R}$.
With respect to a one-level simulation, 
the exponential growth of the required computational effort 
is nevertheless dramatically reduced,
because the exponent now increases only
proportionally to the distance $R$ rather than the area $A$.

\chapter{Statistical error analysis}

The estimation of the statistical errors in numerical lattice QCD 
appears to be an easy topic. However, the physical quantities
often need to be extracted from the simulation data through 
some non-linear procedure, which may involve complicated fits
and extrapolations. 
The correct propagation of the errors becomes a non-trivial
task under these conditions.
Resampling techniques such as the jackknife and bootstrap
methods \shortcite{EfronTibshirani}
allow the errors of the calculated physical quantities 
to be estimated with minimal effort, but are 
a bit of magic and are actually known to be incorrect in certain 
special cases
(the jackknife method, for example, in general 
gives wrong results for the statistical error of the 
median of an observable).

The aim in this chapter is to build up a conceptually
solid framework in which the error propagation is
made transparent. 
In particular, the correctness of the jackknife method can 
then be established for a class of quantities,
which includes most cases of interest.

\section{Primary observables}

Wilson loops, quark-line diagrams and any other 
function of the gauge field, including the stochastic ones
considered in Chapter~3, are the primary observables in 
lattice QCD. Their expectation values are the first 
quantities calculated in a simulation project and
all physical quantities are eventually obtained from these.

\subsection{Correlation functions and data series} 

Let $A_r$ be some real-valued primary observables labeled by an index $r$.
As before, the lattice QCD expectation value of 
an observable $\Obs$ is denoted by $\langle\Obs\rangle$.
The quantities of interest are then 
the $n$-point correlation functions 
$\langle A_{r_1}\ldots A_{r_n}\rangle$.
Independently of any locality properties, their connected parts 
$\langle A_{r_1}\ldots A_{r_n}\rangle_{\con}$ 
may be defined in the familiar way. In particular,
\begin{eqnarray}
  \langle A_r\rangle_{\con}&=&\langle A_r\rangle,
  \\[1.8ex]
  \langle A_rA_s\rangle_{\con}&=&\langle A_rA_s\rangle
  -\langle A_r\rangle\langle A_s\rangle,
  \\[1.8ex]
  \langle A_rA_sA_t\rangle_{\con}&=&\langle A_rA_sA_t\rangle
  -\langle A_r\rangle\langle A_sA_t\rangle
  -\langle A_s\rangle\langle A_tA_r\rangle
  -\langle A_t\rangle\langle A_rA_s\rangle
  \nonumber\\[1.0ex]
  &&+2\kern1pt\langle A_r\rangle\langle A_s\rangle\langle A_t\rangle.
\end{eqnarray}
For an arbitrary number $n$ of observables,
the relation between the connected and the
full correlation functions is given by 
the moment-cumulant transformation
(see Section 4.1.4).

Suppose now that a representative ensemble of $N$ statistically 
independent
gauge-field configuration has been generated in the course of 
a simulation. The evaluation of the primary observable $A_r$
on these configurations yields a series
\begin{equation}
  a_{r,1},a_{r,2},\ldots,a_{r,N}
  \label{DataSeries}
\end{equation}
of ``measured'' values of this observable, whose average
\begin{equation}
  \abar_r={1\over N}\sum_{i=1}^N a_{r,i}
\end{equation}
provides a stochastic estimate of its expectation value
$\langle A_r\rangle$.

If $a_{r,i}$ and $a_{s,i}$, $i=1,\ldots,N$, are the data series 
obtained for the observables $A_r$ and $A_s$, the data series 
for the product $A_rA_s$ is $a_{r,i}a_{s,i}$. 
The product is actually just another primary observable,
even in the case of the stochastic observables
considered in Section 3.2 if the random sources
are included as additional fields
in the ensemble of fields generated by the simulation.
A stochastic estimate of the $n$-point correlation function
$\langle A_{r_1}\ldots A_{r_n}\rangle$ is thus obtained
by calculating the average of
$a_{r_1,i}\ldots a_{r_n,i}$.

\subsection{Simulation statistics}

As previously noted in Section 2.2.4, the question by how much 
$\abar_r$ deviates from $\langle A_r\rangle$ can only
be answered statistically when the simulation is
repeated many times. 
In the following, the average over infinitely many simulations 
of any function $\phi$
of the measured values of the primary observables
is denoted by $\avgavg{\phi}$. 
The standard deviation of $\abar_r$ from $\langle A_r\rangle$, for
example, is given by
$\avgavg{(\abar_r-\langle A_r\rangle)^2}^{1/2}$.

Simulation statistics becomes a useful tool when
the following two assumptions are made.
First it must be guaranteed that the measurements of 
the primary observables are unbiased, i.e.~that
the equality
\begin{equation}
  \avgavg{a_{r,i}}=\langle A_r\rangle
  \label{AvgA}
\end{equation}
holds for all $i$ and all primary observables $A_r$.
The second assumption is that the
fields generated by the simulations are statistically independent.
In particular, 
\begin{equation}
  \avgavg{a_{r,i}a_{s,j}}=\avgavg{a_{r,i}}\avgavg{a_{s,j}}
  \quad\hbox{if}\quad i\neq j.
  \label{StatIndep}
\end{equation}
More generally, the average $\avgavg{a_{r_1,i_1}\ldots a_{r_n,i_n}}$
factorizes into a product of averages, one for each
subset of the factors $a_{r,i}$ with a given value of the 
configuration index $i$. 
As discussed in Section 2.2.4,
both conditions are fulfilled if 
the residual autocorrelations among the fields
in the representative ensembles generated by the simulation
algorithm are negligible.

In practice the statistical independence of
the measured values should be carefully checked
by computing the associated autocorrelation 
functions. However, since the latter is estimated from
the data and is therefore subject to statistical fluctuations, 
a reliable determination
of the autocorrelation times can sometimes be difficult,
particularly so if the available data series is not very long.
The issue of ``calculating the error of the error'' 
was addressed by \shortciteN{MadrasSokal} and in greater detail 
again by \shortciteN{WolffErrors}
(see also
\shortciteN{DDHMC}, Appendix E, for some additional
material).

\subsection{Distribution of the mean values}

The mean values $\abar_r,\abar_s,\ldots$ of the measured values of 
the primary observables are correlated to some extent,
because the underlying ensemble of gauge-field configurations is the same.
It is possible to work out the correlations 
\begin{equation}
  \avgavg{\abar_{r_1}\ldots\abar_{r_k}}=
  {1\over N^k}\sum_{i_1=1}^N\ldots\sum_{i_k=1}^N
  \avgavg{a_{r_1,i_1}\ldots a_{r_k,i_k}}
  \label{AvgCorrelation}
\end{equation}
in terms of
the correlation functions $\langle A_{r_1}\ldots A_{r_n}\rangle$.
In the case of the two-point correlation functions, for example, 
one obtains
\begin{equation}
  \avgavg{\abar_r\abar_s}=
  {1\over N^2}\sum_{i,j=1}^N\avgavg{a_{r,i}a_{s,j}}
  =\langle A_r\rangle\langle A_s\rangle+
  {1\over N}\langle A_rA_s\rangle_{\con},
  \label{TwoPointAvg}
\end{equation}
where use was made of eqns~(\ref{AvgA}) and (\ref{StatIndep}).

For any $k\geq1$, the analogous expression for the
correlation function (\ref{AvgCorrelation}) reads
\begin{equation}
  \avgavg{\abar_{r_1}\ldots\abar_{r_k}}=
  \sum_{l=1}^k{1\over N^{k-l}l!}
  \sum_{P\in{\mathcal P}_{k,l}}
  \langle A_{P_1}\rangle_{\con}\ldots \langle A_{P_l}\rangle_{\con}.
  \label{DataCorrelationI}
\end{equation}
The second sum in this formula runs over the set
${\mathcal P}_{k,l}$ of all partitions $P=(P_1,\ldots,P_l)$
of the set $\{1,\ldots,k\}$ into $l$ non-empty subsets. Furthermore, 
\begin{equation}
  A_{P_i}=\prod_{j\in P_i}A_{r_j}.
\end{equation}
Note that the subsets $P_1,\ldots,P_l$ are ordered and therefore 
distinguished. In particular, at large $N$ the dominant term is
\begin{equation}
  \avgavg{\abar_{r_1}\ldots\abar_{r_k}}=
  \langle A_{r_1}\rangle\ldots \langle A_{r_k}\rangle+\rmO(N^{-1}).
\end{equation}
The proof of eqn~(\ref{DataCorrelationI}) is a bit technical and
is deferred to Section 4.1.4.

The statistical properties of the deviations
\begin{equation}
  \delta\abar_r=\abar_r-\langle A_r\rangle
\end{equation}
can now be determined as follows.
First note that eqn~(\ref{TwoPointAvg}) may be rewritten 
in the form
\begin{equation}
  \avgavg{\delta\abar_r\delta\abar_s}={1\over N}
  \langle A_rA_s\rangle_{\con}.
\end{equation}
On average, the magnitude of the deviations $\delta\abar_r$ is 
thus proportional to $N^{-1/2}$.

A more detailed characterization of the distribution of the 
deviations is obtained by working out their higher-order 
correlations.
Starting from eqn~(\ref{DataCorrelationI}),
it is possible to show (Section 4.1.4) that 
\begin{equation}
  \avgavg{\delta\abar_{r_1}\ldots\delta\abar_{r_k}}=
  \sum_{l=1}^k{1\over N^{k-l}l!}
  \sum_{P\in\tilde{\mathcal P}_{k,l}}
  \langle A_{P_1}\rangle_{\con}\ldots \langle A_{P_l}\rangle_{\con},
  \label{DataCorrelationII}
\end{equation}
where $\tilde{\mathcal P}_{k,l}\subset{\mathcal P}_{k,l}$
denotes the set of partitions of $\{1,\ldots,k\}$ into 
$l$ subsets with two or more elements.
In particular, for all even $k$ 
\begin{equation}
  \avgavg{\delta\abar_{r_1}\ldots\delta\abar_{r_k}}=
  {1\over N^{k/2}}
  \left\{
  \langle A_{r_1}A_{r_2}\rangle_{\con}\ldots
  \langle A_{r_{k-1}}A_{r_k}\rangle_{\con}+\hbox{permutations}\right\}
  +\ldots,
\end{equation}
while for all odd $k$
the leading terms are of order $N^{-(k+1)/2}$, because
the admissible partitions contain at least one subset with 3 or more
elements. Taken together, these results show that the joint
probability distribution of the scaled deviations $\sqrt{N}\delta\abar_r$
is gaussian to leading order in $1/N$, with mean zero and variance 
$\langle A_rA_s\rangle_{\con}$.

\subsection{Appendix: Proof of eqns~(\ref{DataCorrelationI}) and 
            (\ref{DataCorrelationII})}

Let $J_r$ be real-valued sources for the selected primary 
observables $A_r$. The generating function of
the correlation functions $\langle A_{r_1}\ldots A_{r_n}\rangle$
is a formal power series 
\begin{equation}
  \Zfun(J)=1+\sum_{n=1}^{\infty}{1\over n!}
  \sum_{r_1}\ldots\sum_{r_n}
  \langle A_{r_1}\ldots A_{r_n}\rangle J_{r_1}\ldots J_{r_n}
\end{equation}
in these sources. Similarly,
the generating function
of the connected parts of the correlation functions is
given by
\begin{equation}
  \Wfun(J)=\sum_{n=1}^{\infty}{1\over n!}
  \sum_{r_1}\ldots\sum_{r_n}
  \langle A_{r_1}\ldots A_{r_n}\rangle_{\con} J_{r_1}\ldots J_{r_n}.
\end{equation}
The moment-cumulant transformation is then summarized by
the identity
\begin{equation}
  \Zfun(J)=\rme^{\Wfun(J)}
  \label{MomentCumulant}
\end{equation}
among formal power series.

The left-hand side of eqn~(\ref{DataCorrelationI}) is
related to the generating function $\Zfun(J)$ through
\begin{eqnarray}
  \avgavg{\abar_{r_1}\ldots\abar_{r_k}}&=&
  {1\over N^k}
  \left.{\partial^k\avgavg{\exp\{\ssum{r,i}a_{r,i}J_r\}}
  \over\partial J_{r_1}\ldots\partial J_{r_k}}
  \right|_{J=0}
  \nonumber\\[1.8ex]
  &=&
  {1\over N^k}\left.{\partial^k\Zfun(J)^N
  \over\partial J_{r_1}\ldots\partial J_{r_k}}
  \right|_{J=0}.
  \label{DataCorrelationZJ}
\end{eqnarray}
Use has here been made of the statistical independence of the data $a_{r,i}$
at different values of the index $i$ and of the fact that
their products at fixed $i$ 
are unbiased estimators of the correlation functions
of the primary observables. 
The insertion of eqn~(\ref{MomentCumulant}) and the subsequent
expansion of the exponential function $\rme^{\Wfun(J)}$ now
leads to the formula
\begin{equation}
  \avgavg{\abar_{r_1}\ldots\abar_{r_k}}=
  \sum_{l=1}^k{1\over N^{k-l}l!}
  \left.{\partial^k\Wfun(J)^l
  \over\partial J_{r_1}\ldots\partial J_{r_k}}
  \right|_{J=0}.
  \label{DataCorrelationWJ}
\end{equation}
Each derivative in this expression
acts on the $l$ factors $\Wfun(J)\ldots\Wfun(J)$ one by one.
It is then not difficult to convince oneself that the possible
distributions of the derivatives to the factors match
the possible partitions $P\in{\mathcal P}_{k,l}$, thus proving 
eqn~(\ref{DataCorrelationI}).

The proof of eqn~(\ref{DataCorrelationII}) proceeds in the same
way. In order to obtain the correlation functions of the
deviations $\delta\abar_r$, it suffices
to substitute $a_{r,i}\to a_{r,i}-\langle A_r\rangle$ 
on the right of eqn~(\ref{DataCorrelationZJ}).
The generating function 
in eqn~(\ref{DataCorrelationWJ}) 
then gets replaced by
\begin{equation}
  \tilde{\Wfun}(J)=\Wfun(J)-\sum_r\langle A_r\rangle J_r.
\end{equation}
Since at least two derivatives must act on each factor 
$\tilde{\Wfun}(J)$, the possible distributions of the derivatives
to the factors
now match the partitions $P\in\tilde{\mathcal P}_{k,l}$.

\section{Physical quantities}

The correlation functions of the primary observables only rarely
have an immediate physical meaning. In the present
context, any well-defined function of the expectation values
$\langle A_r\rangle,\langle A_rA_s\rangle,\ldots$
is referred to as a physical quantity. Ratios of
the expectation values of Wilson loops, for example, are
considered to be physical quantities as well as the heavy-quark
potential, which is a limit of such ratios.

\subsection{From the primary observables to the physical quantities}

In practice, the correlation functions
required for the calculation of a physical quantity
are approximated by the averages of the measured values of
the appropriate primary observables. 
A prototype computation
of the pion mass $\mpi$, for example, starts from the data series 
for the observables $\Obs_{\pi}(x_0,y)$
at all times $x_0$ and, say, a single source point $y$
(cf.~Section 3.1.1). The so-called effective mass
\begin{equation}
  \meff(x_0)=-a^{-1}\ln{\avg{\Obs}_{\pi}(x_0+a,y)\over
                        \avg{\Obs}_{\pi}(x_0,y)}
  \label{EffectiveMass}
\end{equation}
is then computed and fitted by a constant in a sensible range of $x_0$,
where the contributions of the higher-energy states to the pion propagator
are negligible with respect to the statistical errors. 
As long as this condition is satisfied, 
the fitted constant provides
a stochastic estimate of the pion mass on the given lattice.
 
Once a definite fit procedure is adopted, the so calculated
values of the pion mass become a complicated
but unambiguously defined function of the data series 
for the observables $\Obs_{\pi}(x_0,y)$. It should
be noted, however, that
this function is not simply a function 
of the average values of these observables.
The fit also requires an estimate of the covariance matrix
of the latter as input, which one obtains from 
the available data through the jackknife method or in some
other way. Evidently, when fits of fitted quantities are considered,
as may be the case in studies of the 
quark-mass dependence of the pion mass, the functional dependence
of the calculated quantities on the primary data series
is further obscured.

\subsection{Stochastic estimators}

A stochastic estimator of a physical quantity $Q$
is any function $\phi$ of $N$ and the measured values 
of the primary observables $A_r$ such that
\begin{equation}
   Q=\lim_{N\to\infty}\phi\quad\hbox{with probability 1.}
\end{equation}
In the following, a class of stochastic estimators
with some additional properties will be considered. 
More precisely, it will be assumed that the
asymptotic expansion
\begin{equation}
   \phi\mathrel{\mathop\sim_{N\to\infty}}\sum_{k=0}^{\infty}
   N^{-k}\phi^{(k)}(\abar_{r_1},\abar_{r_2},\ldots)
   \label{PhiExpansion}
\end{equation}
holds, where the coefficients $\phi^{(k)}$ 
are smooth functions of their arguments. The number of arguments
may grow with $k$ but is required to be finite
for all $k$.

It may not be obvious at this point
why one needs to consider stochastic estimators
with an explicit dependence on $N$.
The leading coefficient 
in the expansion (\ref{PhiExpansion}) is in fact a valid
stochastic estimator for $Q$ which depends only on the
averages $\abar_r$ of the primary data series.
However, when a stochastic estimator is implicitly defined
through a fit procedure, for example,
its leading coefficient at large $N$ may be
inaccessible in practice.
In these cases one does not really have the choice and is 
forced to deal with the operationally well-defined
but $N$-dependent estimators.

A simple example of an admissible stochastic estimator is
provided by the effective mass (\ref{EffectiveMass}).
The r\^ole of the
physical quantity $Q$ is here played by 
the effective mass calculated from the
exact pion propagator $\gpi(x_0-y_0)$.
More complicated, $N$-dependent stochastic estimators 
will be discussed in Section 4.3.

\subsection{Bias and covariance matrix}

Let $Q_{\alpha}$ be some physical quantities, labeled by an index $\alpha$,
and $\phi_{\alpha}$ stochastic estimators of these. 
On average, the values of the estimators obtained in a simulation 
approximate the physical quantities up to a deviation
given by
\begin{equation}
   B_{\alpha}=\avgavg{\delta\phi_{\alpha}},
   \qquad
   \delta\phi_{\alpha}=\phi_{\alpha}-Q_{\alpha}.
   \label{Bias}
\end{equation}
$B_{\alpha}$ is referred to as the bias of the chosen estimators.
The statistical fluctuations of the measured values
are described by the covariance matrix
\begin{equation}
   C_{\alpha\beta}=\avgavg{\delta\phi_{\alpha}\delta\phi_{\beta}}
   \label{Covariance}
\end{equation}
and are usually significantly larger than the bias.

For $N\to\infty$, the bias and the covariance matrix can be expanded
in a series in inverse powers of $N$ with coefficients depending
on the correlation functions 
$\langle A_{r_1}\ldots A_{r_n}\rangle$ of the primary observables.
To show this, first recall that
\begin{equation}
   \abar_r=\langle A_r\rangle+\delta\abar_r,
   \qquad
   \delta\abar_r=\rmO(N^{-1/2}).
\end{equation}
The Taylor expansion
of the coefficient functions in eqn~(\ref{PhiExpansion})
in powers of the deviations $\delta\abar_r$
then leads to the expression
\begin{eqnarray}
  \phi_{\alpha}&=&
  \phihat_{\alpha}^{(0)}+\sum_r\partial_r\phihat_{\alpha}^{(0)}\delta\abar_r
  \nonumber\\[1.8ex]
  &&+{1\over N}\phihat_{\alpha}^{(1)}
  +\frac{1}{2}\sum_{r,s}\partial_r\partial_s\phihat_{\alpha}^{(0)}
  \delta\abar_r\delta\abar_s+\rmO(N^{-3/2})
  \label{PhiTaylorExpansion}
\end{eqnarray}
in which 
\begin{equation}
   \phihat^{(k)}_{\alpha}=
   \phi^{(k)}_{\alpha}
   (\langle A_{r_1}\rangle,\langle A_{r_2}\rangle,\ldots),
   \qquad
   \partial_r={\partial\over\partial\langle A_r\rangle}.
\end{equation}
Evidently, since the leading term $\phihat^{(0)}_{\alpha}$ coincides 
with $Q_{\alpha}$,
the deviation $\delta\phi_{\alpha}$ is given by
the sum of all other terms on the right of 
eqn~(\ref{PhiTaylorExpansion}).

It is important to realize that the only stochastic variables in
the expansion (\ref{PhiTaylorExpansion}) are the deviations 
$\delta\abar_r$. The averages (\ref{Bias}) and (\ref{Covariance})
over repeated simulations can therefore be computed 
straightforwardly using the results
obtained in Section 4.1.
In a few lines one then obtains the result
\begin{eqnarray}
  B_{\alpha}&=&
  {1\over N}\Bigl\{\phihat^{(1)}_{\alpha}+
  \frac{1}{2}\sum_{r,s}\partial_r\partial_s\phihat_{\alpha}^{(0)}
  \langle A_rA_s\rangle_{\con}\Bigr\}+\rmO(N^{-2}),
  \label{BiasExpansion}
  \\[1.8ex]
  C_{\alpha\beta}&=&
  {1\over N}\sum_{r,s}\partial_r\phihat_{\alpha}^{(0)} 
  \partial_s\phihat_{\beta}^{(0)}
  \langle A_rA_s\rangle_{\con}+\rmO(N^{-2}).
  \label{CovarianceExpansion}
\end{eqnarray}
In particular, the bias of the stochastic estimators 
is of order $N^{-1}$ while their statistical 
fluctuations are of order $N^{-1/2}$.

\section{Jackknife error estimation}

The bias $B_{\alpha}$ and the covariance matrix
$C_{\alpha\beta}$ usually need to be estimated from the 
data. Such estimates can in principle be 
obtained via eqns~(\ref{BiasExpansion}) and (\ref{CovarianceExpansion})
by calculating the expectation values and two-point functions
of the relevant primary observables. The coefficient functions
$\phi^{(0)}$ and $\phi^{(1)}$ 
must be known explicitly in this case.

The jackknife method allows the bias and covariance matrix to be
estimated even if the coefficient functions are inaccessible
or too complicated to be used directly.
More precisely, the method constructs stochastic estimators
for the large-$N$ limits of $NB_{\alpha}$ and $NC_{\alpha\beta}$
(which are functions of the expectation values of the primary observables
and therefore physical quantities).

\subsection{Jackknife samples}

A jackknife sample of the measured values $a_{r,1},\ldots,a_{r,N}$ 
of a primary observable $A_r$ is obtained by omitting one measurement
from the full series. If, say, the $i$'th measurement is omitted,
the corresponding jackknife sample consists of the measurements
\begin{equation}
  a_{r,1},\ldots,a_{r,i-1},a_{r,i+1},\ldots a_{r,N}.
\end{equation}
Evidently, there are $N$ distinct jackknife samples of 
$N-1$ measurements,
labeled by the number $i$ of the omitted measurement.

The average of the measurements included in the jackknife sample
number $i$ is denoted by
\begin{equation}
  \abar^J_{r,i}={1\over N-1}\sum_{j=1,j\neq i}^N a_{r,j}.
\end{equation}
More generally, a stochastic estimator $\phi$ 
assumes some value $\phi^J_i$ if the $i$'th measurement
of the primary observables is discarded. 
Note that the jackknife samples are treated like any
other measurement series of length $N-1$ in this context.
From the expansion (\ref{PhiExpansion})
one then infers that
\begin{equation}
  \phi^J_{i}
  \mathrel{\mathop\sim_{N\to\infty}}\sum_{k=0}^{\infty}
  (N-1)^{-k}\phi^{(k)}(\abar^J_{r_1,i},\abar^J_{r_2,i},\ldots),
  \label{PhiJExpansion}
\end{equation}
where the coefficient functions $\phi^{(k)}$ are the same as before.

\subsection{Estimators for the bias and the covariance matrix}

The jackknife estimators for the 
bias and the covariance matrix are now given by the 
elegant formulae
\begin{eqnarray}
  B^J_{\alpha}&=&
  \sum_{i=1}^N(\phi^J_{\alpha,i}-\phi_{\alpha}),
  \label{JackBias}
  \\[1.8ex]
  C^J_{\alpha\beta}&=&
  \sum_{i=1}^N(\phi^J_{\alpha,i}-\phi_{\alpha})
              (\phi^J_{\beta,i}-\phi_{\beta}).
  \label{JackCovariance}
\end{eqnarray}
An important result of the discussion below is going to be that 
the scaled expressions
$NB^J_{\alpha}$ and $NC^J_{\alpha\beta}$ are 
stochastic estimators 
in the sense of Section 4.2.2, the associated
physical quantities being
\begin{eqnarray}
  \lim_{N\to\infty}NB_{\alpha}&=&
  \phihat^{(1)}_{\alpha}+
  \frac{1}{2}\sum_{r,s}\partial_r\partial_s\phihat_{\alpha}^{(0)}
  \langle A_rA_s\rangle_{\con},
  \label{LargeNB}
  \\[1.8ex]
  \lim_{N\to\infty}NC_{\alpha\beta}&=&
  \sum_{r,s}\partial_r\phihat_{\alpha}^{(0)} 
  \partial_s\phihat_{\beta}^{(0)}
  \langle A_rA_s\rangle_{\con}.
  \label{LargeNC}
\end{eqnarray}
The bias $B_{\alpha}$ and the covariance matrix
$C_{\alpha\beta}$ are therefore approximated by the estimators
$B^J_{\alpha}$ and $C^J_{\alpha\beta}$ 
up to statistical fluctuations
of order $N^{-3/2}$ and further terms of order $N^{-2}$.

The expansion of the jackknife estimators
(\ref{JackBias}) and (\ref{JackCovariance})
for $N\to\infty$
is obtained straightforwardly by substituting
\begin{equation}
  \abar^J_{r,i}=\abar_r+{1\over N-1}(\abar_r-a_{r,i})
\end{equation}
in eqn~(\ref{PhiJExpansion}) and by systematically expanding all terms
in powers of $N^{-1}$.
In the case of the covariance matrix, for example, 
this leads to the expression
\begin{equation}
  C^J_{\alpha\beta}={1\over N}\sum_{r,s}
  \bar{\partial}_r\phi^{(0)}_{\alpha}(\abar_{r_1},\ldots)
  \bar{\partial}_s\phi^{(0)}_{\beta}(\abar_{r_1},\ldots)
  (\abar_{rs}-\abar_r\abar_s)+\rmO(N^{-2}),
  \label{CJExpansion}
\end{equation}
where the notation has been simplified by setting
\begin{equation}
  \abar_{rs}={1\over N}\sum_{i=1}^N a_{r,i}a_{s,i},
  \qquad
  \bar{\partial}_r={\partial\over\partial\abar_r}.
\end{equation}
The higher-order terms in eqn~(\ref{CJExpansion})
involve averages of increasingly longer products of the measured
values $a_{r,i}$, but are otherwise of the same structure as 
the leading one. Since the corresponding products of the 
primary observables are primary observables too,
the expansion shows that $NC^J_{\alpha\beta}$ is 
a stochastic estimator of the kind specified in Section 4.2.2. 
Moreover, proceeding as in Section 4.2.3, the physical quantity
approximated by $NC^J_{\alpha\beta}$ is seen to coincide
with the expression on the right of eqn~(\ref{LargeNC}),
as asserted above.

An interesting feature of the jackknife method 
is the fact that it does not require 
the derivatives of the coefficient function
$\phi^{(0)}$ in the asymptotic expressions
(\ref{LargeNB}) and (\ref{LargeNC}) to be calculated.
As is evident from the derivation of eqn~(\ref{CJExpansion}),
the method effectively performs a numerical differentiation by
probing the stochastic estimators $\phi_{\alpha}$
through the jackknife samples. 
The estimators are thus implicitly assumed to vary 
smoothly on the scale
of the deviations $\abar^J_{r,i}-\abar_{r}=\rmO(N^{-1})$.

\subsection{Error propagation}

The jackknife method is easy to apply, because it requires no more
than the evaluation of the stochastic estimators of interest
for the full sample and the jackknife samples of the measured values of the
primary observables.
Note that 
$\psi^J_i=\psi(\phi^J_{1,i},\ldots,\phi^J_{m,i})$
if $\psi$ is a function of the stochastic estimators
$\phi_1,\ldots,\phi_m$.
The calculation can therefore often be organized
in a hierarchical manner, where one proceeds from 
the primary observables to more and more complicated stochastic
estimators.

In the case of the computation of the pion mass sketched in
Section 4.2.1, for example, 
the estimation of the statistical errors
starts from the jackknife averages $\avg{\Obs}_{\pi}(x_0,y)^J_i$
of the primary observables $\Obs_{\pi}(x_0,y)$. The 
bias and covariance matrix of the effective mass are then
obtained from the jackknife sample values
\begin{equation}
  \meff(x_0)^J_i=-a^{-1}\ln{\avg{\Obs}_{\pi}(x_0+a,y)^J_i\over
                            \avg{\Obs}_{\pi}(x_0,y)^J_i}
\end{equation}
using eqns~(\ref{JackBias}) and (\ref{JackCovariance}).
In particular, the jackknife estimator of the covariance matrix
is given by
\begin{equation}
  C^J_{x_0x_0'}=\sum_{i=1}^N\{\meff(x_0)^J_i-\meff(x_0)\}
                            \{\meff(x_0')^J_i-\meff(x_0')\}.
\end{equation}
The statistical error of $\meff(x_0)$, for example, 
is estimated to be $(C^J_{x_0x_0})^{1/2}$.

In the next step, the pion mass $\mpi$ is computed by minimizing
the $\chi^2$-statistic
\begin{equation}
  \chi^2=\sum_{x_0=t_0}^{t_1}\sum_{x_0'=t_0}^{t_1}
  \{\mpi-\meff(x_0)\}[(C^J)^{-1}]_{x_0x_0'}\{\mpi-\meff(x_0')\}
\end{equation}
in some range $[t_0,t_1]$ of time. 
The pion mass
\begin{equation}
  \mpi={\sum_{x_0,x_0'=t_0}^{t_1}
  [(C^J)^{-1}]_{x_0x_0'}\meff(x_0')\over 
  \sum_{x_0,x_0'=t_0}^{t_1}
  [(C^J)^{-1}]_{x_0x_0'}}
\end{equation}
thus calculated
is algebraically expressed through the effective mass 
and the scaled jackknife estimator of the associated
covariance matrix. In particular, since all of these are
stochastic estimators of the kind specified in Section 4.2.2,
the same is the case for the pion mass determined in this way.
Its statistical error can therefore be computed using
the jackknife method.

Following the general rules, the application of the jackknife method
requires the pion mass to be recomputed for each jackknife sample of 
the primary data. In particular, the covariance matrix of the 
effective mass must be recomputed, which requires 
the jackknife samples of the jackknife samples to be considered,
i.e.~samples where a second measurement is discarded from the 
full data series. The computational effort thus grows like $N^2$,
but the calculation is otherwise entirely straightforward.

In practice a simplified procedure is often adopted, where
the statistical error of the covariance matrix
is ignored. The matrix is
calculated for the full sample in this case and is held fixed
when the errors of the effective mass are propagated
to those of the pion mass. An advantage of this procedure is
that the computational effort increases like $N$ rather than $N^2$,
but the correctness of the error estimation
can only be shown for asymptotically large values of $N$ and if
the systematic deviation of the effective mass from being constant
in time is much smaller than the statistical errors.
Moreover, the bias of the pion mass is no longer correctly given
by the jackknife formula (\ref{JackBias}). It is therefore advisable
to pass to the simplified procedure only after having checked
its consistency with the results of the full procedure.


\bibliographystyle{OUPnamed_notitle}
\bibliography{refs}

\end{document}